\documentclass[useAMS,usenatbib]{mn2e}

\usepackage{epsfig}
\usepackage{graphicx,color}
\usepackage{times}
\usepackage{natbib}
\usepackage{setspace}
\usepackage[usenames,dvipsnames]{xcolor}
\newif\ifAMStwofonts
\AMStwofontstrue

\newcommand{\be}{\begin{equation}}
\newcommand{\ee}{\end{equation}}
\newcommand{\ba}{\begin{eqnarray}}
\newcommand{\ea}{\end{eqnarray}}
\newcommand{\brr}{\begin{array}}
\newcommand{\err}{\end{array}}
\newcommand{\bc}{\begin{center}}
\newcommand{\ec}{\end{center}}

\newcommand{\msun}{M$_\odot$}

\newcommand{\surf}{M$_\odot$ pc$^{-2}$}

\newcommand{\muppi}{{\it MUPPI}}
\newcommand{\sffb}{SF\&FB}

\newcommand{\mincir}{\raise
  -2.truept\hbox{\rlap{\hbox{$\sim$}}\raise5.truept \hbox{$<$}\ }}
\newcommand{\magcir}{\raise
  -2.truept\hbox{\rlap{\hbox{$\sim$}}\raise5.truept \hbox{$>$}\ }}
\newcommand{\siml}{\raise
  -2.truept\hbox{\rlap{\hbox{$\sim$}}\raise5.truept \hbox{$<$}\ }}
\newcommand{\simg}{\raise
  -2.truept\hbox{\rlap{\hbox{$\sim$}}\raise5.truept \hbox{$>$}\ }}

\title{Simulating realistic disk galaxies with a novel sub-resolution ISM model}

\author[Murante et al.] {
Giuseppe  Murante$^1$, 
Pierluigi Monaco$^{2,1}$, 
Stefano Borgani$^{2,1,3}$,
Luca Tornatore$^1$,
\and
Klaus Dolag$^4$,
David Goz$^2$\\
$^1$ INAF - Osservatorio Astronomico di Trieste, Via Tiepolo 11,
I-34131 Trieste (Italy) (murante, tornatore, goz @oats.inaf.it)\\
$^2$ Dipartimento di Fisica,  Universit\`a di Trieste, via
Tiepolo 11, I- 34131 Trieste, Italy (borgani, monaco@oats.inaf.it)\\ 
$^3$ INFN, Istituto Nazionale di Fisica Nuclare, Trieste (Italy) \\ 
$^4$ University Observatory M\"unchen, Scheinerstr. 1,
81679,M\"unchen, Germany (dolag@usm.uni-muenchen.de)
}

\begin{document}

\maketitle

\label{firstpage}

\begin{abstract}
  We present results of cosmological simulations of disk galaxies
  carried out with the GADGET-3 TreePM+SPH code, where star formation
  and stellar feedback are described using our MUlti Phase Particle
  Integrator (\muppi) model. This description is based on a simple
  multi-phase model of the interstellar medium at unresolved scales,
  where mass and energy flows among the components are explicitly
  followed by solving a system of ordinary differential equations.
  Thermal energy from SNe is injected into the local hot phase, so as
  to avoid that it is promptly radiated away.  A kinetic feedback
  prescription generates the massive outflows needed to avoid the
  over-production of stars.  We use two sets of zoomed-in initial
  conditions of isolated cosmological halos with masses $(2-3)\cdot
  10^{12}$ {\msun}, both available at several resolution levels.  In
  all cases we obtain spiral galaxies with small bulge-over-total
  stellar mass ratios ($B/T\sim 0.2$), extended stellar and gas disks,
  flat rotation curves and realistic values of stellar masses. Gas
  profiles are relatively flat, molecular gas is found to dominate at
  the centre of galaxies, with star formation rates following the
  observed Schmidt-Kennicutt relation. Stars kinematically belonging
  to the bulge form early, while disk stars show a clear inside-out
  formation pattern and mostly form after redshift $z=2$.  However,
  the baryon conversion efficiencies in our simulations differ from
  the relation given by \cite{Moster10} at a $3\sigma$ level, thus
  indicating that our stellar disks are still too massive for the Dark
  Matter halo in which they reside.  Results are found to be
  remarkably stable against resolution. This further demonstrates the
  feasibility of carrying out simulations producing a realistic
  population of galaxies within representative cosmological volumes,
  at a relatively modest resolution.
\end{abstract}


\section{Introduction}
\label{section:introduction}

The study of galaxy formation in a cosmological framework represents
since more than two decades a challenge for hydrodynamic simulations
aimed at describing the evolution of cosmic structures. This is
especially true when addressing the problem of formation of disk
galaxies \citep[e.g.][for a review; a short updated review of
past and present efforts in this field is presented it in Section
\ref{section:review}]{MayerGovernato}.

The recent history of numerical studies of the formation of galaxies
demonstrated that the most crucial ingredient for a successful
simulation of a disk-dominated galaxy is proper modelling of star
formation and stellar feedback (hereafter {\sffb}).  This history can
be schematically divided into three phases.  In a first pioneeristic
phase the simplest models of {\sffb}, based on a Schmidt-like law for
star formation and supernova (SN) feedback in the form of thermal
energy resulted in a cooling catastrophe, with too many baryons
condensing into galaxy and most angular momentum being lost.  Galaxy
disks, when present, were very compact and with exceedingly high
rotation velocities.  Kinetic feedback was found to improve the
results by producing outflows and reducing overcooling.

During the first decade of 2000, much emphasis was given to the
solution of the angular momentum problem.  It was fully recognised
that structure in the Inter-Stellar Medium (ISM) below the resolution
limit achievable in cosmological simulations was crucial in
determining the efficiency with which SN energy is able to heat the
surrounding gas and produce massive outflows, in the form of fountains
or prominent galaxy winds.

In this paper we will adopt a conservative definition of
``sub-resolution physics'': every process that is not explicitly
resolved in a simulation implementing only fundamental laws of physics
is defined to be sub-resolution.  We make no difference between models
that combine in a simple way the resolved hydrodynamical properties of
gas particles and models that provide an explicit treatment of the
unresolved structure of the ISM.  In this sense, any star-formation
prescription is sub-resolution, since the formation of a single star
is not explicitly resolved. This also applies to any form of energetic
feedback, as long as individual SN blasts are not resolved since their
free expansion phase (or as long as radiative transfer of light from
massive stars is not tracked), and to chemical evolution of gas and
stars.

Several approaches to model the sub-resolution behaviour of gas were
proposed (see Section~\ref{section:review}). These approaches provided
significant improvements in the description of disk galaxy
formation. In spite of these improvements, the {\it Aquila} comparison
project showed that, when running many codes on the same set of
initial conditions of an isolated ``Milky Way'' sized halo, no code
was able to produce a completely realistic spiral galaxy with low
bulge-over-total ($B/T$) stellar mass ratio and flat rotation
curve. Moreover, different {\sffb} prescriptions gave very different
results.

A third phase is taking place at present: thanks to a careful tuning
of models and the introduction of more refined forms of kinetic and
``early'' stellar feedback, several independent groups are now
succeeding in better regulating overcooling and the loss of angular
momentum.  This was done by using several different approaches to
{\sffb}, that may be broadly grouped into two categories. One approach
is based on reaching the highest numerical resolution affordable with
the present generation of supercomputers, thus resolving higher gas
densities and pushing the need of sub-resolution modelling from
$\la$kpc toward $\la$10 pc scales.  Other groups prefer to improve and
refine their sub-resolution models so as to be able to work at
resolutions in the range from $\sim$kpc to $\sim$100 pc. This latter
approach mainly focuses on the modelling of feedback and, in some case,
on refining the sub--resolution description of the ISM structure.

In this paper we follow the latter approach. In \cite{M10} we proposed
a model for {\sffb} called MUlti Phase Particle Integrator (\muppi).
While being inspired to the analytic model of \cite{Monaco04a}, it has
several points of contact with the effective model of
\cite{SpringHern03}. In {\muppi}, each gas particle eligible to host
star formation is treated as a multi-phase portion of the ISM, made by
cold and hot gas in thermal pressure equilibrium and a stellar
reservoir.  Mass and energy flows among the various phases are
described by a suitable system of ordinary differential equations
(ODE), and no assumption of self-regulation is made. For each gas
particle, the system of ODE is solved at each time step, also taking
into account the effect of the hydrodynamics. SN energy is distributed
to the gas particles surrounding the star forming ones both in thermal
and kinetic form.

In \cite{M10} we tested a first version of {\muppi}, which described
primordial gas composition and thermal feedback only, on isolated
galaxies and rotating halos. We showed that simulations naturally
reproduce the Schmidt-Kennicutt relation, instead of imposing it in
the model \citep[see also][]{Monaco12}, and the main properties of the
ISM.  In the absence of kinetic feedback, the model generated galactic
fountains and weak galactic winds, but strong galactic outflows were
absent.  Our model was included in the Aquila comparison project
\cite{Scannapieco12}, and shared virtues and weaknesses of several
other {\sffb} models tested in that paper.

In this work, we describe an updated version of {\muppi} and test it
on cosmological halos. We implemented a kinetic feedback scheme and
included the description of chemical evolution developed by
\cite{Tornatore07}, with metal-dependent cooling described as in
\cite{Wiersma09a}.  Here we present results of zoomed-in cosmological
simulations of Milky-Way sized DM halos.  We use two sets of initial
conditions, one of which is the same used in the Aquila comparison
project, at different numerical resolutions. As a main result, we will
show that our simulations produce realistic, disk-dominated galaxies,
with flat rotation curves, low galactic baryon fractions and low value
of the bulge-to-total stellar mass ratio ($B/T$), in good agreement
with the Tully-Fisher and the stellar mass - halo mass
relations. Since we use one of the Aquarius halo, which has also been
recently simulated by \cite{Aumer13} and \cite{Marinacci13}, we can
show how, two years after the ``Aquila'' comparison project, simulated
galaxy properties from different groups, using different codes and
{\sffb} algorithms, agree now much better with each other. Finally, we
demonstrate that our simulations shows very good convergence with
resolution. In a companion paper (Goz {\it et al.} 2014) we will
present an analysis of the properties of bars found in the simulations
presented here, while a forthcoming paper (Monaco {\it et al.}, in
preparation) we will present a more detailed discussion of the
implementation of {\sffb} and the optimisation of the choice of the
model parameters.

The plan of the paper is as follows. In Section~\ref{section:review}
we provide an overview of the literature on cosmological simulations
of disk galaxies.  The numerical implementation of the updated
{\muppi} model is described in Section~\ref{section:model}.  The
properties of the simulations presented in the paper are given
in~\ref{section:simulations}. Results are presented and commented in
Section~\ref{section:results}, including a discussion on the effect of
resolution and on numerical
convergence. Section~\ref{section:conclusions} summarises the main
conclusions of our analysis.

\section{Overview of simulations of disk galaxy formation}
\label{section:review}
In this section, we provide a concise review of the results presented
in the literature concerning simulations of disk galaxies in a
cosmological context. As mentioned in the Introduction, past attempts
to simulate disk galaxies can be divided in three distinct phases.

\subsection{Pioneeristic phase}
Starting from a first generations of pioneering analyses
(e.g. \citealt{Evrard88,Hernquist89,HernquistKatz89,BarnesHernquist91,Hiotelis91,KatzGunn91,Katz92,Thomas92,Cen93,Navarro94,Steinmetz95,Mihos96,Walker96,Navarro97,Carraro98,Steinmetz99,SommerLarsen99,Lia2000}),
simulating a realistic spiral galaxy in a cosmological DM halo has
been recognised as a tough problem to solve. The basic reason for this
is that radiative gas cooling at high redshift produces a runaway
condensation in the central parts of newly forming Dark Matter (DM)
halos, the so-called ``cooling catastrophe''
(\citealt{NavarroBenz91}, \citealt{NavarroWhite94}). As a result,
baryonic matter loses orbital angular momentum along with the central
parts of the host DM haloes \citep[see also][]{Elena06}, thereby
producing by $z=0$ galaxies that are too concentrated, compact and
rapidly spinning.

Early attempts to form realistic disk galaxies relied a simple
prescription for forming stars within a hydrodynamical cosmological
simulation (e.g. \citealt{SteinmetzMuller94}, \citealt{Katz96};
hereafter ``simple star formation model'').  In this model, a dense
cold gas particle forms stars at a rate depending on $\beta
\rho_g/t_{dyn}$, the ratio between its density and dynamical time,
times an efficiency $\beta$ that is a free parameter of the model.
This is a three-dimensional relation analogous to the two-dimensional
one of Schmidt and Kennicutt
(\citealt{Schmidt59},\citealt{1998ApJ...498..541K}), with the free
parameter to be chosen so as to reproduce the observational relation,
which is recovered when projecting gas density of a thin rotating
disc.

In this scheme, some form of energetic feedback is needed to regulate
star formation in galaxies.  Since gas circulating inside a DM halo
falls back to the galaxy in a few dynamical times, feedback must be
violent enough to eject gas from the halos. The right amount of energy
is required to allow a fraction of this expelled gas to fall back at
low redshift. At the same time, this feedback needs not to be too
violent when star formation takes place in galaxy discs, with a low
velocity dispersion. Supernovae (SNe) were recognised as the most
plausible candidates as a driver of such feedback.  SNe can supply
energy to surrounding gas particles in two different forms, kinetic
and thermal.  Whenever the internal structure of star-forming
molecular clouds is not resolved, thermal energy feedback is not
efficient. In fact, star formation takes place where gas reaches high
density and, therefore, has short cooling time.  As a consequence,
energy given to the gas surrounding a star-forming region is promptly
radiated away (see e.g. \citealt{Katz92}).  Kinetic energy is much
more resilient to radiative losses, so an implementation of kinetic
feedback can easily produce massive outflows
\citep{NavarroSteinmetz00}.  When some form of kinetic feedback was
used, experiments succeeded in producing realistic disk galaxies, but
failed to produce bulge-less late-type spirals
\citep{Abadi03,Governato04}.

\subsection{The importance of the ISM physics}
\cite{SpringHern03} introduced a new, more refined model for
describing the process of star formation (hereafter ``effective
model'').  They treated gas particles eligible to form stars as a
multi-phase medium, composed by a cold and a hot phase in thermal
pressure equilibrium. The cold gas forms stars at a given
efficiency. This model describes mass and energy flows between the
phases with a system of ODE, with
equilibrium solutions that depend on average density and pressure of
the gas. These equilibrium solutions are used to predict the star formation rate
of a given star-forming gas element.  This effective model has the
following features: {\it (i) } it assumes quiescent, self-regulated
star formation; {\it (ii) } as a consequence, thermal energy from SNe
only establishes the equilibrium temperature of the hot gas phase,
and thus thermal feedback cannot drive massive outflows; {\it (iii) }
a (three-dimensional) Schmidt-like relation is imposed, not obtained
as a result of the model; {\it (iv) } kinetic feedback is implemented
using a phenomenological prescription, that is added to the model;
{\it (v) } in order to guarantee the onset of galactic winds, gas
particles subject to kinetic feedback become non-collisional for some
time, during which they do not interact with the surrounding gas.

This model aims at providing a realistic description of the physical
properties of the ISM at scales well below the numerical resolution
limit.  Such physics is considered to be the cause of two phenomena,
that are necessary ingredients for a successful description of
observed late-type spirals: quenching of early star formation and
expulsion of significant amount of gas mass from the high-redshift DM
halos. Part of the expelled gas must fall back in DM halos at low
redshifts, thus allowing late, quiescent, ongoing star formation. The
inside-out growth of stellar disks is due to this mechanism.

The effective model by \cite{SpringHern03} was used by
\cite{Robertson04} to perform simulations of the formation of a disk
galaxy, and by \cite{Nagamine04} and \cite{Night06} to study the
formation of Lyman-break galaxies in cosmological volumes.
\cite{Okamoto05} used the same model to study various regimes of
feedback for quiescent and starburst star formation, triggered by high
gas densities or strong shocks.  They claimed that the latter trigger
leads to an improvement in the production of extended disks. While
these numerical experiments provided an improvement in the description
of galaxy formation in a cosmological context, they were still not
able to produce a fully realistic late-time spiral galaxy.  Apart from
having too large bulge masses, the fraction of baryons in the
resulting galaxies were still too high when compared to the observed
relation between halo mass and stellar mass relation.

The implementation of kinetic feedback by \cite{SpringHern03}, where a
fraction of the SN energy budget is given to the outflow particle and
wind velocity is assumed to be constant, is usually referred to as
{\it energy-driven} kinetic feedback.  As an alternative,
\cite{Oppen06} proposed an implementation of {\it momentum-driven}
winds where, following \cite{Murray05}, the outflow is driven by
radiation pressure of massive stars more than by SNe.  In this model,
the wind terminal velocity scales with the galaxy circular velocity, a
behaviour supported by observations \citep[e.g.][and references
therein]{Martin05,Oppen06}. In the numerical implementation of
momentum-driven outflows, square root of the gravitational potential
or velocity dispersion of DM particles can be used as proxies of the
galaxy circular velocity
\citep[e.g.][]{Tescari09,Okamoto10,Oser10,Tescari11,Puchwein13}.
Other variants of this models for galactic winds were presented by
\cite{Choi11} and \cite{Barai13}, that also provided detailed comparisons
of the prediction of different outflow models \citep[see
also][]{Schaye10,Hirschmann13a}.

\cite{Gov07} adopted a feedback model previously suggested by
\cite{GerritsenIcke97} and \cite{ThacCouch00}.  In this model SN
thermal energy is assigned to the neighbouring gas particles; these
particles are then not allowed to cool for a given amount of time, so
as to mimic the effect of SNe blast waves. This prescription evolved
into the blast-wave feedback recipe by \cite{Stinson06}. These authors
claimed that, to successfully tackle the angular momentum problem in
disk-galaxy formation, high numerical resolution is needed.

\cite{2007MNRAS.376.1588B} proposed a star formation model in which
molecular clouds form through radiative cooling, and subsequently evolve
ballistically and coagulate whenever colliding. 
SPH was used to describe the ambient hot gas, with the effect of
thermal SNe feedback modelled using solutions of Sedov blasts. They
called their model ``sticky particles'' and showed that it is able to
reproduce a number of observed properties of the ISM in simulations of
isolated disk galaxies. \cite{Kobayashi07} adopted a simple SF model
and pure thermal feedback, but included the effect of hypernovae, that
release ten times the energy of a normal SN-II. They focused on
studying the impact of hypernovae feedback on star formation history
and enrichment of diffuse baryons, but did not provide results on the
morphological properties of their simulated galaxies.

\cite{Schaye08a} pointed out that, if gas in a galaxy disc obeys an
effective equation of state, as in the effective model of
\cite{SpringHern03}, then it obeys a Schmidt-Kennicutt relation. Based
on this, they argued that it is easy to control star formation
without the need of making assumptions about the unresolved
ISM. \cite{Schaye08b} also suggested that outflowing gas particles
should not be hydrodynamically decoupled, thus at variance with
\citealt{SpringHern03}. These prescriptions were used to simulate
cosmological volumes in the GIMIC \citep{Crain09} and in the OWLS
\citep{OWLS} projects. Later on \cite{DallaVecchia12} suggested that
thermal energy should be distributed in a more selective way: imposing
a minimum temperature at which each gas particle must be heated,
cooling times become longer than the sound-crossing time, thereby
allowing heated particles to expand and produce outflows before energy
is radiated away.  

Following \cite{Marri03}, \cite{Scannapieco09} revised the SPH scheme
to prevent overcooling of a hot phase which is spatially coexisting
with cold gas: in this prescription, the search of neighbours of a hot
gas particle is limited to those particles whose entropy is within a
given range of entropy.  They used a simple star formation
prescription: SN thermal energy distributed to {\it
  hot} gas is not immediately radiated away, because of its lower
density.  Cold gas particles cumulate SN energy until they can be {\it
  promoted} to become hot particles. 
They simulated eight DM halos taken from
the Aquarius project \citep{Springel08}, with a mass similar to that
of the Milky Way halo.  None of their simulated galaxies had a disk
stellar mass larger than 20 per cent of the total stellar mass of the
galaxy.  However, they emphasised that the alignment of the angular
momentum of gas accreting on the galaxy is quite important for the
formation of stable disks.

\cite{Ceverino09} studied the role of SNe feedback on the multiphase
ISM by combining high resolution, small scale simulations of the ISM
and cosmological simulations. Their simulations were based on the
Adaptive-Mesh Refinement (AMR) ART code \citep{Kravtsov97}.  They
first carried out parsec-scale simulations of portions of a disk
galaxy, then used them to build a sub-resolution model for \sffb~ in
cosmological simulations. As a result of their analysis, they claimed
that very high resolution is needed in this approach, so that they had
to stop their cosmological simulation at high redshift, $z=3$.
\cite{Colin10} also used the ART code to study the effect of varying
the sub-resolution model parameters on simulated low-mass galaxy
properties. They implemented a simple star formation model, but 
stopped the cooling of gas receiving energy from SNe. They found that
galaxy properties are very sensitive to these parameters: even tuning
them, they were not able to reproduce observed properties of low-mass
galaxies.

Increasing resolution and using a high value for star formation
density threshold in their blast-wave SN feedback model,
\cite{Governato10} succeeded in producing a bulgeless dwarf
galaxy. This galaxy was analysed in detail by \cite{Brook11} and
\cite{Christensen12}. They also added a prescription to estimate the
amount of molecular hydrogen formed in the simulation and linked the
star formation to it. \cite{Christensen14} studied, with the same
prescription, the scaling laws of galactic bulges.

\cite{Stinson10} simulated a set of nine galaxies, with halo masses
ranging from $5 \times 10^{11}$ to $2 \times 10^{12}$ M$_\odot$, with
blast-wave feedback but a lower density threshold for star formation.
They successfully reproduced the Tully-Fisher relation, but reported
that their simulated galaxies still are too centrally
concentrated. Using a similar {\sffb} scheme, \cite{Piontek11}
confirmed that this implementation of feedback is able to alleviate
the angular momentum problem, while varying the numerical mass
resolution over four order of magnitude {\it does not} impact on the
angular momentum loss.

With the ERIS simulation, \cite{Guedes11} successfully produced a
Milky-Way-like galaxy, with an extended disk, a flat rotation curve
and a $B/D$ ratio as low as $B/D=0.35$ in the {\it i}-band.  They used
the same {\sffb} model of \cite{Stinson06}, and obtained this result
by using very high numerical resolution and a high density threshold
for star formation. \cite{Agertz11} used the RAMSES Eulerian AMR code
\citep{RAMSES}, with a simple star formation model. They also turned
off cooling for gas receiving SNe energy. Using a {\it low} star
formation {\it efficiency}, but also a {\it low} density threshold,
they successfully reproduced several observed properties of Milky-Way
like galaxies. However, their circular velocities usually show a large
peak at small radii. This problem is alleviated in their simulations
based on the lowest efficiency SF efficiency and lowest density. We
note that their mass resolution was approximately four times worse
than that of ERIS simulations. \cite{Sales10} carried out simulations
of cosmological volumes, instead of a zoom-in simulation of a single
galaxy, and found that a {\it high} density threshold for SF is
required to produce realistic disks.

Another direction of investigation concerns the description of other
sources of energy feedback, in addition to SNe.  The effect of Active
Galactic Nuclei (AGN) feedback is usually considered not to be very
important in the formation of disk galaxies. However, it was
included by some groups (\citealt{DiMatteo03},
\citealt{Springel05b}, \citealt{Booth09}, \citealt{Hirschmann13b}).
Besides SNe and AGN, cosmic ray pressure could also represent an
important and known source of feedback, that can help in driving
massive galaxy winds.  Attempts to implement cosmic rays feedback in
cosmological simulations were presented, e.g., by
\cite{Jubelgas08}, \cite{Wadepuhl11} and \cite{Uhlig2012}. The latter
two works focused on the relevance of cosmic rays feedback for
satellite and small galaxies.

\subsection{Towards realistic disk galaxy simulations}
While progress was achieved in the ability to produce disk galaxies,
no consensus still emerged neither on the nature of feedback required,
nor on the details of its numerical implementation. In the ``Aquila
comparison project'', \cite{Scannapieco12} presented a comparison
among 12 different Lagrangian and Eulerian codes implementing
different {\sffb} prescriptions.  An earlier version of {\muppi}, not
including chemical evolution and kinetic feedback, also took part of
this comparison project. Nine of such models were implemented in the
same TreePM+SPH code GADGET-3 (non-public evolution of the code
GADGET-2, \citealt{Springel05b}).  The conclusion of the comparison was that
better agreement with observations, both in terms of fraction of halo
mass in the galaxy and in terms of conservation of angular momentum,
was obtained with {\sffb} models that have more effective feedback.
As a general result, all the simulated galaxies tended to be too massive, too compact
and centrally concentrated. Also, the models that are most successful
in producing a flat rotation curve had to resort to such a strong feedback
that the disk component was destroyed or very thick.  The results were
presented at two different resolutions, and the numerical convergence
was generally not particularly good. To cite the conclusion of the
paper by \cite{Scannapieco12} :
``state-of-the-art simulations cannot yet uniquely predict the
properties of the baryonic component of a galaxy, even when the
assembly history of its host halo is fully specified.''

This work triggered a burst of efforts to improve the different models
in the direction of resolving the discrepancies with
observations.  

\cite{McCarthy12} used the same {\sffb} model of \cite{Crain09} and
showed that, in their cosmological run, many disk galaxies with flat
rotation curves and low baryon fractions were present, even if their
resolution was low ($M_{DM} \approx 7 \cdot 10^7$ M$_\odot$).  Their
model was included in the Aquila comparison project, and indeed their
results were among the best, although with rather peaked rotation
curves.

At smaller halo masses, \cite{Stinson13} 
obtained a realistic late-type galaxy, with a moderate mass resolution
(mass of the dark matter particle $m_{DM}=1.1 \times 10^6$ M$_\odot$)
and using the simple SF model and the blastwave feedback.  In this
work they introduced a form of ``early stellar feedback'', motivated
by the expectation that the UV radiation of young stars can quench
the star formation rate in cold molecular clouds. This concept was
already introduced by \cite{Hopkins11}; the difference is that while
\cite{Hopkins11} used isolated idealised galaxy model, that allowed them
to reach high resolutions and directly model the kinetic radiation
pressure from young stars, \cite{Stinson13} modelled the same process
as thermal feedback.

\cite{Aumer13} simulated a subset of the halos from the Aquarius
project, using an improved version of the model by
\cite{Scannapieco09}.  They added a non-decoupled kinetic feedback,
along with feedback from radiation pressure of young massive stars,
that was already experimented in \cite{Hopkins11}. Also their results
improved considerably over those reported by the Aquila comparison
project. With the above prescriptions, their model is able to produce
realistic late-type spiral galaxies.

\cite{Vogelsberger13} used the moving-mesh hydro code AREPO
\citep{AREPO}, with a modified version of the effective model. They
performed cosmological simulations with various prescriptions of
kinetic feedback; in one of them, the wind speed depends on the mass
of the host DM halo, in line with \cite{Oppen06} and
\cite{Puchwein13}. Their simulations included also AGN radiation
feedback, i.e. the possibility of AGN radiation to destroy molecular
clouds.  They successfully matched a number of observational
properties of the galaxy population, such as the Tully-Fisher relation
and the stellar mass-halo mass relation. Unfortunately they did not
discuss galaxy morphologies.

\cite{Marinacci13} used AREPO and the same sub-resolution model of
\cite{Vogelsberger13} to simulated again the Aquarius set of initial
conditions.  Also in this case, they produced realistic late-type
galaxies, with low $B/T$ ratios and low baryon fractions.  To obtain
this result they scaled their kinetic feedback with halo mass,
similarly to the momentum-driven wind mechanism explained above, but
increasing by a factor of three the standard value of $10^{51}$ erg
associated to each SN explosion.

\cite{Hopkins13} presented a series of simulations of individual
galaxies, spanning a wide range of halo masses. They employed high
resolution (up to 
$\approx 2.4 \times 10^4$ M$_\odot$ in cosmological runs, for the DM
particles), a simple SF model with a high density threshold ($n_h=100$
cm$^{-3}$), early stellar feedback both in the form of radiation
pressure and photo-ionisation of molecular clouds.  SN feedback was
implemented as energy- or momentum-driven, depending on whether the
shock of the supernova bubble is energy- or momentum- conserving at
the resolved scales. They were able to achieve such high densities
thanks to an extremely small gravitational softening length for
baryonic particles.  They predicted a relation between stellar mass
and halo mass in good agreement with observational results, but
provided no information on the morphology of galaxies.

\cite{Vogel14a} presented a large, well resolved cosmological
  simulation, dubbed ``Illustris'', performed using the same {\sffb}
  of \cite{Vogelsberger13}. The box size is 106.5 Mpc
  ($h=0.704$), simulated at several resolutions. At their
  highest resolution, the DM particle mass is $1.26 \cdot 10^6$
  M$_\odot$. The galaxy population was analyzed in \cite{Vogel14b},
  \cite{Genel14} and \cite{Vogel14c}. Several properties of simulated
  galaxies agree well with observations, including the morphological
  classifications and the shape of late-type galaxy rotation curves.
  Some residual tensions remain between their simulation results and
  low-redshift observations, e.g. in the TF relation, the baryon
  convertion efficiency relation, and the stellar mass function;
  \cite{Vogel14b} point out that some other properties, namely colours
  of intermediate-mass galaxies and age of dwarf ones, are
  still not in agreement with observations.

\cite{Cen14} studied the colour bi-modality of galaxies at low redshift
($z=0.62$) using a large cosmological simulation named ``LAOZI''. They
used the Eulerian AMR code ENZO \citep{Bryan00,Joung09} with a simple
star formation prescription. They took into account the amount of UV
photons produced by young stars, with SN kinetic energy assigned to
the 27 cells centred around the exploding star. Also in this case, no
detailed morphological analysis of simulated galaxies was carried out.

{\cite{EAGLE} presented a suite of cosmological simulations, named
  ``EAGLE''. The included sub-resolution physics is that described in
  \cite{OWLS}. Their larger simulation has a size of $100$ Mpc
  ($h=0.68$), a mass resolution of $9.7 \cdot 10^6$ M$_\odot$
  and a force resolution of $0.7$ kpc. They paid particular attention
  to the calibration of sub-resolution model parameters. These
  simulations reproduce a number of observational properties of galaxies,
  in particular the stellar mass function, though the authors state
  that gas and stellar metallicities of their dwarf galaxies are still
  too high.

In summary, different groups claim at present to be able to produce
late-type spiral galaxies with properties similar to the observed
ones. In general, some consensus emerges as for the need to
significantly heat up or kick gas at early times through galactic
winds, and to have efficient feedback to form disk galaxies (see
\cite{Ubler14} for a recent work on this subject).  On the other hand,
no consensus has been reached on the source of the required feedback
energy: are SNe alone sufficient to provide this energy, or do we need
early stellar radiation, or AGN feedback, or cosmic ray feedback, or a
combination of them? Moreover, a number of open questions, concerning
technical points, have been raised and not yet fully solved. Is
extremely high numerical resolution mandatory to simulate realistic
disk galaxies? Do we need models with a high density threshold for the
star forming gas?  Is there a unique combination of threshold and star
formation efficiency that allows us to simulate realistic disk
galaxies? How important is the choice of hydrodynamic scheme on which
simulations are based?

In the following, we present cosmological simulations of individual
disk galaxies, carried out with GADGET3 including our sub-resolution
model MUPPI. Our model describes the behaviour of the ISM at
unresolved scales, but does not include early stellar feedback, high
density threshold for star formation and AGN feedback. We will show
that with this prescription we obtain late-type galaxies with
properties in broad agreement with observation without the need of
reaching extremely high resolution. A comparison with results from
other groups, that simulated one of the halos we present here, shows
that simulation results are quite close to each other, even if {\sffb}
models and hydro schemes are significantly different. In this sense,
our results further demonstrate that possible avenues exist to produce
realistic disk galaxies, relying on a relatively low resolution that
can be afforded in large-scale simulations of representative volumes
of the universe.

\section{The sub-resolution model}
\label{section:model}

In this Section, we describe the updated implementation of the
{\muppi} algorithm for star formation and stellar feedback.  {\muppi}
is implemented within the TreePM+SPH GADGET3 code, which represents
an evolution of the TreePM+SPH GADGET2 \citep{GADGET2}. Our version
of GADGET-3 includes a uniform UV background from
\cite{HM01}, chemical evolution, metal cooling and kinetic feedback.
A brief account of the original version of the algorithm, described in
full detail in \cite{M10}, is given in Section~\ref{section:muppistd},
while the changes introduced in this work, in particular chemical
evolution, metal cooling and kinetic feedback, are described in
Sections~\ref{section:muppimet} and \ref{section:muppikin}.

\subsection{The {\muppi} algorithm}
\label{section:muppistd}

We assume that our simulations work at a typical force resolution in
the range from 100 pc to 1 kpc, and at a mass resolution from from
$10^4$ to $10^7$ M$_\odot$.  In these conditions, the ISM of a
star-forming region will have much structure at unresolved scales.
The aim of our sub-grid model is to provide a description of this
multi-phase gas which is accurate enough to represent in a realistic
way the emergent effects of star formation and stellar feedback on
resolved scales.

Following \cite{M04}, we assume that each SPH particle that is
eligible to host star formation (under a set of conditions that will
be specified below) represents a multi-phase ISM that is composed by a
hot, tenuous and pervasive gas phase and a cold phase that is
fragmented into clouds with a low filling factor. These two phases are
assumed to be in pressure equilibrium.  A fraction of the cold phase,
that we call molecular gas fraction, provides the reservoir for star
formation.  Spawning of collisionless star particles from the
resulting stellar component of star-forming particles takes place
according to a standard stochastic star formation algorithm
\citep{Katz92,SpringHern03}.  The hot phase is heated by the emerging
energy from massive and dying stars and radiatively cools.

This setting is quantified as follows.  Within each SPH particle of
mass $M_{\rm P}$, the masses of the hot, cold and stellar components
are $M_{\rm h}$, $M_{\rm c}$ and $M_\star$.  If $n_{\rm h}$ and
$n_{\rm c}$ are the particle number densities of the two gas phases, and
$T_{\rm h}$ and $T_{\rm c}$ their respective temperatures, the
condition of pressure equilibrium translates into:
\be n_{\rm h}\cdot T_{\rm h} = n_{\rm c} \cdot T_{\rm c}\, .
\label{eq:pressure_eq}\ee 
Here $T_{\rm c}$ should be considered as an effective temperature,
that also takes into account the effect of kinetic pressure;  this is left as a
free parameter in the model.

Densities of cold and hot phases are computed starting from their filling factors: calling $F_{\rm h}$ the fraction of gas mass in the hot phase (the cold phase having a mass fraction of $F_{\rm c}=1-F_{\rm h}$), its filling factor $f_{\rm h}$ is:

\be
f_{\rm h} = 1-f_{\rm c} = \frac{1}{ 1 +  \frac{F_{\rm c}}{F_{\rm h}}\cdot \frac{\mu_{\rm h}}{\mu_{\rm c}} \cdot \frac{T_{\rm c}}{T_{\rm h}}}\, .
\ee

\noindent
Then, if $\rho$ is the average gas density, for the two phases we have

\be
n_{\rm h,c}=\rho F_{\rm h,c} / f_{\rm h,c} \mu_{\rm h,c} m_p\, ,
\ee

\noindent
$\mu_{h,c}$ being the corresponding molecular weights.

The molecular fraction is computed using the phenomenological
relation by \cite{Blitz06} between the ratio of surface
densities of molecular and atomic gas, and the estimated external
pressure exerted on molecular clouds.  Using the hydrodynamic
pressure of the particle in place of the external pressure, we obtain
a simple way to estimate the molecular fraction $f_{\rm mol}$:
\be
f_{\rm mol} = \frac{1}{1 + P_0/P}
\label{eq:fcoll}
\ee 
A gas particle enters the multi-phase regime whenever its temperature
drops below $10^5$ K and its density is higher than a threshold value
$\rho_{\rm thr}$.  The multi-phase system is initialised with all the
mass in the ``hot'' component, $M_{\rm h} = M_{\rm P}$ and $T_{\rm
  h}=T_{\rm P}$, where $T_{\rm P}$ is the gas temperature.  Its evolution
is described by a system of four ordinary differential equations
(see below), in which the variables are the masses of the
three components, $M_{\rm h}$, $M_{\rm c}$ and $M_\star$, and the
thermal energy of the hot phase $E_{\rm h}$.  At each SPH time-step,
this system is integrated with a Runge-Kutta integrator with adaptive
time-steps.  This means that the integration time-step is much
shorter than the SPH one.  To mimic the
disruption of molecular clouds due to the activity of massive stars,
and to limit the entrainment of the cold phase in a particle dominated
(in volume) by the hot phase, a multi-phase cycle lasts at most a time
$t_{\rm clock}$, that is set to be proportional to the dynamical time of the
cold phase, $t_{\rm dyn}$; this quantity is defined below and used to
compute the star formation rate.  A particle exits the multi-phase
regime also when its density becomes lower that $1/5$ of the entrance
density threshold $\rho_{\rm th}$.  Moreover, at low densities it can
happen that the energy from SNe is not sufficient to sustain a hot
phase.  This results in a hot phase temperature that does not raise
above $10^5$ K but remains very low.  In this case the particle is
forced to exit the multi-phase regime. A particle that has exited the
multi-phase regime can enter it at the next time-step, if it meets the
required conditions.

Matter flows among the three components as follows: cooling deposits
hot gas into the cold phase; evaporation brings cold gas back to the
hot phase; star formation moves mass from the cold gas to the stellar
component; restoration moves mass from stars back to the hot phase.
This is represented through the following system of differential
equations:
\begin{eqnarray}
\dot{M}_{\star} & = & \dot{M}_{\rm sf} - \dot{M}_{\rm re} \label{eq:sf} \\
\dot{M}_{\rm c} & = & \dot{M}_{\rm cool} - \dot{M}_{\rm sf} - \dot{M}_{\rm ev} \label{eq:mc} \\
\dot{M}_{\rm h} & = & -\dot{M}_{\rm cool} + \dot{M}_{\rm re} + \dot{M}_{\rm ev} \label{eq:mh}
\end{eqnarray}
The various terms of the system (\ref{eq:sf}-\ref{eq:mh}) are computed as
follows.  The cooling flow $\dot{M}_{\rm cool}$ is
\be
\dot{M}_{\rm cool} =  \frac{M_{\rm h}}{t_{\rm cool}}\,.
\label{eq:mcool}
\ee
This implies that cooling leads to the deposition of mass in the
cold phase, at constant $T_{\rm h}$, and not to a decrease of $T_{\rm
  h}$. The cooling time is computed using hot phase density and
temperature, $n_{\rm h}$ and $T_{\rm h}$, so that cooling times are
relatively long whenever the hot phase has a low mass fraction and a
high filling factor.  The star formation rate is
\be
\dot{M}_{\rm sf} = f_{\star} \cdot \frac{f_{\rm mol}\cdot M_{\rm  c}}{t_{\rm dyn}}\, , 
\label{eq:sfr}
\ee
where the dynamical time of the cold gas phase is given by
\be
t_{\rm dyn} = \sqrt{\frac{3\pi}{32G\rho_{\rm c}}} \ {\rm yr} \,.
\label{eq:tdyn}
\ee
As explained and commented in \cite{M10}, during each multi-phase
cycle the dynamical time is computed as soon as 90 per cent of mass is
accumulated in the cold phase, and is left constant for the rest of
the cycle.  Following \cite{M04}, evaporation is assumed to be
proportional to the star formation rate:
\be
\dot{M}_{\rm ev} = f_{\rm ev} \cdot \dot{M}_{\rm sf}   
\label{eq:ev}
\ee
In the absence of chemical evolution, the restoration term
$\dot{M}_{\rm re}$ is computed in the Instantaneous Recycling
Approximation (IRA):
\be 
\dot{M}_{\rm re} = f_{\rm re} \cdot \dot{M}_{\rm sf}\, ,
\label{eq:ira}
\ee
otherwise the modelling of this mass (and metal) flow is performed by
the chemical evolution code, as described in the next Section.

The fourth variable of the model is the thermal energy of the hot
phase, $E_{\rm h}$.  Its evolution is described as
\be
\dot{E}_{\rm h} = \dot{E}_{\rm heat,local} - \dot{E}_{\rm cool} + \dot{E}_{\rm hydro}
\label{eq:eh}
\ee
The first heating term reads
\be
\dot{E}_{\rm heat,local} = E_{\rm SN} \cdot f_{\rm fb,local} \cdot \frac{\dot{M}_{\rm sf}}{M_{\rm\star,SN}} \,.
\label{eq:esn}
\ee
and accounts for the contribution of SN energy from stars formed in
the same multi-phase particle.  Here $E_{\rm SN}$ is the energy
supplied by a
single supernova, and $M_{\rm\star,SN}$ the stellar mass associated to
each SN event, 
while $f_{\rm fb,local}$ is the fraction of SN energy that is deposited in the hot phase of the particle itself.
Consistently with the mass cooling flow, energy losses by
cooling are expressed as
\be \dot{E}_{\rm cool} = \frac{E_{\rm h}}{t_{\rm cool}}\, .
\label{eq:cool}
\ee

The $\dot{E}_{\rm hydro}$ term takes into account the energy due to
interactions with the neighbour particles. It is computed as the
energy accumulated during the last hydrodynamical timestep, divided by
the timestep itself. This energy accounts for the change in entropy
given by SPH hydrodynamics, i.e. the $PdV$ work on the particle plus
the effect of numerical viscosity, and for the SN energy coming from
neighbouring particles.

Apart from the small fraction $f_{\rm fb,local}$ given to the
star-forming particle itself, the energy budget from SNe is
distributed in the form of thermal and kinetic energy. The thermal
energy distributed to neighbours by each star-forming particle can be
written as
\be 
\Delta E_{\rm heat,o} = E_{\rm SN} \cdot f_{\rm fb,out}\cdot \frac{\Delta
  M_\star}{M_{\rm\star,SN}}\,. 
\label{eq:en}
\ee 
To mimic the blow-out of superbubbles along the least Resistance path
\citep[see][]{M04}, each multi-phase particle distributes its thermal
energy to particles that are within its SPH smoothing length and lie
within a cone whose axis is aligned along the direction of minus the
local density gradient, and whose semi-aperture angle is $\theta$.
Energy contributions are weighted with the SPH kernel, but using the
distance from the cone axis in place of the radial distance.  This
thermal feedback scheme is relatively effective even at high
densities.  In fact cooling has the effect of depositing hot gas into
the cold phase, and cooling times are computed using density and
temperature of the hot phase, so they are relatively long.  The main
effect of cooling is then to reduce the mass fraction of hot gas,
while its temperature is kept high by SN feedback; when the particle
is taken in isolation ($\dot{E}_{\rm hydro}=0$), energy injection from
SNe and the relation between pressure and molecular fraction
(Equation~\ref{eq:fcoll}) create a runaway of star formation, until
the molecular fraction saturates to unity.  The particle can lower its
pressure by expanding and performing $PdV$ work on other
particles. Therefore, it is the hydrodynamic interaction with
neighbouring particles that halts the runaway.  This non-equilibrium
dynamics is strong enough to avoid the formation of very cold and
dense blobs.  However, the effectiveness of thermal feedback is 
more limited when the contribution from metal lines is included.

At the end of the integration, we use the new state of the multi-phase
system to recompute the entropy of the gas particle; the entropy
change will thus include the effect of thermal energy from SNe.  The
entropy determines the internal energy and pressure of the particle,
so this change will be self-consistently accounted for by the SPH
hydrodynamical integrator.

As mentioned above, the creation of star particles is implemented with
the stochastic algorithm described by \cite{SpringHern03}.  The
probability of a gas particle to spawn a new star particle is
proportional to the (virtual) stellar mass formed during the last
hydrodynamical time-step.  When a star particle is spawned, its mass
is taken from $M_\star$; if this is not sufficient, the remaining mass
is taken from the cold phase, or from the hot phase in the unlikely
case the mass is still insufficient.

In summary, the parameters of the model are $T_{\rm c}$, $P_0$,
$f_{\rm re}$ (when chemical evolution is not implemented), $ f_{\rm
  ev}$, $f_{\star}$, $t_{\rm clock}$, $\rho_{\rm thr}$, $\theta$ and
the energy fractions $f_{\rm fb,local}$ and $f_{\rm fb,out}$.  Their
values are given and commented in \cite{M10}, and have been slightly
adjusted to the newest version including chemical evolution, taking
values as reported in Table~\ref{table:MUPPIparam}.  As a remark on
the density threshold $\rho_{\rm thr}$, we remind that it should not be
confused with the star formation density threshold used, e.g., in the
effective model of \cite{SpringHern03}.  In our model a gas particle
can be in the multi-phase regime, but have very low star formation
rate; this happens when the molecular fraction is low, due to low pressure.
For example, if we consider the ``star formation threshold'' as the value
of the numerical density of the cold phase where $f_{mol} =1/2$,
taking reference values of $P_0/k_B=20000$ K cm$^{-3}$, $T_c=300$ K, in
Eq. (\ref{eq:fcoll}) we have $n_c T_c= P_0/k_B$, from which the threshold
would be $n_c=66.6$ cm$^{-3}$.  This high density is however modelled
at the sub-resolution level, without the need of resolving it.

\begin{table*}
  \caption{Parameters of {\muppi}.
    Column 1: Cold phase temperature (K).
    Column 2: Pressure at which $f_{\rm mol}=0.5$ (K cm$^{-3}$).
    Column 3: restoration fraction (now computed by the chemical
    evolution model).
    Column 4: evaporation fraction.
    Column 5: star formation efficiency (referred to the {\it
      molecular} gas).
    Column 6: duration of a multi-phase cycle in dynamical times.
    Column 7: density threshold for multi-phase particles (cm$^{-3}$).
    Column 8: Semi-aperture of the cone (degrees).
    Column 9: SN thermal energy given to local hot gas.
    Column 10: SN thermal energy given to neighbouring hot gas.
    Column 11: SN kinetic energy given to wind particles.
    Column 12: Probability for a gas particle to be converted in a wind particle.
    Parameters marked with $\star$ have been revised with respect to
    what reported by Murante {\it et al.} (2010).
  }
\begin{tabular}{c c || c c c c c c || c c c c }
\hline\hline $T_{\rm c}$ & $P_{\rm 0}$ & $f_{\rm re}$& $f_{\rm ev}$
&$ f_{\star}$ & $t_{\rm clock}/t_{\rm dyn}$ & $rho_{\rm thr}$ &
$\theta $ & $f_{\rm fb,local}$ & $f_{\rm fb.out}$ & $f_{\rm fb,kin}$ &
$P_{\rm kin}$  \\
\hline
300$^\star$ & 20000$^\star$&  ---$^\star$ & 0.1 
& 0.02 & 1$^\star$ & 0.01 & 60 & 0.02 &
0.2$^\star$ & 0.6$^\star$ & 0.03$^\star$  \\
\hline
\end{tabular}
\label{table:MUPPIparam}
\end{table*}

\subsection{Chemical evolution, metal cooling}
\label{section:muppimet}

We have merged {\muppi} with the chemical evolution code originally
presented by \cite{Luca07}.  In this code each star particle is
treated as a Simple Stellar Population (SSP), whose evolution is
followed starting from the time at which it has been spawned from a
gas particle.  Given a stellar Initial Mass Function (IMF), the mass
of the SSP is varied in time following the death of stars, and
accounting for stellar mass losses. We follow the production of
several elements through SNII, SNIa and AGB stars.  The enriched material is
spread among the neighbouring gas particles with weights given by the
SPH kernel. For each gas particle the code tracks the mass in
$H$, $He$ and in several other chemical species, namely $C$, $Ca$,
$O$, $N$, $Ne$, $Mg$, $S$, $Si$, $Fe$, plus that in generic other
elements.  When a star particle is spawned, its composition is taken
as the one of the parent gas particle.

The code allows us to flexibly choose the stellar Initial Mass
Function (IMF), the minimum stellar mass for SNII, metal yields, stellar
lifetimes, and the elements to follow in detail.  In this paper,
stellar lifetimes are taken from \cite{PadMat93}, we assume the IMF
proposed by \cite{Kroupa}, in the range 0.1-100 M$\odot$. This IMF is
similar to that proposed by \cite{Chabrier}  ​in the same range.  SNII
are assumed to be originated by stars more massive than 8 {\msun}, while stars more massive
than 40 {\msun} are assumed to implode into black holes and not to
contribute to chemical enrichment.  Metal yields are taken from
\cite{WW95} for SNII, \cite{Thielemann03} for SNIa and \cite{vdH97}
for AGB stars.  We follow the production of ten metal species, namely
$C$, $Ca$, $O$, $N$, $Ne$, $Mg$, $S$, $Si$, $Fe$, plus $He$.

The implementation of the chemical evolution model by \cite{Luca07}
follows the injection of energy from SNe along with mass ejection. In
its integration with {\muppi} we have decoupled the treatment of
energy and mass.  Energy injection is treated by {\muppi} in the IRA
as explained above, while the energy from SNIa is not implemented in
the present version of the code.  Mass restoration from dying stars is
not treated in the IRA, as in Equation~\ref{eq:ira}; the same mass
flow $\dot{M}_{\rm re}$ is used to inject into the hot phase, at each
SPH time-step, the mass (including metals) acquired from nearby star
particles during the previous time-step.  Because this mass is
connected to dying stars, we assume that the energy necessary to keep
it hot comes from the SN thermal energy budget.  Finally, the
parameter $M_{\rm \star,SN}$ is computed from the assumed IMF as the
inverse of the number of stars more massive than 8 {\msun} per unit
mass of stars formed.

The contribution of metals to gas cooling is computed by following the
procedure of \cite{Wiersma09a} \citep[as in][]{Planelles13,Barai13}.  The
emissivity of an optically thin gas of a specified composition, under
the influence of a uniform ionising background \citep[from][]{HM01} is
computed without assuming fixed (solar) abundance ratios of elements
or collisional ionisation equilibrium.  

\subsection{Kinetic feedback}
\label{section:muppikin}

The original version of our {\sffb} model included SN feedback only in
the form of thermal energy.  As illustrated in Monaco et al. (in
prep), thermal feedback alone is able to efficiently suppress star
formation when cooling is included for a gas of primordial
compositions, although at the cost of creating a very hot
circum-galactic halo. On the other hand, the efficiency of thermal
feedback is much weaker when metal cooling is included, and much
overcooling takes place.  The reason for this is that thermal feedback
can trigger fountain-like outflows of $\sim50$ km s$^{-1}$, so that
suppression of star formation for a Hubble time is obtained not by
ejecting gas from the halo but by keeping it hot. This is no longer
possible when cooling is boosted by a factor of 10 or more due to the
higher metallicity.  On the other hand, SNe inject both thermal and
kinetic energy in the ISM, so it is natural to expect an emergence of
energy at the $\sim$kpc scale both in thermal and in kinetic form.

To implement a kinetic feedback scheme, we broadly follow the scheme
by \cite{SpringHern03}, but with several differences to make it
compatible with the thermal feedback scheme.  When a particle exits a
multi-phase cycle,
we assign it a probability $P_{\rm kin}$ to become a ``wind''
particle.  For some time $t_{\rm wind}$, such particles can 
receive kinetic energy from neighbouring multi-phase particles.
Because outflows are driven by SNII exploding after the destruction of
the molecular cloud, this time is set equal to the stellar lifetime of
an 8 {\msun} star, $t_8$, minus the duration $t_{\rm clock}$ of the
past multi-phase cycle:
\be
t_{\rm wind} = t_8 - t_{\rm clock}\, . \label{eq:twind}
\ee
However, the wind phase stops earlier than $t_{\rm wind}$ whenever
the particle achieves low density, set to $0.3 \rho_{\rm thr}$.
For each star-forming particle, the available kinetic energy budget
is
\be
E_{\rm kin}=f_{\rm fb,kin} E_{\rm SN}\, .
\label{eq:kinenergy}
\ee
This energy is distributed from multi-phase particles to wind
particles with the same scheme of thermal energy: eligible wind
particles are those within the SPH kernel and within a cone of
semi-aperture $\theta$, anti-aligned with the density gradient, with
relative contribution weighted by the distance from the cone axis.
These particles receive ``velocity kicks'' as follows. For each wind
particle we compute the energy contributions from all kicking
particles and the energy-weighted average vector from kicking
particles to the wind one. Then the kinetic energy of the wind
particle is increased\footnote{In the reference frame of the particle
  itself.} and the increase in velocity is in the direction defined
above.  The emergence of the wind, presumably due to the blow-out of
an SN-driven super-bubble, takes place at scales that are still
smaller than the ones that are typically resolved. Therefore, in order
to avoid hydrodynamical coupling at $\sim$kpc scale, the wind particle
is decoupled from the surrounding gas as long as it receives kinetic
energy. However, we have verified that, when the hydrodynamical
  decoupling is not implemented, the resulting galaxy has very similar
  properties compared to the one obtained with decoupled winds; gas discs
  are slightly more perturbed, but this perturbation does not
  propagate into an appreciable thickening of the stellar disc. 

In our prescription, the free parameters are $P_{\rm kin}$ and $f_{\rm
  fb,kin}$. Our tests suggest for them the values of 0.03 and 0.6,
respectively, as specified in Table~\ref{table:MUPPIparam}.  At
variance with other kinetic wind prescriptions, neither wind mass-load
nor wind velocity are fixed.  Nevertheless, typical values of these
quantities can be estimated as follows.  In a time interval $\Delta
t$, the gas mass that will be uploaded in wind reads 
\be \Delta M_{\rm
  wind}=P_{\rm kin}N_{\rm gas} \frac{\Delta t}{\langle t_{\rm
    dyn}\rangle} {\langle m_{\rm gas}\rangle}\,, 
\label{eq:mwind}
\ee
where$M_{\rm gas}$ is the mass of gas in multi-phase in a given
region,  $N_{\rm gas}=M_{\rm gas}/ {\langle m_{\rm gas}\rangle}$ is the
total number of multi-phase gas particles within the same
region, $\langle t_{\rm dyn}\rangle$ the average dynamical time of the
cold phase, and $ \langle m_{\rm gas}\rangle$ the average gas particle
mass (our particle have can have variable mass). The mass load can
then be cast in the form
\be \dot{M}_{\rm wind} = P_{\rm kin}  {1 \over <t_{\rm dyn}>}
N_{\rm gas} <m_{\rm gas}> = P_{\rm kin} \frac{M_{\rm gas}}{ <t_{\rm
    dyn}>}\ee  
As such, it depends on the cold phase density, and then on pressure
through $\langle t_{\rm dyn}\rangle$.  
Since in one time step, using Eq. \ref{eq:sfr}, we have
\be
\Delta M_* = f_{\rm cold} f_{\rm mol} f_* {M_{\rm gas}\over \langle
  t_{\rm dyn} \rangle} \Delta t
\label{eq:deltamstar}
\ee
where we defined $f_{\rm cold} = M_c/M_{\rm gas}$, from the definition
of mass load factor $\eta = \dot{M_{\rm wind}}/ \dot{M_{\rm sfr}}$ we
obtain
\be
\eta =  {P_{\rm kin} \over  f_{\rm cold} f_{\rm mol} f_*}
\label{eq:massload}
\ee
With our choice of parameters, we obtain $\eta \simeq 1.5$.

Conversely, the mass-weighted
average wind velocity at the end of the wind phase is
\be
<v_{\rm wind}> = 2 ( \frac{<e_{\rm kin}>}{P_{\rm kin}M_{\rm gas}} )^{1/2}\,,
\ee
where $e_{\rm kin}$ is the mass-weighted kinetic energy deposited in
the gas mass $M_{\rm gas}$ in one dynamical time.
Similarly, defining $v_{\rm SN}^2=E_{\rm SN}/M_{*,SN}$ it is easy to
show that
\be
v^2_{\rm wind} = {f_{\rm fb,kin} \over \eta} v_{\rm SN}^2
\label{eq:velwind}
\ee With the chosen parameters, we obtain $v^2_{\rm wind} \simeq 600$
km s$^{-1}$. Note that, since these values depend on $\langle t_{\rm
  dyn}\rangle$, that varies depending on the local gas properties, the
exact values of the mass load and wind velocity will also depend on
such local properties.  We will present a detailed
analysis of our feedback model in a forthcoming paper (Monaco {\it
  et. al.}, in preparation).

\section{Simulations}
\label{section:simulations}

In this work, we simulated two sets of cosmological zoomed-in halos.
The standard zoom-in technique consists in isolating the object of
interest in a low resolution, usually DM only simulation, at redshift
$z=0$.  Particles are then traced back to their Lagrangian
coordinates.  The region occupied by these particles, the
``Lagrangian'' region of the forming halo, is then resampled at higher
resolution and with the addition of gas particles, taking care of
conserving both amplitudes and phases of the original Fourier-space
linear density.  The refined Lagrangian region is chosen to include,
at the final redshift, a volume that is larger than the virial radius
of the selected halo, in order to avoid that lower-resolution
particles affect the evolution of the halo.  The resolution is
degraded at larger and larger distances from the Lagrangian region of
interest.  This technique allows to greatly increase the resolution in
the chosen halo, while correctly describing the effect of the
large-scale tidal field.  As a side note, we remark that the evolution
of a resimulated object {\it changes} when resolution is varied: in
fact, power added at small scales modifies the halo's accretion
history, e.g. the timing of the mergers and the distribution of the
angular momentum.

The two sets of initial conditions describe the evolution of two
isolated halos with mass in the range $\sim(1-2) \cdot 10^{12}$
$h^{-1}$ M$_\odot$, both having a quiet merging history since
$z\sim2$.  Due to the lack of recent major mergers, these halos are
expected to host an $M_*$ disk galaxy at $z=0$.\footnote{
  This is however
  not guaranteed, and can only be judged a-posteriori from the results
  of the numerical simulations, better if comparing results from
  different groups. }
In both cases initial
conditions are available at several resolutions.  The first set of
initial conditions, called GA in this paper, has been presented in
\cite{Stoehr02}.  The second set, called AqC in this paper, has been
presented by \cite{Springel08} and used, among other papers, in the
Aquila comparison project \citep{Scannapieco12}.  We refer to the
original papers for more details on the construction of the two sets
of initial conditions.

We used three different resolutions for the GA set and two for the AqC
set. Table \ref{table:runs_std} shows mass and force
resolutions\footnote{Note that in this Section we give lengths in
  units of h$^{-1}$ kpc and masses in units of M$_\odot$ h$^{-1}$, at
  variance with the rest of the paper where lengths are expressed in
  kpc and masses in M$_\odot$. This choice is to ease the comparison
  with other works in literature, where numerical details are often
  given in terms of $h$. Softenings will always be expressed in
  h$^{-1}$kpc.}  used for our simulations, together with the virial
mass and radius \footnote{We define virial quantities as those
  computed in a sphere centred on the minimum potential particle of
  the halo and encompassing an overdensity of 200 times the {\it
    critical} cosmic density.}.  Virial masses and radii are quite
stable against resolution; there is a slight increase in virial
mass in the GA set, going from our intermediate to our highest
resolution.  Our chosen ICs span a factor of 87 in mass for
the GA set and a factor of 8 for the AqC set.  Force softening was
chosen to be constant in physical coordinates since $z=6$, while it is
comoving with Hubble expansion at higher redshift.  We scaled the
softening of the GA set with the cubic root of the mass
resolution. For the AqC5 and AqC6 simulations we used the same
softenings as for GA2 and GA1, respectively.

For comparison, our highest mass resolution is the same used by
\cite{Marinacci13} and \cite{Aumer13} for the AqC5 halo. Our
intermediate mass resolution is a factor $\approx 5$ better than the
resolution used by \cite{Vogelsberger13}. Comparing with the Eris
simulation presented in \cite{Guedes11},  resolution in the
GA2 and AqC5 simulations are coarser
by a factor 21.4 in mass and by a factor 3.9 in gravitational
softening. 

The GA set used a $\Lambda$CDM cosmology with
$\Omega_m=0.3$,$\Omega_\Lambda=0.7$, $\Omega_{baryon}=0.043$ and
$H_0=70$ km s$^{-1}$. Also AqC was run with a $\Lambda$CDM cosmology,
but with $\Omega_m=0.25$,$\Omega_\Lambda=0.75$, $\Omega_{baryon}=0.04$
and $H_0=73$ km s$^{-1}$.

\begin{table*}
  \caption{Basic characteristics of the different runs 
    Column 1: simulation name; 
    Column 2: mass of the DM particles; 
    Column 3: initial mass of the gas particles; 
    Column 4: Plummer-equivalent softening length for gravitational force.
    Column 5: Virial mass of the DM halo at $z=0$;
    Column 6: Virial radius of the DM halo at $z=0$;
    Column 7: number of DM particles within the virial radius at $z=0$; 
    Column 8: number of gas particles within the virial radius at $z=0$; 
    Column 9: number of star particles within the virial radius at
    $z=0$; 
    Masses are expressed in units of h$^{-1}$M$_\odot$  and  
    softening lengths in units of h$^{-1}$ kpc.   
  }
\begin{tabular}{c || c c c || c c c c c||}
\hline\hline Simulation & $M_{\rm DM}$ & $M_{\rm gas}$& $\epsilon_{\rm Pl}$
&$ M_{\rm Vir}$ & $R_{\rm Vir}$ & $N_{\rm DM}$ & $N_{\rm gas}$ &
$N_{\rm star}$\\
\hline
GA0 & $1.4 \cdot 10^8$ &  $2.6 \cdot  10^7$ & 1.4  & $2.28 \cdot 10^{12}$ &
212.72 & 13998 & 5873 &  24021\\
GA1 & $1.5 \cdot 10^7$ &  $2.8 \cdot  10^6$ & 0.65 & $2.30 \cdot 10^{12}$ &
212.56 & 133066  & 68130 & 178626\\
GA2 
& $1.6 \cdot 10^6$ &  $3.0 \cdot  10^5$ & 0.325 & $2.20 \cdot 10^{12}$ &
209.89 & 1214958 & 534567 & 1429204\\
AqC6 & $1.3 \cdot 10^7$ &  $4.8 \cdot  10^6$ & 0.65 & $1.21 \cdot 10^{12}$ &
169.03 & 87933 & 40362  & 123307\\
AqC5 & $1.6 \cdot 10^6$ &  $3.0 \cdot  10^5$ & 0.325 & $1.16 \cdot 10^{12}$ &
166.75 & 687003 &276881  & 898777\\

\hline
\end{tabular}
\label{table:runs_std}
\end{table*}


\section{Results}
\label{section:results}

In this Section, we first present results obtained from the highest
resolution GA2 and AqC5 halos at $z=0$.  The evolution of
galaxies with redshift is discussed in Section \ref{section:evolution},
while the effect of resolution is presented in Section
\ref{section:resolution}.

\subsection{Galaxy Morphology}
\label{section:morphology}

In Figure \ref{fig:MapGA2}, we show face-on and edge-on maps of gas
and stellar density for the GA2 simulation. We rotate the coordinate
system so that its z-axis is aligned with the vector of angular
momentum of star and cold or multi-phase gas particles within 8 kpc
from the position of the minimum of the gravitational potential, and
centered on it. This is the reference system with respect to which all
of our analyses have been performed.  The presence of an extended disk
is evident in both gas and stellar components. The gas disk shows a
complex spiral pattern and is warped in the outer regions.  A spiral
pattern is visible also in the outer part of the stellar disk. At the
centre, a bar is visible both in the gas and in the stellar
component. A full analysis of the bar features in our simulated
  galaxies is presented in the companion paper by Goz {\it et al.}
  (2014).The absence of a prominent bulge is already clear at a
first glance of the stellar maps.

In Figure \ref{fig:MapAqC5} we show the same maps for the simulation
AqC5. The appearance of the galaxy is similar to that of GA2, though
here the disk is smaller, the warp in the gas disk is less evident and
the distribution of stars shows even clearer spiral pattern and bar.

\begin{figure*}
\centerline{
\includegraphics[scale=0.5]{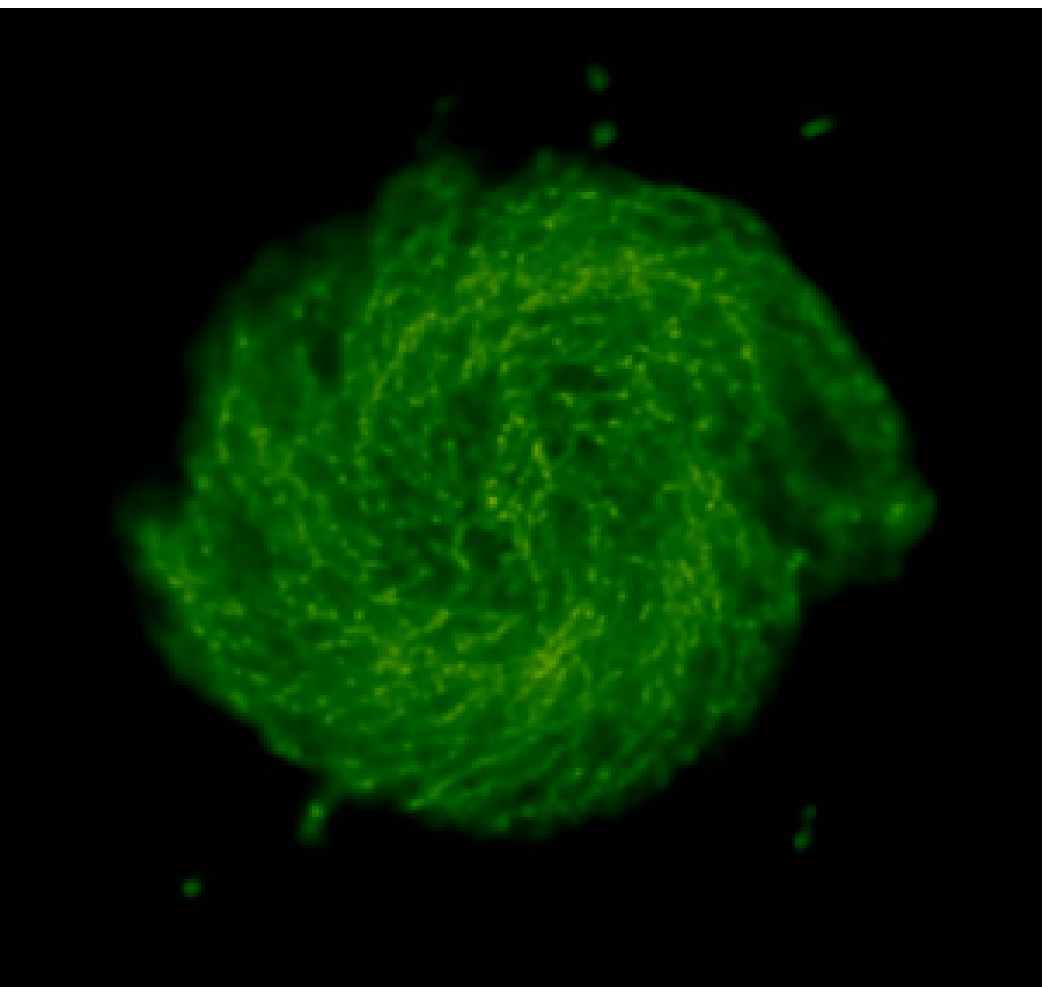}
\includegraphics[scale=0.5]{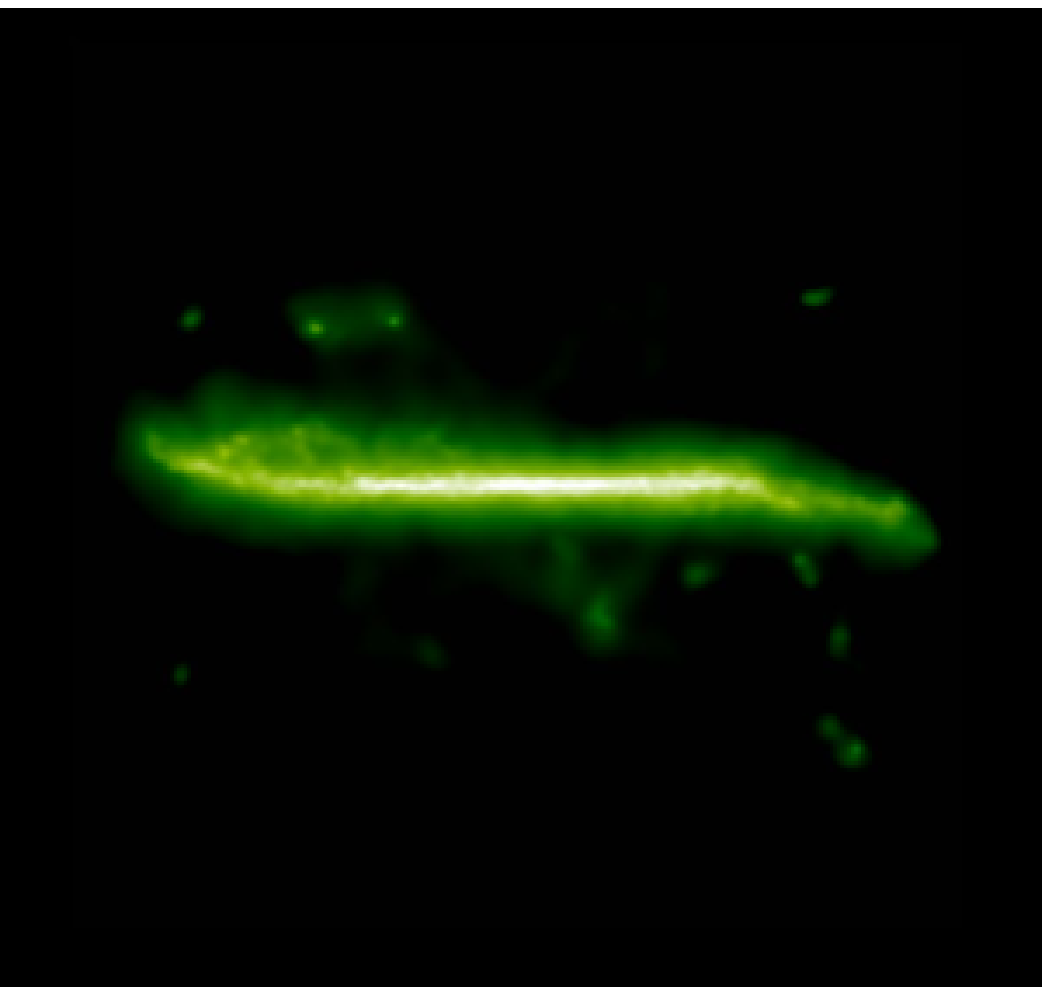}
}
\centerline{
\includegraphics[scale=0.5]{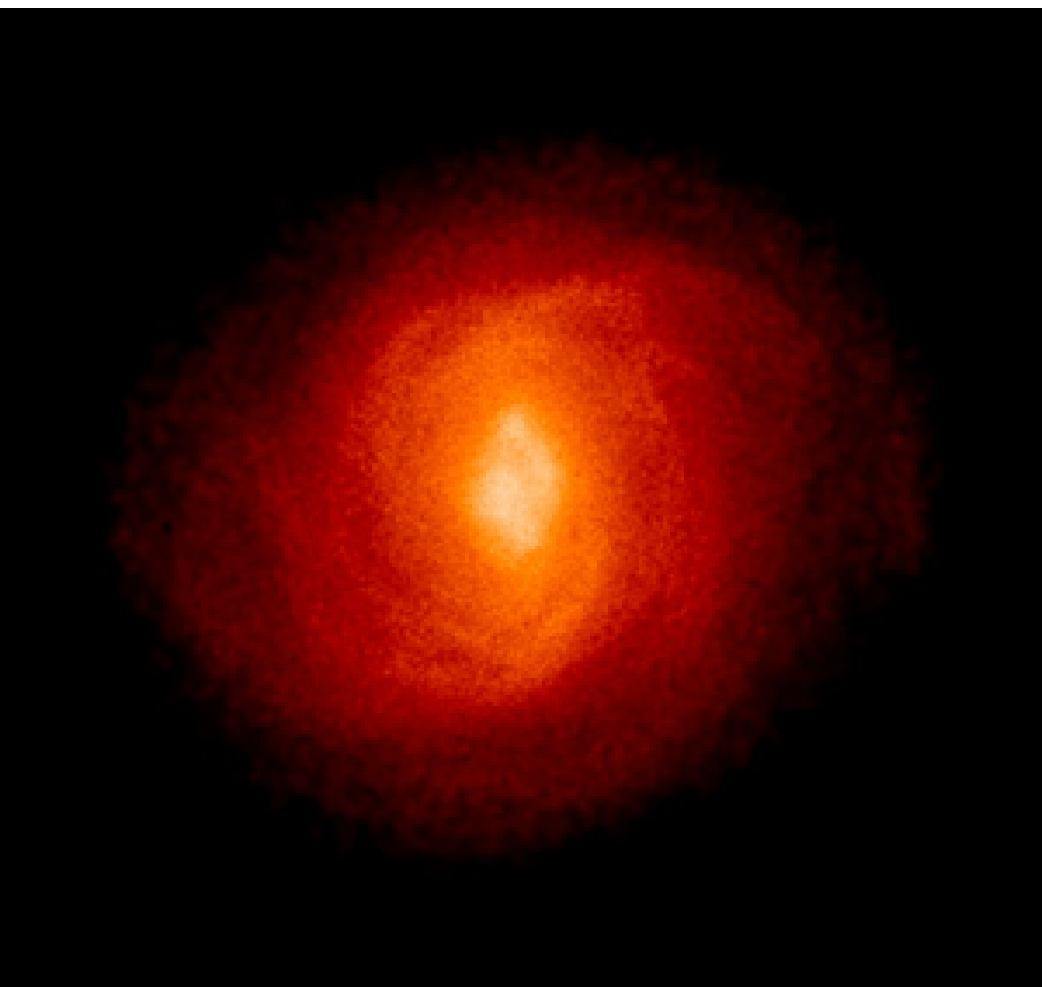}
\includegraphics[scale=0.5]{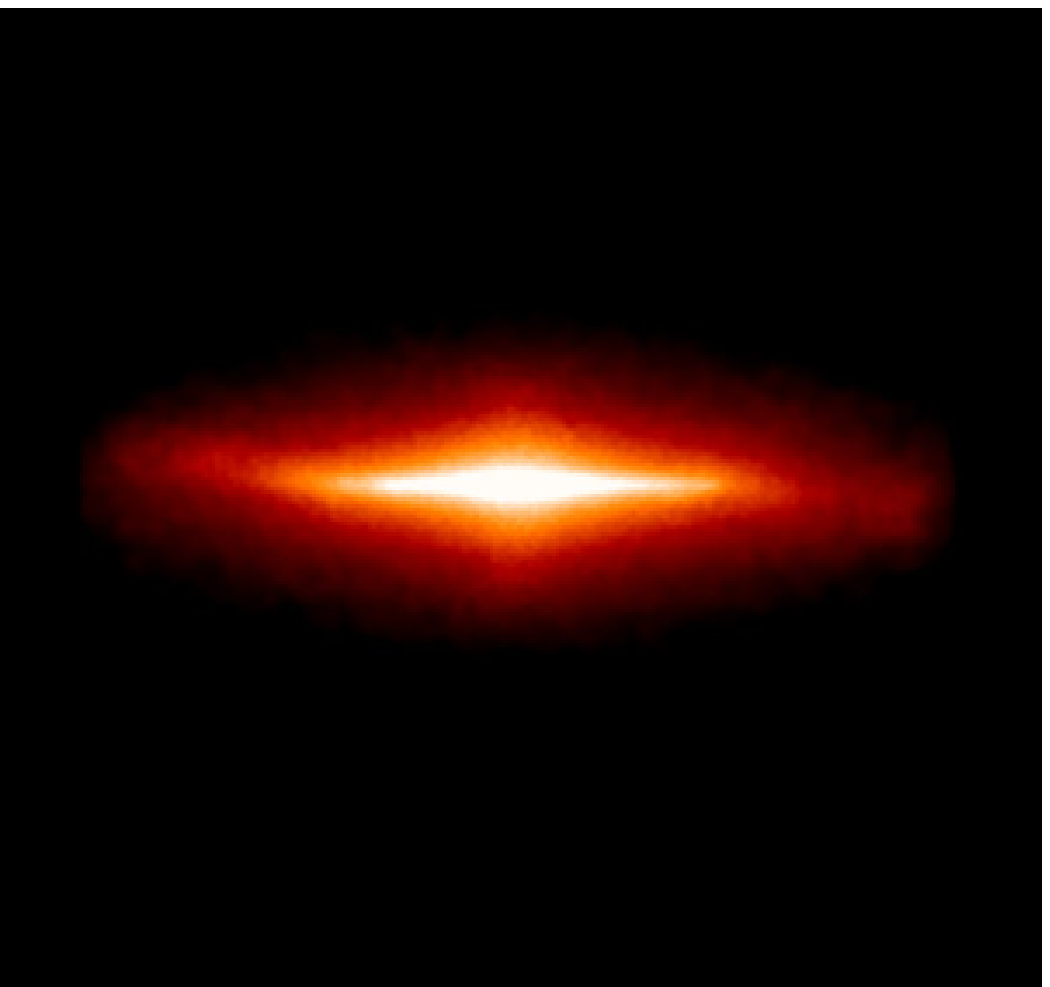}
}
\caption{ Projected gas (upper panels) and stellar (lower panels)
  density for the GA2 simulation. The z-axis of the coordinate system
  is aligned with the angular momentum vector of gas and stars
  enclosed within  8 kpc from the position of the minimum of the
    gravitational potential. Left panels show face-on densities,
  right column shows edge-on densities.  Box size is $57$ kpc. }
\label{fig:MapGA2}
\end{figure*}

\begin{figure*}
\centerline{
\includegraphics[scale=0.5]{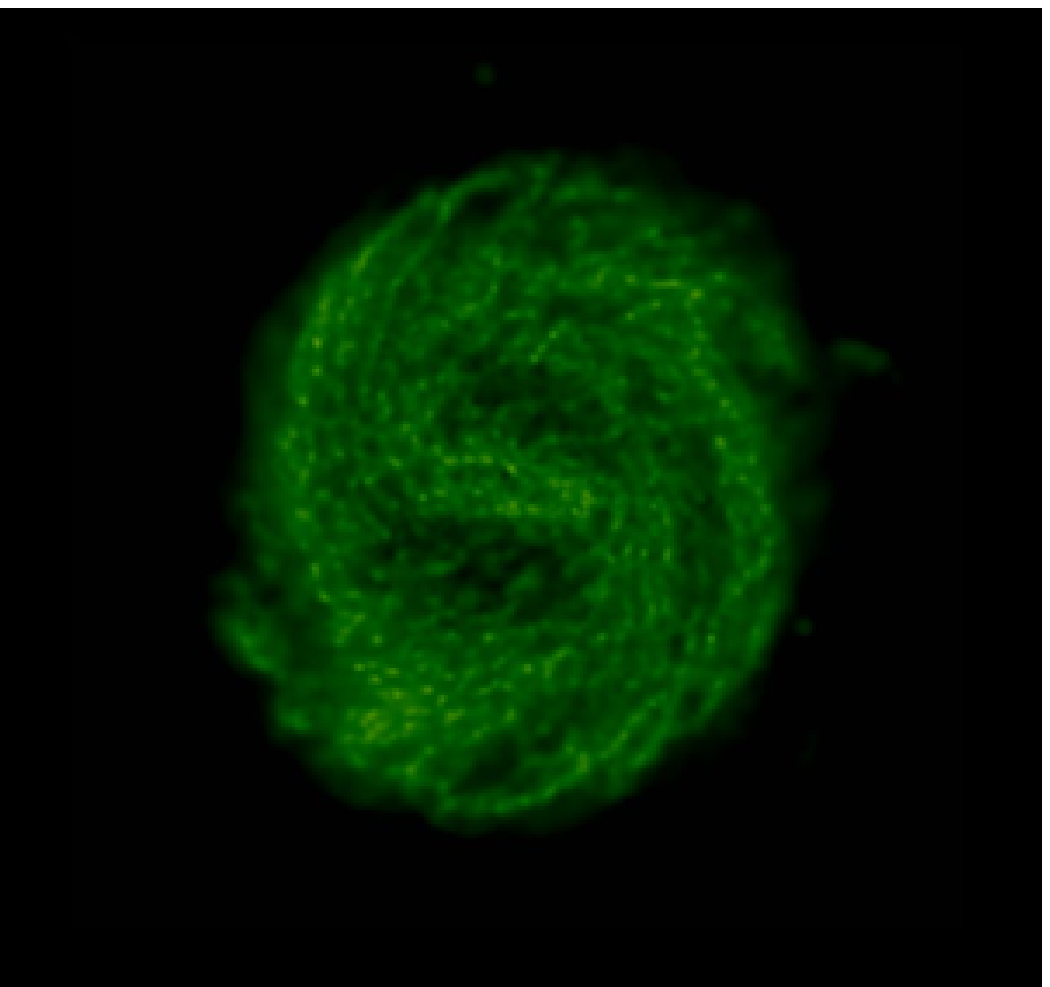}
\includegraphics[scale=0.5]{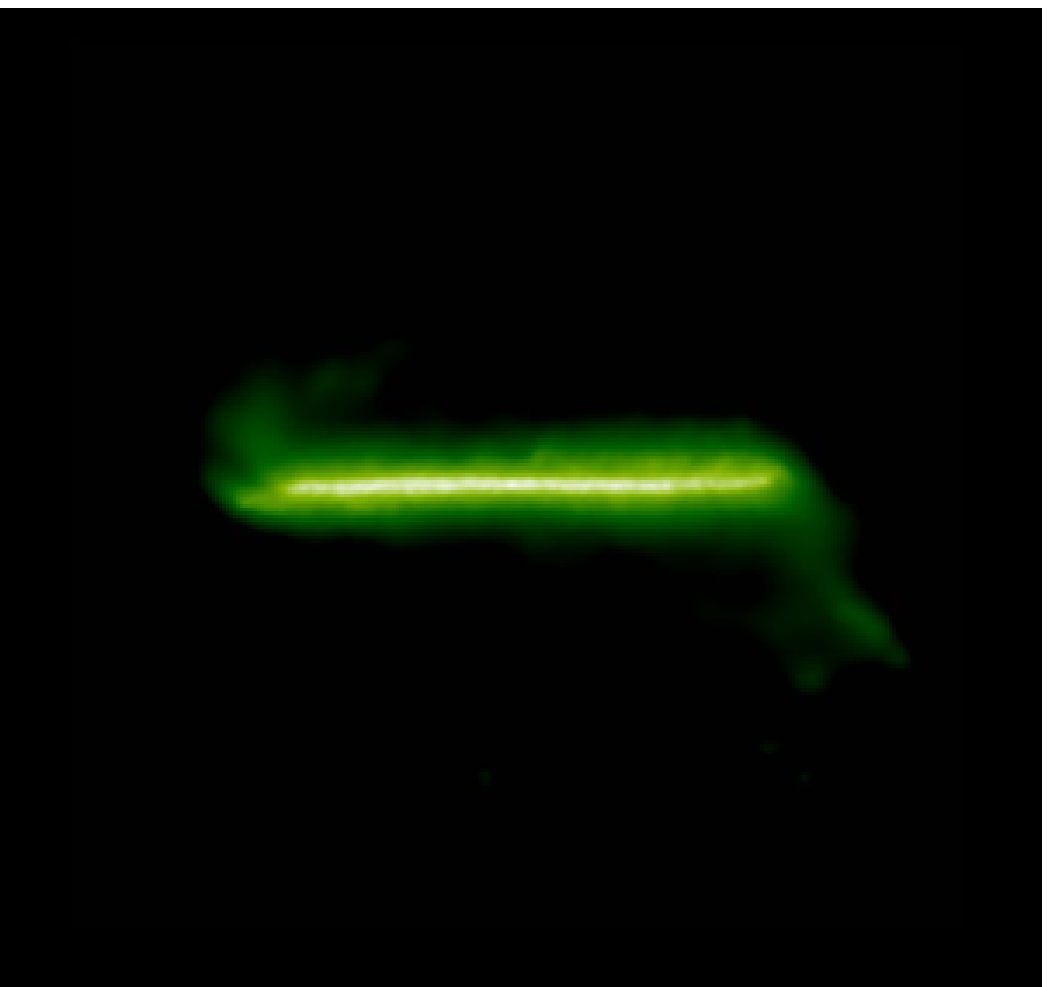}
}
\centerline{
\includegraphics[scale=0.5]{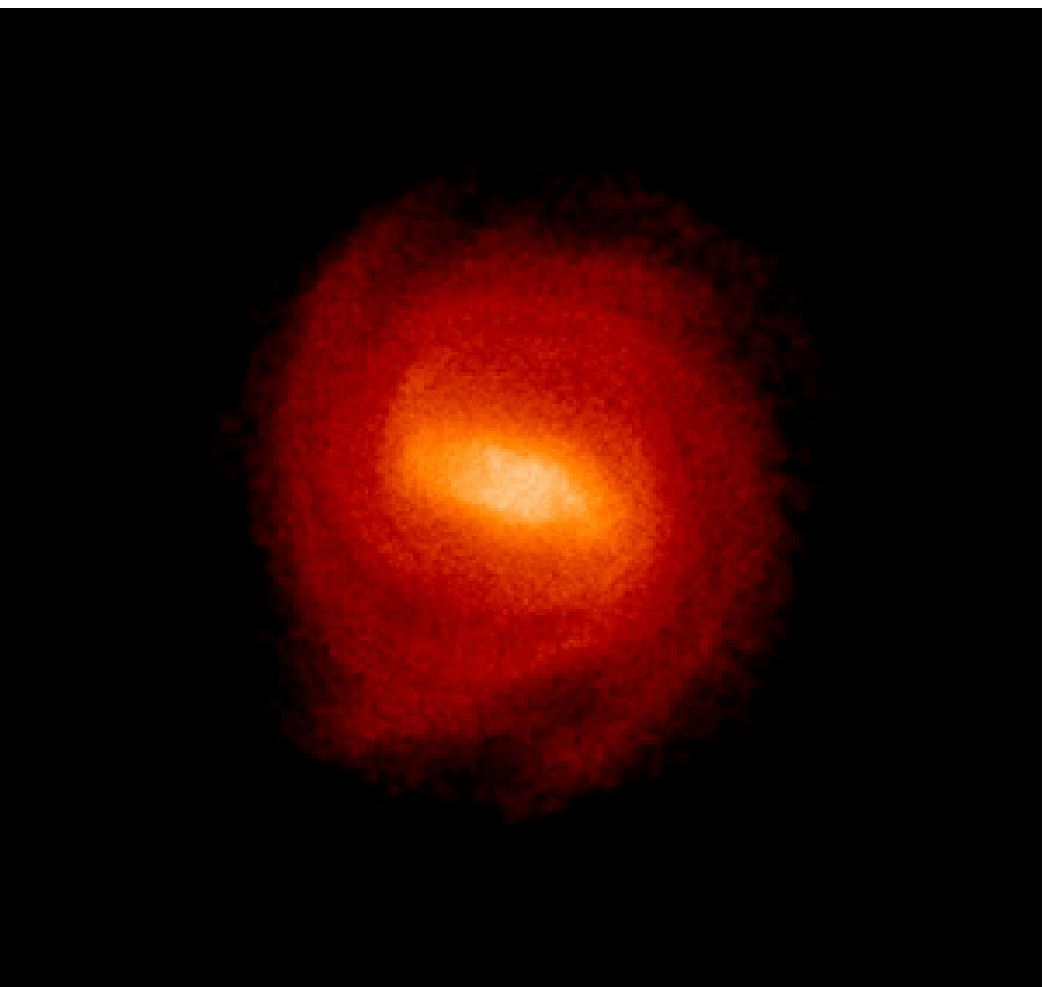}
\includegraphics[scale=0.5]{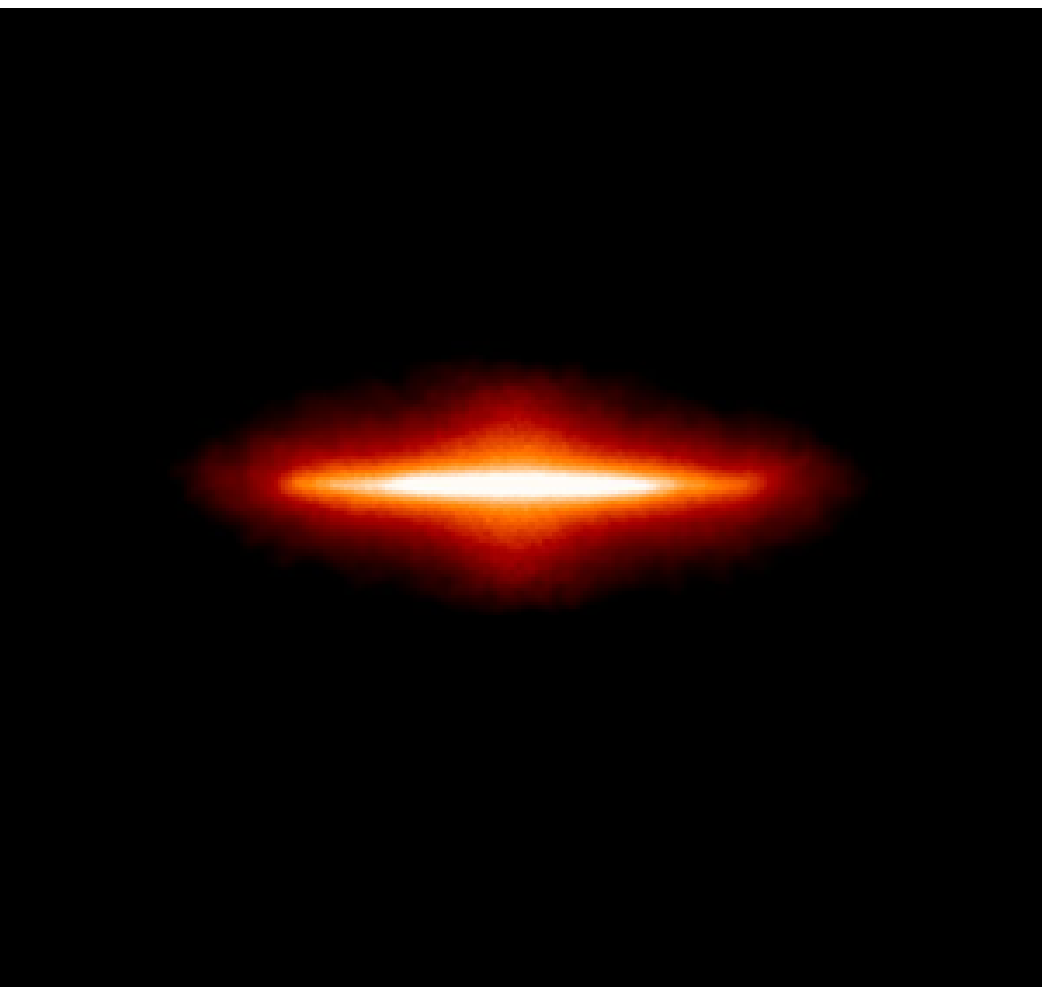}
}
\caption{The same as in
  Fig. \ref{fig:MapGA2}, but for the AqC5 simulation.  Box size is $57$ kpc.  }
\label{fig:MapAqC5}
\end{figure*}

 To quantify the kinematics of a galaxy it is customary to
  consider the distribution of orbit circularities of the star
  particles. The circularity $\epsilon$ of an orbit is defined as the
  ratio of the specific angular momentum in the direction
  perpendicular to the disk, on the specific angular momentum of a
  reference circular orbit: $\epsilon=J_z / J_{\rm circ}$.  \cite{Scannapieco09} computed the latter
  quantity as $J_{\rm circ} = r \cdot v_c (r) = r \sqrt(GM(< r)/r),$
  where r is the distance of each star from the centre and $v_c (r)$
  the circular velocity at that position. Abadi et al. (2003) instead
  define $J_{\rm circ}=J(E)$, where $J(E)$ is the maximum specific
  angular momentum allowed given the specific binding energy $E$ of
  each star; in this way $\epsilon < 1.$

  In Figure \ref{fig:jhist}, we show the histograms of
  circularities of all star particles within $R_{\rm vir}/10$ for our
  GA2 and AqC5 simulations, using both methods outlined above.

We will use the second method in the rest of the present work, but
show the results for both of them in this Figure to facilitate
comparison with other works in literature that use the first one.  The
visual impression given by Figures~\ref{fig:MapGA2} and
\ref{fig:MapAqC5} is confirmed by these distributions: at redshift
$z=0$ both histograms show a prominent peak at $\epsilon\sim1$, where
stars rotating on a disk are expected to lie.  The bulge 
component, corresponding to the peak at $\epsilon\sim0$, is
quite small in both cases, somewhat larger for GA2 than for AqC5.  We
estimate $B/T$, the ratio of bulge over stellar mass within $R_{\rm
  gal}$, by simply counting the counter-rotating stars and doubling
their mass, under the hypothesis that the bulge is supported by
velocity dispersion and thus has an equal amount of co- and
counter-rotating stars.  This kinematical condition selects both halo
and bulge stars. Since our definition is based on the sign of the
quantity $J_z/J_{\rm circ}$, it does not depend on the method used to
evaluate the circularity distributions.  The resulting ratios are
$B/T=0.20$ for GA2 and $B/T=0.23$ for AqC5.  
As a matter of fact, \cite{Scannapieco10} analysed a synthetic image
of a simulated spiral galaxy with standard data analysis tools and
showed that the definition of $B/T$ based on all counter-rotating
stars overestimates what would be measured by an observer.

Even if the peak at
$\epsilon\sim0$ is higher for GA2, the total counter-rotating stellar
mass is larger for AqC5. This is due to the larger stellar halo
component of GA2, also visible in Figure~\ref{fig:MapGA2}.

In Figure \ref{fig:vrot}, we show the rotation curves of GA2 and AqC5
at redshift $z=0$.  We show the total rotation curve, and the
contribution of DM, gas and stars separately for both simulations.
Both galaxies have a remarkably flat rotation curve, reaching their
maxima at 11.3 (AqC5) and 11.7 (GA2) kpc, after which they gently
decline.  The maximum rotation velocities are 270 and 299 km/s
respectively, about 20 per cent higher than their circular velocities
at the virial radius (and, incidentally, 20 per cent higher than the
rotation velocity of the Milky Way at the solar radius, consistent
with e.g. \citealt{Papa11}).

From the visual appearance, from the shape of the rotation curves and
from the distribution of circularities, it is clear that both
simulated galaxies are disk-like and with a modest central mass
concentration. This finding is at variance with respect to our earlier
results published within the Aquila Comparison project
\citep{Scannapieco12}. The inclusion of metal-dependent gas cooling
and, more important, the inclusion of kinetic feedback are the reason
for this improvement.

\begin{figure}
\centerline{
\includegraphics[scale=0.25]{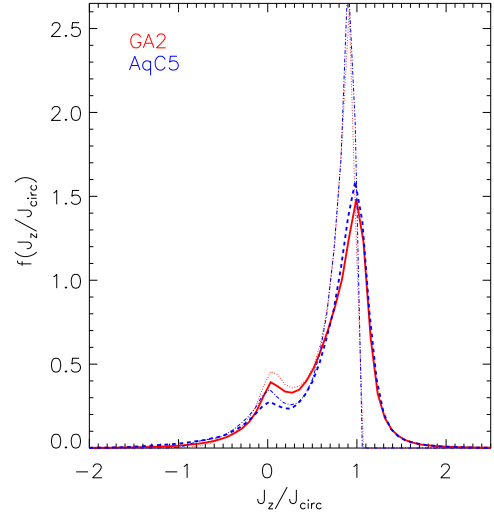}
}
\caption{  Distribution $f(\epsilon)/d\epsilon$ of the circularity
  parameter for the GA2 (red continuous  and dotted lines) and AqC5
  (blue dashed and dash-dotted lines) simulations at $z=0$, for the stellar component,
  as a function of the circularity $\epsilon=J_z/J_{\rm circ}$. 
  Thick (continuous and dashed) lines show circularity evaluated as in Scannapieco et al
  (2009).  Dotted and dash-dotted thin lines, as in Abadi et al. (2003). Distributions
  are normalized so that $\int f(\epsilon) d\epsilon =1$ }
\label{fig:jhist}
\end{figure}

\begin{figure}
\centerline{
\includegraphics[scale=0.25]{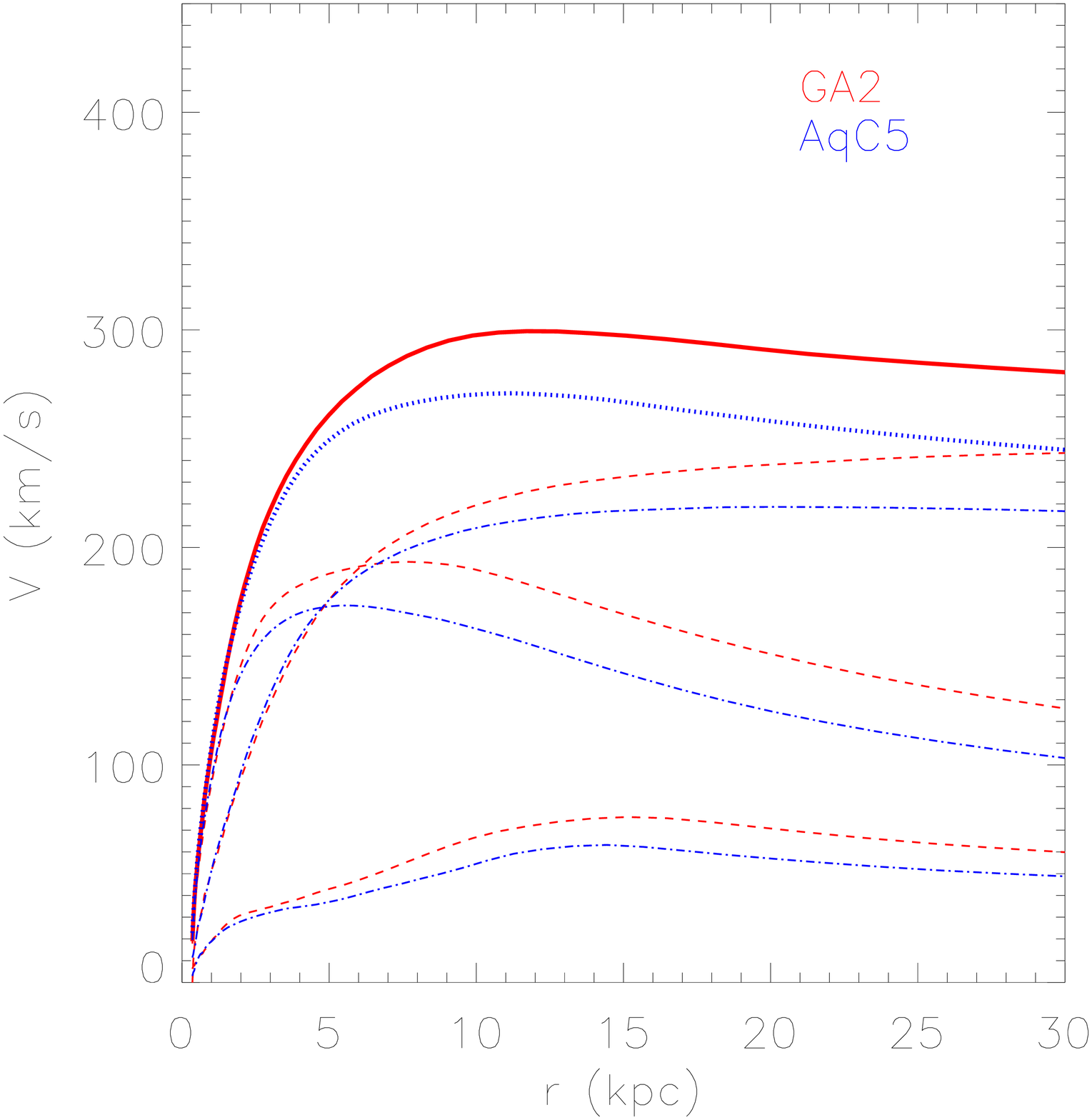}
}
\caption{
Rotation curves for the  GA2 (red) and AqC5 (blue) simulations. Thick lines
show the total curve, thin dashed lines show the contribution of the
DM, thin dotted lines that of the gas, and thin dotted-dashed lines
that of the stellar component.
}
\label{fig:vrot}
\end{figure}

\begin{table*}
  \caption{Basic characteristics of the different runs, done with MUPPI-std. 
    Column 1: Simulation name; 
    Column 2: Galaxy radius (kpc), set to $1/10$ of the virial radius; 
    Column 3: Disk scale radius $R_d$ (kpc), obtained using and
    exponential fit fo the surface density profile of stars (See
    Figure 7) ; 
    Column 4: B/T ratio inside the galaxy radius; 
    Column 5:  mass fraction associated to stars having
    circularity $\epsilon>0.8$ ; $\epsilon$ is calculated using the method
    of Scannapieco {\it et al} (2009), while we report in parenthesis the values obtained
    using the method by Abadi  {\it et al.} (2003);
    Column 6: Total  galaxy stellar mass, inside the galaxy radius (M$_\odot$);
    Column 7: Stellar mass in the bulge component inside the
      galaxy radius (M$_\odot$); 
    Column 8: Stellar mass in the disk component inside the
      galaxy radius (M$_\odot$);
    Column 9: Stellar mass within the virial radius (M$_\odot$);
    Column 10: Cold gas mass (M$_\odot$);
    Column 11: Baryon fraction within virial radius;
    Column 12: Fraction of total baryon mass in the galaxy;
    Column 13: Ratio between the specific angular momentum of
    stars and cold gas within the galaxy radius  and the specific angular momentum
    of the dark matter within the virial
    radius.
    The virial radius is defined as the radius of a sphere centered on
    the most bound particle of the halo, and encompassing and
    overdensity $\delta=200$ with respect to the {\it critical}
    density. The bulge mass is defined as twice the mass of
    counter-rotating stars, those having $j_z/j_{\rm circ}<0$ in
    Figure \ref{fig:jhist} and within 5 kpc from the galaxy center. 
  }
\begin{tabular}{c || c c || c c || c c c c c || c c || c ||}
\hline\hline Simulation & $R_{\rm gal}$ & $R_d$ & $B/T$&
$f(\epsilon>0.8)$& $M_\star$ & $M_{\rm *, bulge}$
&$ M_{\rm *, disk}$ & $ M_{\rm*,vir}$ & $M_{\rm cold}$ & $f_{\rm b,vir}$ & $f_{gal}$ & $f_J$\\
\hline
GA0 & 30.39 & 1.98 & 0.30 & 0.24 (0.30) & $1.78 \cdot 10^{11}$ & $5.37 \cdot 10^{10}$ & $1.25 \cdot 10^{11}$
& $ 2.11 \cdot 10^{11}$ & $7.17 \cdot 10^9$ & 0.13 &  0.06 & 0.28\\
GA1 & 30.37 & 3.93 & 0.22 & 0.47 (0.30) & $1.35 \cdot 10^{11}$ & $2.91 \cdot 10^{10}$  & $1.06 \cdot 10^{11}$ 
& $ 1.59 \cdot 10^{11}$ & $1.95 \cdot 10^{11}$ & 0.11 &  0.049 & 0.56 \\
GA2 & 29.98 & 4.45 & 0.20 & 0.51 (0.45) & $1.11 \cdot 10^{11}$ & $2.26 \cdot 10^{10}$  & $8.83 \cdot 10^{10}$ 
& $ 1.31 \cdot 10^{11}$ & $3.49 \cdot 10^{10}$ & 0.10 &  0.043 & 0.54 \\
\hline
AQ-C-6 & 24.15 & 4.08 & 0.24 & 0.51 (0.46) & $8.26 \cdot 10^{10}$ & $2.01 \cdot 10^{10}$  & $6.25 \cdot 10^{10}$ 
& $ 0.85 \cdot 10^{11}$ & $1.97 \cdot 10^9$ & 0.11 &  0.057 & 1.18 \\
AQ-C-5 & 23.82 & 3.42 & 0.23 & 0.55 (0.54) & $7.32 \cdot 10^{10}$ & $1.67 \cdot 10^{10}$  & $5.66 \cdot 10^{10}$ 
& $ 0.77 \cdot 10^{11}$ & $2.04 \cdot 10^{10}$ & 0.10 &  0.054 & 0.92 \\
\hline
\hline
\end{tabular}
\label{table:baryons_z0}
\end{table*}

In Table \ref{table:baryons_z0}, we list the main characteristics of
the simulated galaxies at $z=0$.  Here the disk scale radius $R_d$ is
estimated by fitting an exponential profiles to the stellar surface
density from 4 to 12 kpc. Stellar masses are reported within $R_{\rm
  gal}=R_{\rm Vir}/10$\footnote{Here and in the following, stellar
  masses include bulge, halo and disk components.}, while cold gas
includes multi-phase gas particles and single-phase ones with
temperature lower than $10^5$ K.  We also report the ratio of specific
angular momenta of baryons in the disk (stars and cold
gas) over that of the DM within the virial radius. We use all stars and
cold gas within  our galactic radius to evaluate such a ratio.
  This is a rough way to estimate the amount of loss of angular
  momentum suffered by ``galaxy'' particles: in case of perfect
  conservation, we would expect the specific angular momentum of these
  particles (condensed in the central region within $R_{\rm gal}$) to be the same as that of
  the dark matter halo (within $R_{\rm vir}$), so a value near unity
  is a sign of modest loss of angular momentum. Because we include
  all stars in the computation, we do expect to find some angular
  momentum loss.
From Table \ref{table:baryons_z0}, we note
the following characteristics:
\begin{itemize}
\item Both halos host massive disk galaxies.  The total
  stellar mass in the GA2 simulation is $1.02 \cdot 10^{11}$
  M$_\odot$, while it is $M_\star=6.77 \cdot 10^{10}$ M$_\odot$ for
  AqC5. As such, and as also witnessed by their circular velocities,
  these galaxies are more massive than the Milky Way.
\item The cold gas mass, that is assumed to be in the disk, is 28 per cent
  (GA2) and 26 per cent (AqC5) of the total disk mass, a value which is higher
  by a factor $2-3$ than for the Milky Way.
\item Our feedback scheme is efficient in expelling baryons from the
  halo. For GA2, the baryon fraction within the virial radius is 10
  per cent, compared to the cosmic 14.3 per cent, while the baryon
  mass of the galaxy (stars and cold gas) is 4.3 per cent of the total
  halo mass. These values are not far from those estimated for disk
  galaxies like the Milky Way. For AqC5 the baryon fraction within the
  virial radius is again 10 per cent, compared to the cosmic value of
  16 per cent, and the fraction of galaxy mass to total mass is 5.4
  per cent.  
\item The specific angular momentum of galactic baryons
  (cold gas and stars) in GA2 is 54 per cent of the specific angular
  momentum of the DM within the virial radius.  In the AqC5
  simulation, this fraction exceeds 1.  This shows that in our
    simulations, baryons in the galaxy retain a fair share of their
    initial angular momentum.  Thanks to our feedback scheme, we are
    thus able to prevent an excessive angular momentum loss, thereby
    allowing the formation of extended gaseous and stellar disks.

\end{itemize}

\begin{figure}
\centerline{
\includegraphics[scale=0.25]{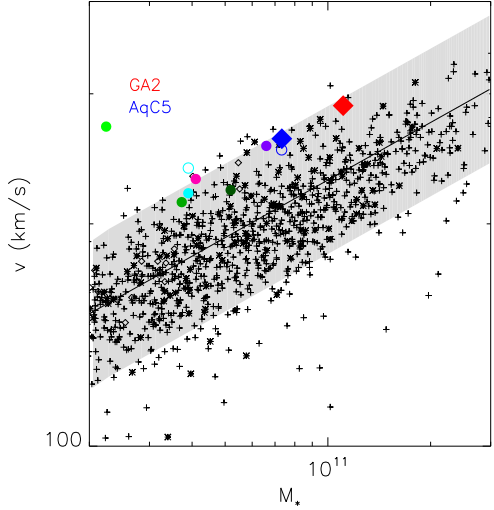}
}
\caption{ Tully-Fisher relation for our two simulated galaxies, GA2
  (red diamond) and AqC5 (blue diamond). Galaxy mass is the stellar
  mass inside $R_{\rm gal}$; velocities are estimated from the
  circular velocity profile at $2.2 R_d$. The line is the fit to
  observations of disk galaxies done by Dutton {\it et al.}(2011). The
  grey area shows an interval of 0.1 dex around the fit. Plus symbols
  are data points from Courteau {\it et al.} (2007), diamonds from
  Verheijen (2001) and asterixes from Pizagno {\it et al.}  (2007).
  We also show for reference the position in the plot of AqC5 from
  Marinacci {\it et al.} (2013), where we took $V$ as the circular
  velocity at 2.2 times their quoted disk radius from their Figure 18
  (purple circle); and the position of ERIS simulation, using their
  quoted $V_{\rm peak}$ (open cyan circle) and the circular velocity
  at 2.2 times their quoted i-band disk scale lenght, from their
  Figure 1 (filled cyan circle). The magenta circle shows the position
  of AqC5 simulated by Aumer {\it et al.}  (courtesy of M. Aumer and
  C. Scannapieco). Green filled circles show the position in the
    diagram of AqC5 simulations G3-TO, G3-CS and R-AGN from
    Scannapieco {\it et al.} 2012 (light green: G3-CS; medium green:
    G3-TO; dark green: R-AGN). For these, circular velocity is
    evaluated at the radius containing half of the galaxy stellar
    mass; blue empty circle refer to our AqC5 simulations, when using
    the same definition of circular velocity. }
\label{fig:tf}
\end{figure}

\begin{figure}
\centerline{
\includegraphics[scale=0.25]{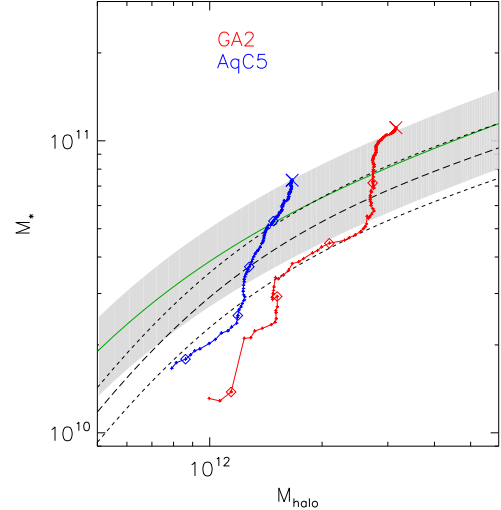}
}
\caption{
Evolution of the relation between stellar mass and halo mass ratio of
GA2 (red) and AqC5 (blue). Crosses correspond to results at $z=0$,
with the lines (with plus signs) showing the corresponding
evolutions. Diamonds show redshifts $z=2$,$1.5$,$1.0$,$0.5$.
The continuous green line shows the fit to the star formation
efficiency of DM halos obtained by Guo et al. (2010) with the
abundance matching technique. The grey area is an interval of 0.2 dex
around it. The dashed black lines is the same fit, as given by Moster
et al. (2010), while the dotted lines show an interval of 1$\sigma$ on
the normalisation of the fit.  }
\label{fig:sfeff}
\end{figure}

Figures \ref{fig:tf} and \ref{fig:sfeff} show the stellar Tully-Fisher
relation and the stellar mass {\it vs} halo mass relation for our two
simulations. Galaxy mass is the stellar mass inside $R_{\rm gal}$;
velocities are taken from the circular velocity profile, at $2.2
R_d$. As for the Tully-Fisher relation, we also plot the fit to
observations of disk galaxies presented \cite{Dutton11}; in grey, we
plot an interval $\pm 0.1$ dex around the fit for reference, and we
overplot observations from \cite{Ver01},\cite{Pizagno07} and
\cite{Courteau07}.  Symbols represent the position of our simulated
galaxies in the plot. As in \cite{Dutton11} we used the circular
velocity at 2.2 times the disk radius of our galaxies (see Table
\ref{table:baryons_z0}).

For reference we also show the position in the same plot of AqC5
simulated by \cite{Marinacci13}, by \cite{Guedes11} and by
\cite{Aumer13}. We also show the position of three among the AqC5
simulations performed in \cite{Scannapieco12}, namely models G3-TO,
G3-CS and R-AGN. Both our simulated galaxies tend to lie on the high
side of the range allowed by observational results.  We note that this
is a common trend in recent simulated disk galaxies. This could be
related to some remaining limitations shared in all {\sffb} models
used, or to the way in which simulations are compared to observations.
Our simulated AqC5 galaxy has a stellar mass in good agreement with
the finding of \cite{Marinacci13}, but higher than that found by other
groups.  In Table~\ref{table:baryons_z0}, we report the mass fraction
of stars having a circularity larger than $\epsilon=0.8$; this
quantity can be considered as a rough estimate of the prominence of
the ``thin'' disk. Our higher resolution runs have fractions $f=0.51$
for GA2 and $f=0.55$ for AqC5, showing that the disk component of our
simulations is significantly more important than that of most runs
showed in the Aquila comparison project\footnote{Runs in the Aquila
  comparison paper having $f(\epsilon)>0.35$ also have very high peak
  velocities, $v_{1/2} > 390$ km s$^{-1}$.}, and similar to that
reported e.g. by \cite{Aumer13}, $f(\epsilon)=0.55$. On the other
hand, our low-redshift SFR, shown in Figure \ref{fig:sfr}, lies on the
high side of the values shown e.g. in \cite{Aumer13} for the same halo
mass range. This suggests that the stellar mass excess we found could
be due to late-time gas infall and its conversion in stars. Taken
together, these data suggests that, with the parameter values used in
this paper, our feedback scheme could be still slightly inefficient,
either in quenching star formation at low redshift, or in expelling a
sufficient amount of gas from haloes at higher redshift. But we have
not yet performed a full sampling of the parameter space of our model,
a task that would require the use of a much more extended set of ICs.

Figure \ref{fig:sfeff} shows the relation between stellar mass in the
galaxy and virial mass of the DM haloes. The green solid line shows the
estimate obtained by \cite{Guo10} using the abundance matching
technique. Following \cite{Marinacci13}, the grey area marks an
interval of $\pm$0.2 dex around it. Black dashed line gives the
relation obtained by \cite{Moster10}, with dotted lines corresponding
to their 1$\sigma$ error on the normalisation.  Symbols represent the position of our galaxies
on this plot, while the lines give the evolution in time of the baryon
formation efficiency during the simulations.  
Again, we tend to lie on the high side of the allowed range of stellar
masses.  Both the position in the Tully-Fisher relation and that in
the stellar {\it vs} halo mass suggest that our simulated galaxies
still are slightly too massive, given the DM halo in which they
reside.  As for the baryon conversion efficiency, we follow the
  definition provided by \cite{Guedes11}: $\eta= (M_*/M_{\rm
    vir})\,(\Omega_m/\Omega_{\rm baryon})$. We obtain $\eta=0.25$ for
  GA2 and $\eta=0.29$ for AqC5. \cite{Guedes11} quote $\eta=0.23$ for
  Eris, but they defined their $M_{\rm vir}$ as the mass contained in
  a sphere having an overdensity of $\sim 93$ times the critical
  density, while we use $\delta_{200}$ (using their definition we
  would get 0.20 for GA2 and 0.26 for AqC5). Moreover, their halo mass
  is smaller than ours, with $M_{\rm eris, 93}=7.9 \cdot 10^{11}$
  M$_\odot$. \cite{Aumer13} also simulated AqC5 and found a
  significantly lower barion conversion efficiency, $\eta=0.15$. In
  fact, while their simulation stays within $1\sigma$ from the fit by
  \cite{Moster10}, our runs are within $3\sigma$ (2.79 and 2.75 for
  GA2 and AqC5 respectively). Note however that the exact relation
  describing the mass dependence of the baryon conversion efficiency
  also depends on the chosen cosmology; for instance, the cosmological
  model used by \cite{Guo10} is the same as for the Aquila series of
  simulations, and at the half mass scale of our runs, it gives a
  significantly higher stellar mass. As for the Tully-Fisher relation,
  this indicates that our predicted stellar mass are still slightly
  too high.

\subsection{ISM properties and the Schmidt-Kennicutt relation}
\label{section:ism}

In Figure~\ref{fig:surfprof} we show the surface density profiles of
various baryonic components, namely stars, total cold gas, atomic,
molecular, and hot gas, for the GA2 and AqC5 simulations (left and
right panel, respectively). Both galaxies exhibit an exponential
profile for the stellar surface density, with some excess in the
centre due to the small bulge component and a break in the external
part, which is more evident for GA2.  The black lines show exponential
fits to the stellar density profiles, performed in the range from 4 to
12 kpc, so as to exclude both the bulge region and the external
regions where the exponential profile breaks.  The resulting scale
radii $R_d$, reported in Table~\ref{table:baryons_z0}, are $R_d=4.45$
kpc for GA2 and $R_d=3.42$ kpc for AqC5.  \footnote{ If we change
  the radial range for the fit, e.g. to $r=(0,20)$ or $(4,20)$, or we
  perform the fit in linear-logarithmic rather than linear-linear
  scales, the values of $R_d$ remain in the range $(3.55,4.46)$ kpc
  for GA2 and $(2.92, 3.41)$ kpc for AqC5. The radial range $(4-12)$
  kpc always gives the lowest chi squared. Given the flatness of
    the rotation curve, such a change in $R_d$ is too small to
  significantly affect the resulting Tully-Fisher relation.}  The hot gas
profile, that includes both particles hotter than $10^5$ K and the hot
component of multi-phase particles, is rather flat around values of
about (1--3) {\surf}, thus describing a pervasive hot corona.  Cold
gas shows a rather flat profile within $\sim10$ kpc.  For both
galaxies it shows a central concentration followed by a minimum at
$\sim3-4$ kpc.  Gas densities are rather flat, with values of
$\sim20-30$ {\surf}, then dropping beyond 12-15 kpc.  The gas fraction
is then a strong function of radius, though we do not see a transition
to gas-dominated disks.  Atomic gas dominates the external regions and
flattens to values of 10 {\surf} \citep[see][]{Monaco12}, while
molecular gas dominates in the inner regions.  We verified that the
flatness of the profiles is typical of the feedback scheme adopted in
our simulations and for the chosen values of the model parameters.

Figure~\ref{fig:ism} shows further properties of the ISM, namely
pressure $P/k_B$, hot phase mass-weighted temperature $T_{\rm
  h}$ and average cold gas temperature $T$.  As for the
  latter, it is computed by considering only cold ($T<10^5$K) and
  multi-phase gas particles, and weighted the contribution of the
  multi-phase particles using their cold gas mass.  Despite of the
flatness of gas profiles, $P/k_B$ and $T_{\rm h}$ have exponential
profiles steeper than those of the gas density, with a slight
drop at the centre where the bar dominates.  This is due to the
stronger gravity of the stellar disc. Hot phase temperature raises
from $2\cdot10^6$ K to $10^7$ K towards the centre.  The average
temperature of disk particles drops with a steeper slope in the outer
regions, ranging from $2\cdot10^5$ K to $\sim10^4$ K at the disk
edge. These values correspond to sound speeds from 40 km/s to 10 km/s.
Given the multi-phase nature of these gas particle, it is not obvious
to decide to which observed phase these temperatures should be
compared with.  A sensible choice would be to compare these thermal
velocities with the velocity dispersion of the warm component visible
in 21 cm observations. In fact, these should correspond to the average
between cold and molecular phase on the one side, and hot phase heated
by SNe on the other side.  As shown by \cite{Tamburro09}, HI
velocities at the centre of galaxies can raise to $\sim$20 km/s.
Using stacking techniques on data on 21cm observations,
\cite{Ianjamasimanana12} robustly identified cold and warm components
in 21 cm emission lines, obtaining for the warm component velocity
dispersions from 10 to 24 km/s. These are lower by almost a factor of
two with respect to our velocities.  Our gas disks are thus likely too
warm and thick, and this may be a result of the entrainment of cold
gas by the hot phase.  On the other hand, in the companion paper by
Goz {\it et al.} (2014) we show that stellar velocity dispersion in
disks is in line with observational estimates, so the relative
thickness of gas disc does not propagate to stellar disks.  As a final
warning, we will discuss in Section \ref{section:resolution} how disk
thickness is strongly influenced by numerical two-body heating. This
indicates that a conservative (i.e. not aggressive) choice of the
softening is recommendable to reproduce disks with the correct
vertical scale height.

\begin{figure*}
\centerline{
\includegraphics[scale=0.25]{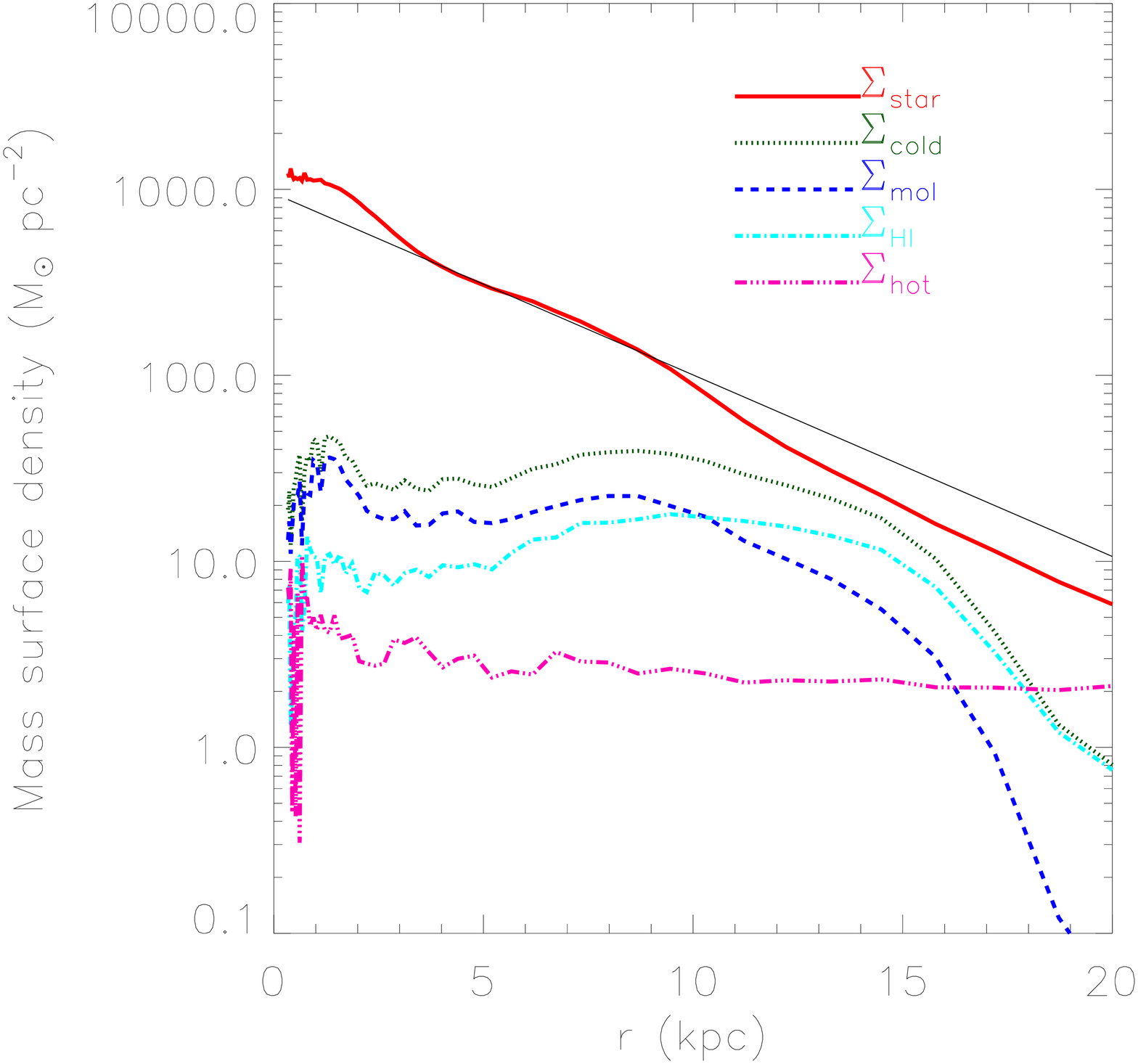}
\includegraphics[scale=0.25]{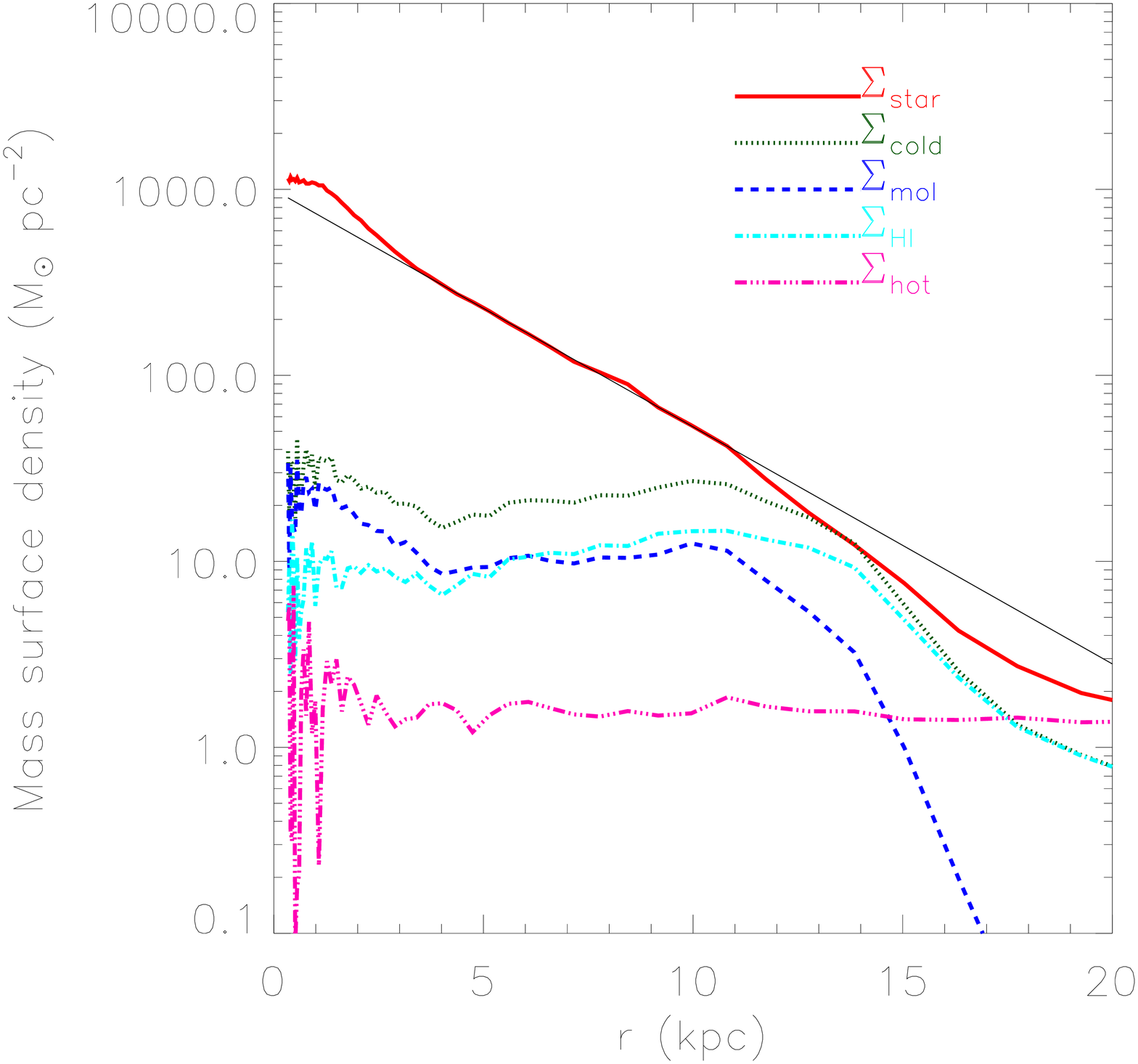}
}
\caption{ Surface density profiles for the GA2 (left panel) and AqC5
  (right panel) simulations, at redshift $z=0$. Red solid line is the
  stellar profile, gray dotted line the cold gas profile, blue dashed
  line the molecular hydrogen profile, cyan dotted-dashed line the
  neutral hydrogen profile, pink triple dotted-dashed line the hot gas
  profile. The black solid line is the exponential fit to the stellar
  surface density profile within the (4--12) kpc radial range.  }
\label{fig:surfprof}
\end{figure*}

\begin{figure*}
\centerline{
\includegraphics[scale=0.25]{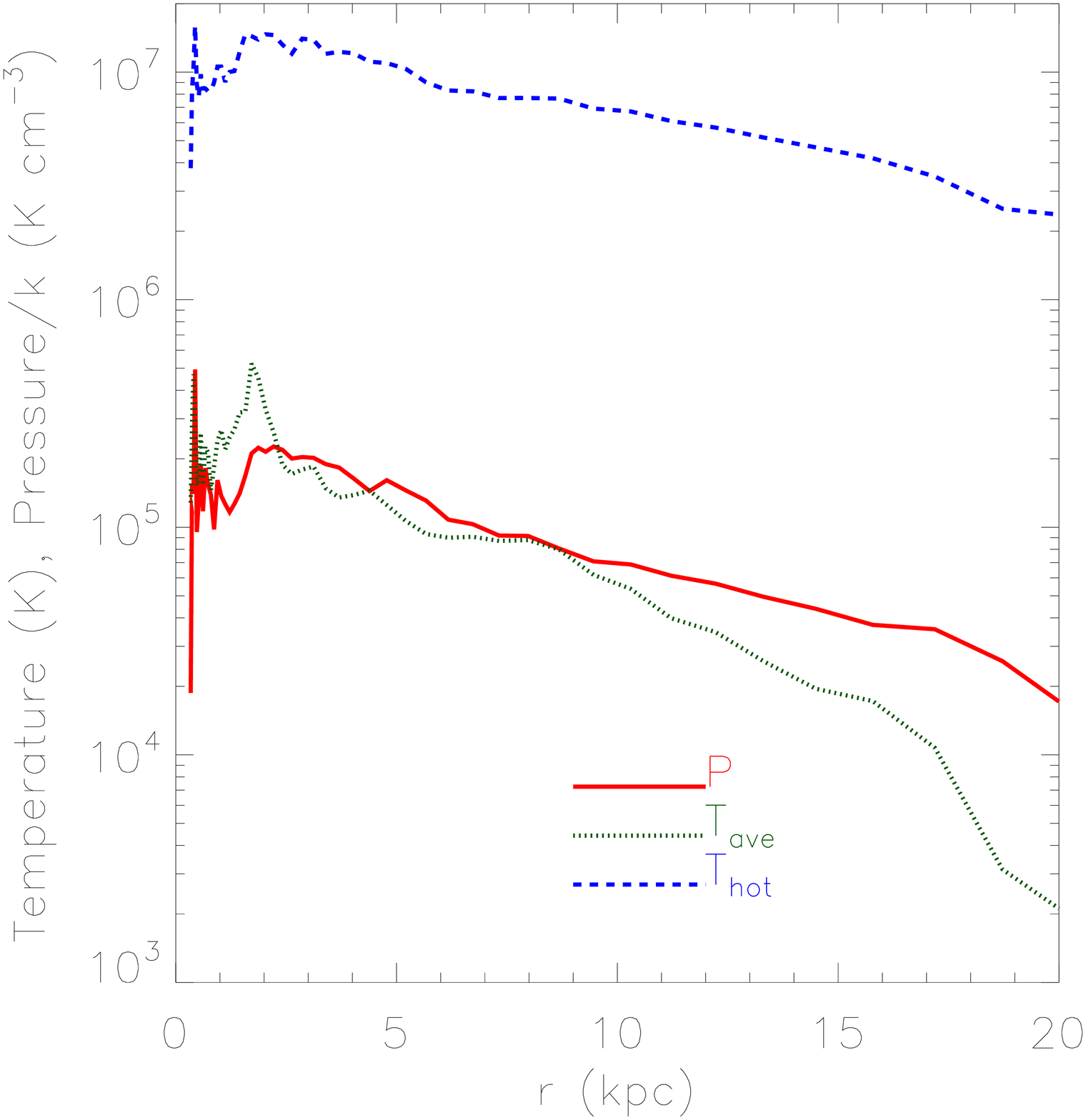}
\includegraphics[scale=0.25]{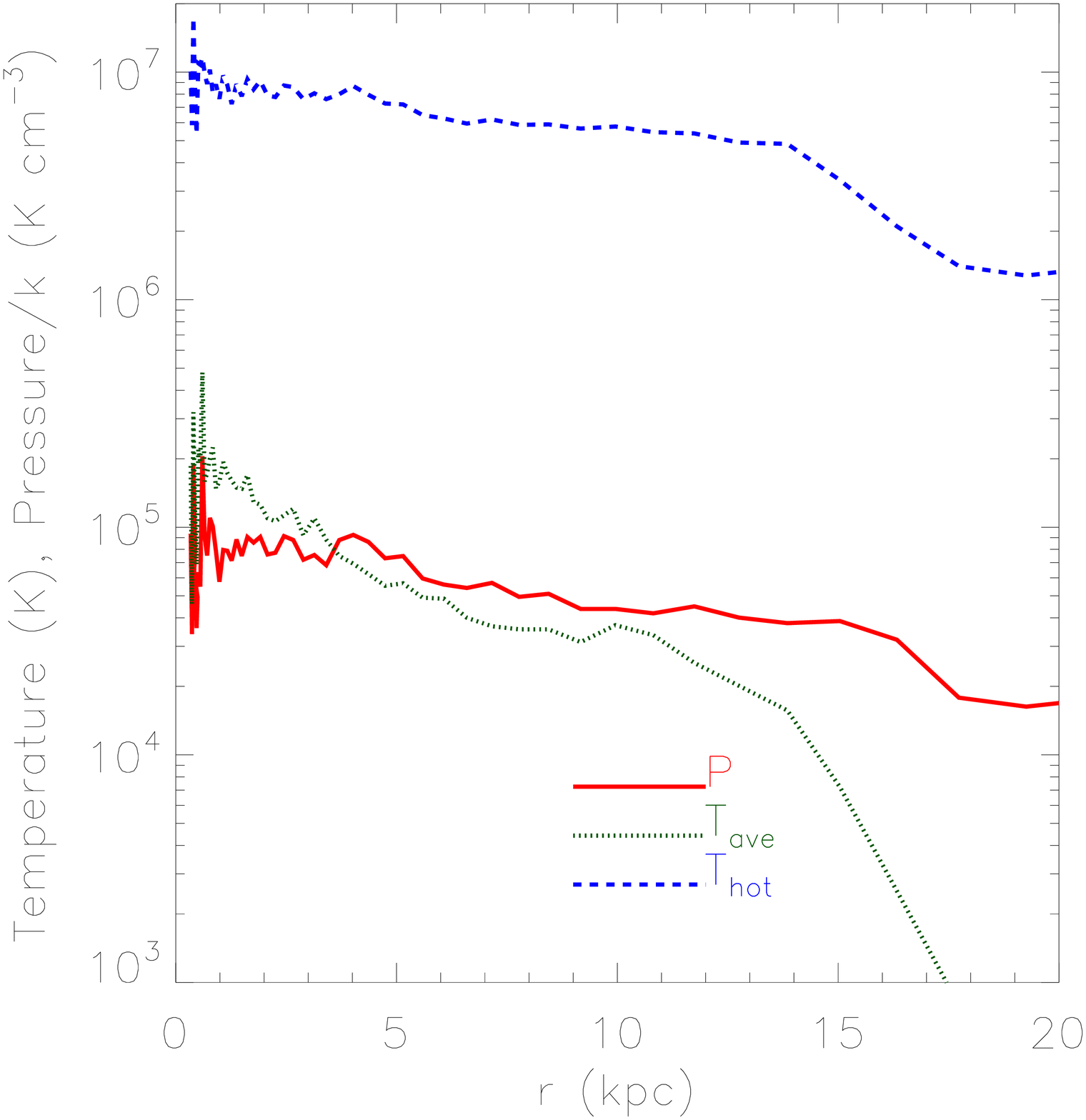}
}
\caption{
Properties of the ISM in the GA2 (left panel) and AqC5 (right
panel) simulations, at redshift $z=0$. In each panel, red continuous line shows the pressure
profile; grey dotted line the average cold gas  temperature profile; blue dashed
line the hot gas temperature profile. The average cold gas temperature
is computed using only cold ($T<10^5$K) and multi-phase gas particles,
and it is weighted using the cold gas mass. Hot gas temperature is mass-weighted. 
}
\label{fig:ism}
\end{figure*}

Figure~\ref{fig:sk} shows the standard Schmidt-Kennicutt (SK)
relation, gas total surface density versus SFR surface density, for
our simulated galaxies, again at redshift $z=0$.  As discussed in
\cite{Monaco12}, we do not impose the SK relation to our star
formation prescription, but obtain it as a natural prediction of our
model.  Contours show observational results from the THINGS galaxies
by \cite{Bigiel08}; lines refer to the two simulations.  Because gas
surface density profile is flat, points tend to cluster at large
values of surface densities.  Simulations stay well within the
observational relation. As also pointed out in \cite{Monaco12}, the
simulated relation tends to have a slope of 1.4, which is somewhat
steeper than that of $1-1.2$ found for the THINGS
galaxies. Furthermore, the external regions of simulated galaxies tend
to assume relatively low values of $\Sigma_{\rm sfr}$.

\begin{figure}
\centerline{
\includegraphics[scale=0.25]{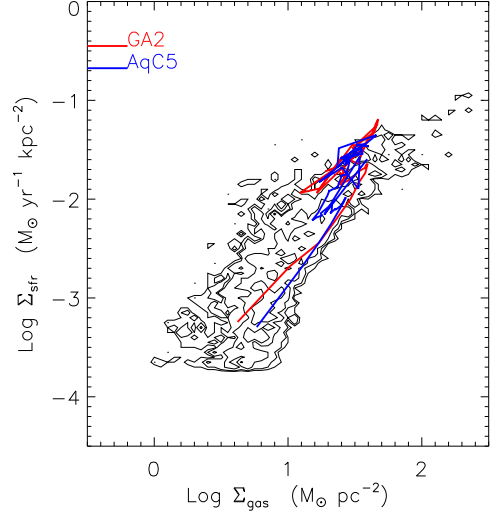}
}
\caption{
Schmidt-Kennicut relation for the GA2 (red) and AqC5
(blue) simulations. Contours show observational results from THINGS (Bigiel et al. 2008).
}
\label{fig:sk}
\end{figure}

\begin{figure}
\centerline{
\includegraphics[scale=0.25]{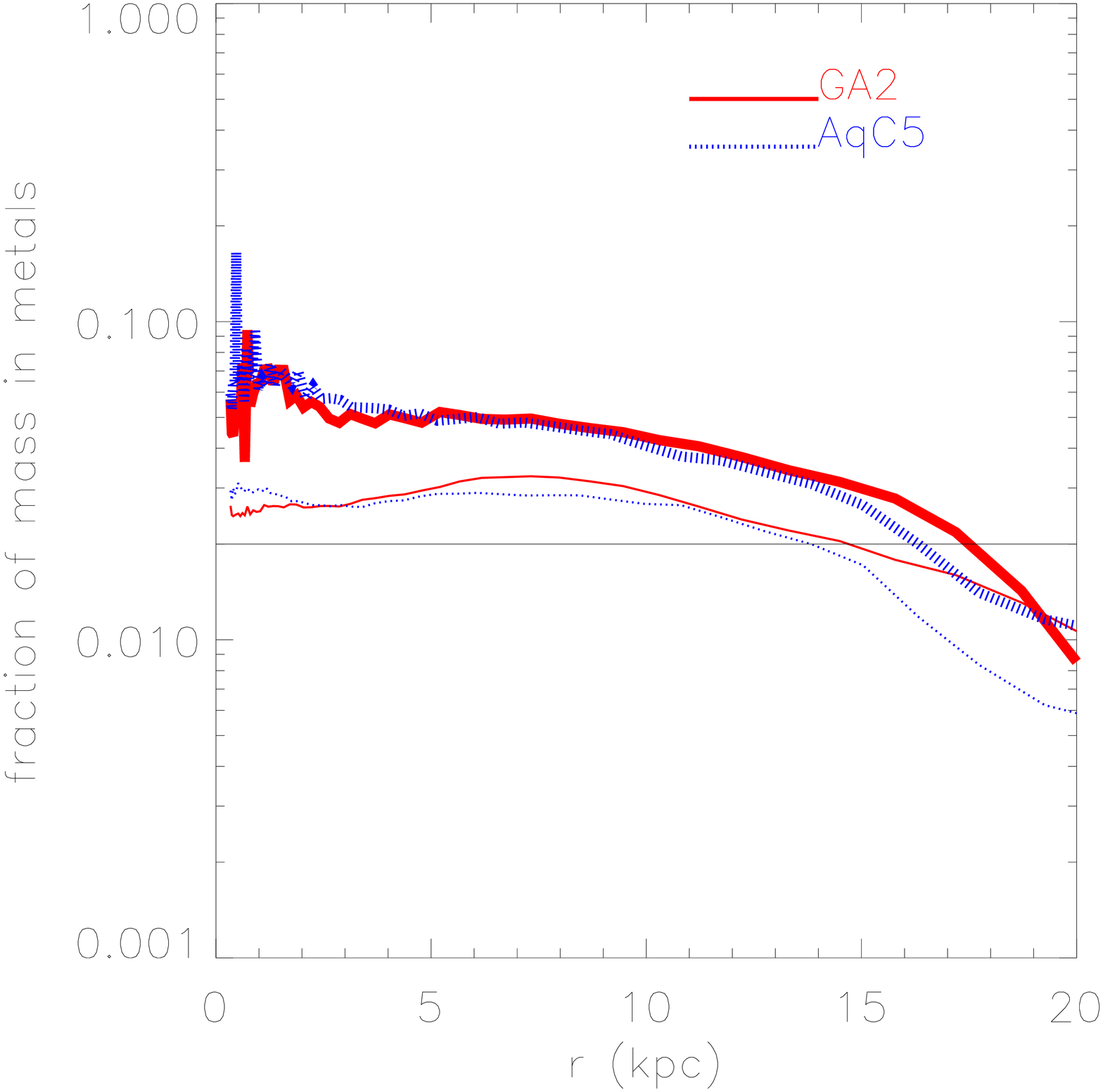}
}
\caption{
Metallicity profiles for our GA2 (red continuous and dashed lines) and
AqC5 (blue dotted and dotted-dashed lines) simulations at $z=0$. Thick
and thin lines refer to gas metallicity and stellar total metallicity,
respectively.  The horizontal thin line is the solar value of the
metallicity.}
\label{fig:metals}
\end{figure}

Figure~\ref{fig:metals} shows the total metallicity profiles for our
simulated galaxies, both for the stellar (thin lines) and for the gas
(thick lines) component. We used both cold and hot gas for the
latter profile. 
These profiles are very similar for the
two galaxies.  Stellar metallicity profiles are rather flat in the inner
10 kpc, with values of about $1.5Z_\odot$. On the contrary, gas
metallicities profiles,
that can be more directly compared with observations, 
get values of $\sim (3-4)Z_\odot$ at the center and have gradients of 
$\sim0.02$ dex/kpc.  These values are relatively flat if compared to
the Milky Way \citep[$\sim0.06$ dex/kpc, e.g.][and references
  therein]{Mott13} but are similar to those of M31 \citep[][and
  references therein]{Matteucci14}.  
The tendency of simulations to produce relatively flat abundance
profiles was already noticed by
\cite{Kobayashi11} and \cite{Pilkington12}.

Finally, Figure \ref{fig:densprof} shows the 
volume density profile for the stellar, gaseous and DM components.
The black line denotes a power law of slope $-1$.  In both simulated
galaxies, stars dominate over the DM at small radii, $r < 3$ kpc. The
depletion of gas in the inner 10 kpc, due to both star formation and
feedback, produces the sharp decrease of the corresponding density
profile.  We note that, in the inner 3 kpc, the profile of the DM
is shallower than the $-1$ slope predicted by \cite{Navarro96}. This
flattening could be due to baryonic processes, e.g the SNe feedback,
as suggested also recently e.g. by \cite{Governato12},
\cite{Pontzen13} and \cite{Zolotov12} who carried out simulations at
significantly higher resolution. In fact, as a caveat, we remind that
our resolution is formally just sufficient to resolve the scales where
a flatter DM density profiles is detected.

\begin{figure*}
\centerline{
\includegraphics[scale=0.25]{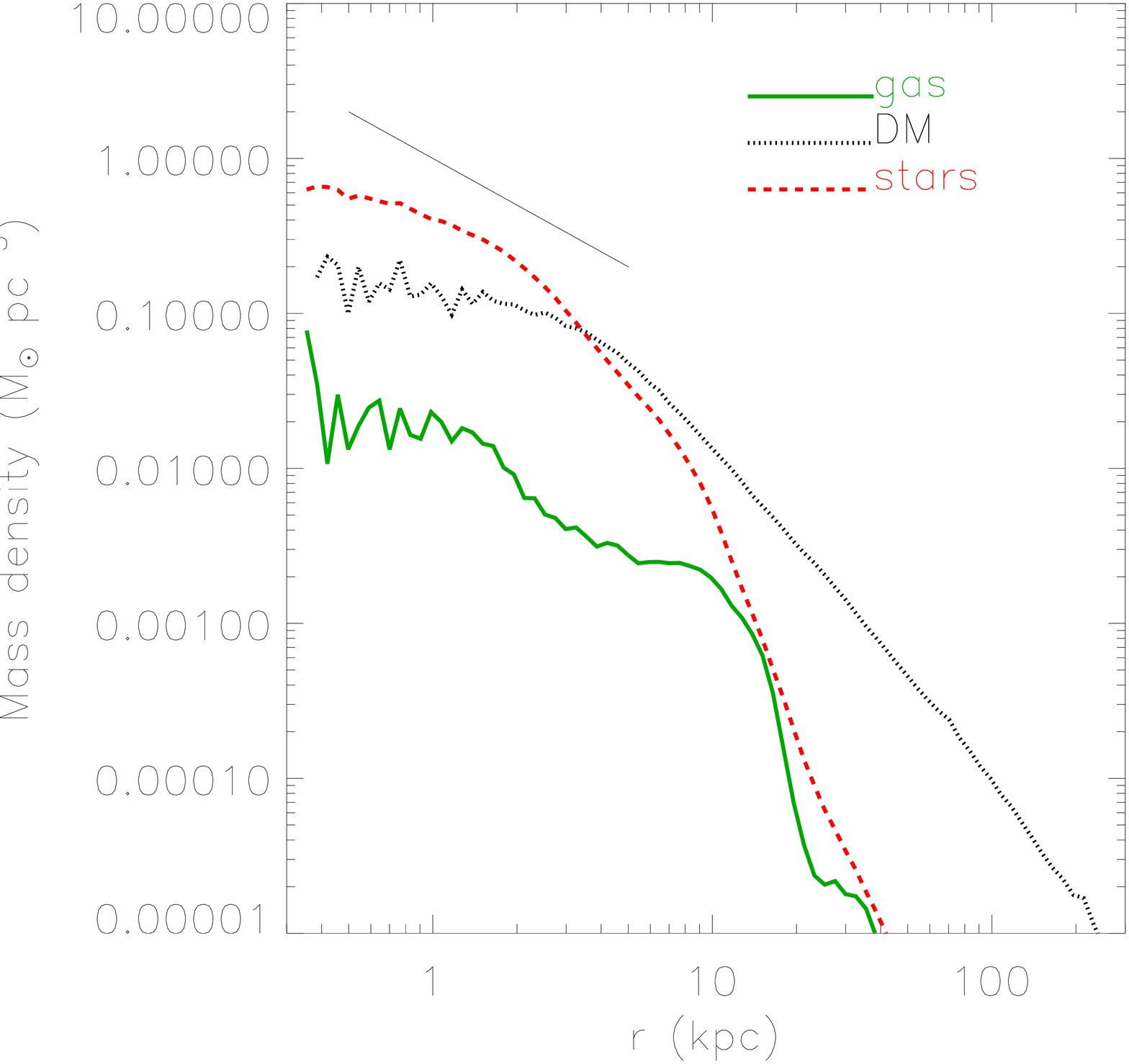}
\includegraphics[scale=0.25]{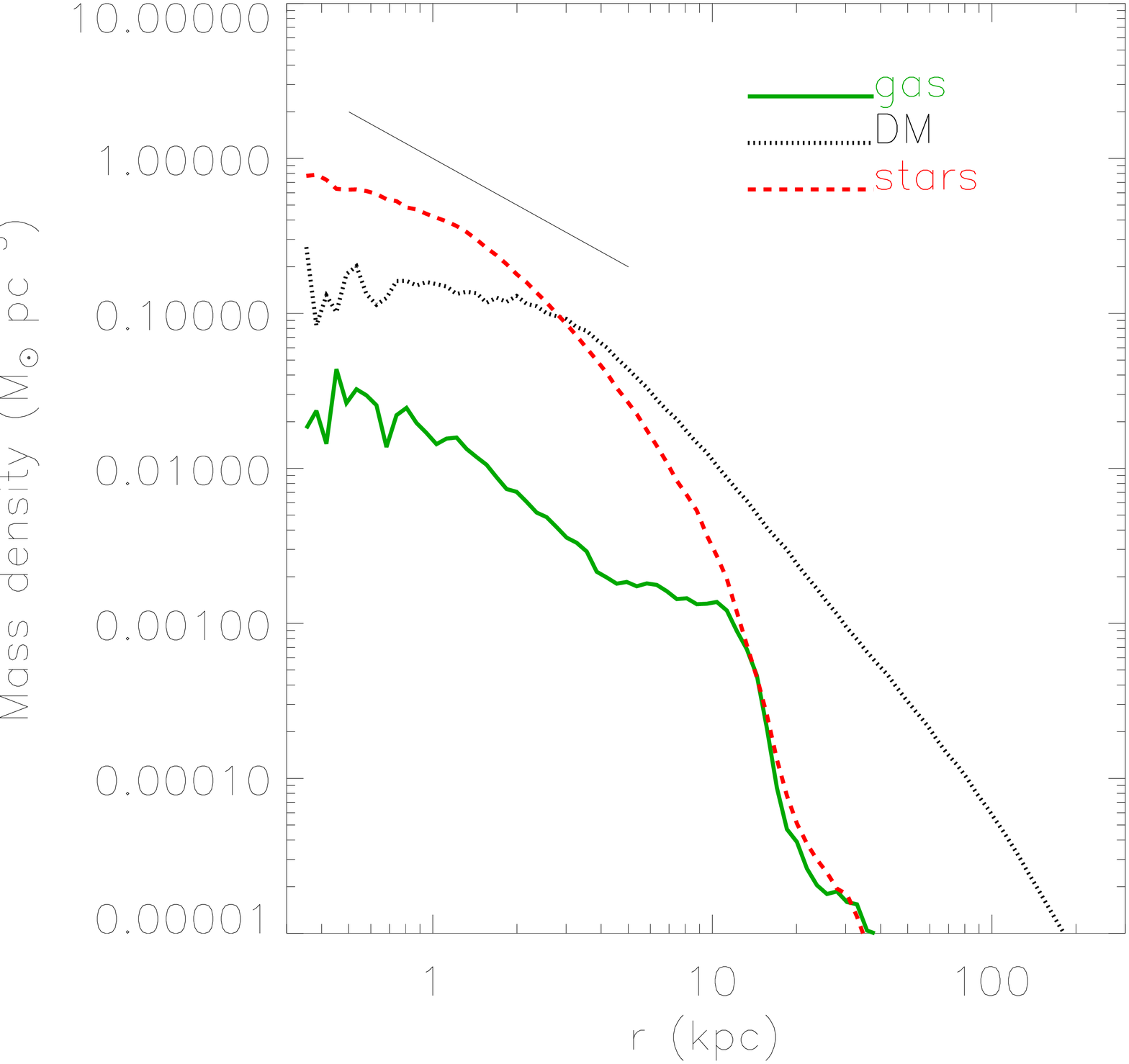}
}
\caption{Density profiles of various components for GA2 (left panel)
  and AqC5 (right panel) simulations, at $z=0$. Green solid lines show
  the gas profile, grey dotted lines the DM ones, red dashed lines the
  stellar ones. The black lines marks the slope $\rho \propto
  r^{-1}$.}
\label{fig:densprof}
\end{figure*}

\subsection{Evolution}
\label{section:evolution}

As a first diagnostic of the evolution of our simulated galaxies, we
show in Figure \ref{fig:sfr} the corresponding star formation
rates. Here we only plot the SFR relative to star particles which lie
inside the galaxy radius $R_{\rm gal}$ at redshift $z=0$.  Both
simulations have higher star formation rates at higher redshift, as
expected. They show a roughly bimodal distribution with a first,
relatively narrow peak and a further broad component.  GA2 has a more
peaky SFR and a substained rate of about 15 M$_\odot$ yr$^{-1}$
between redshift 0.85 and 0.25 (corresponding to a cosmic time of 5
and 8 Gyr respectively), then followed by a slow decline. AqC5 shows a
similar behaviour of its SFR, but slightly anticipated. The SFRs at
redshift $z=0$ is of about 9 (GA2) and 5 (AqC5) M$_\odot$ yr$^{-1}$.
These values are slightly larger than that measured for the
  Milky Way ($[2-5]$ M$_\odot$ pc$^{-2}$; see
  e.g. \citealt{Robitaille10}), but these galaxies are slightly more
  massive than our Galaxy as well.  Using the analytic fit of the main
  sequence of local star-forming galaxies proposed by
  \cite{Schiminovich07}, the expected SFRs would be 4.7 (GA2) and 3.6
  (AqC5) M$_\odot$ yr$^{-1}$, so these galaxies are well within the
  rather broad main sequence, though both of them are on the high-SFR
  side.  

\begin{figure}
\centerline{
\includegraphics[scale=0.26]{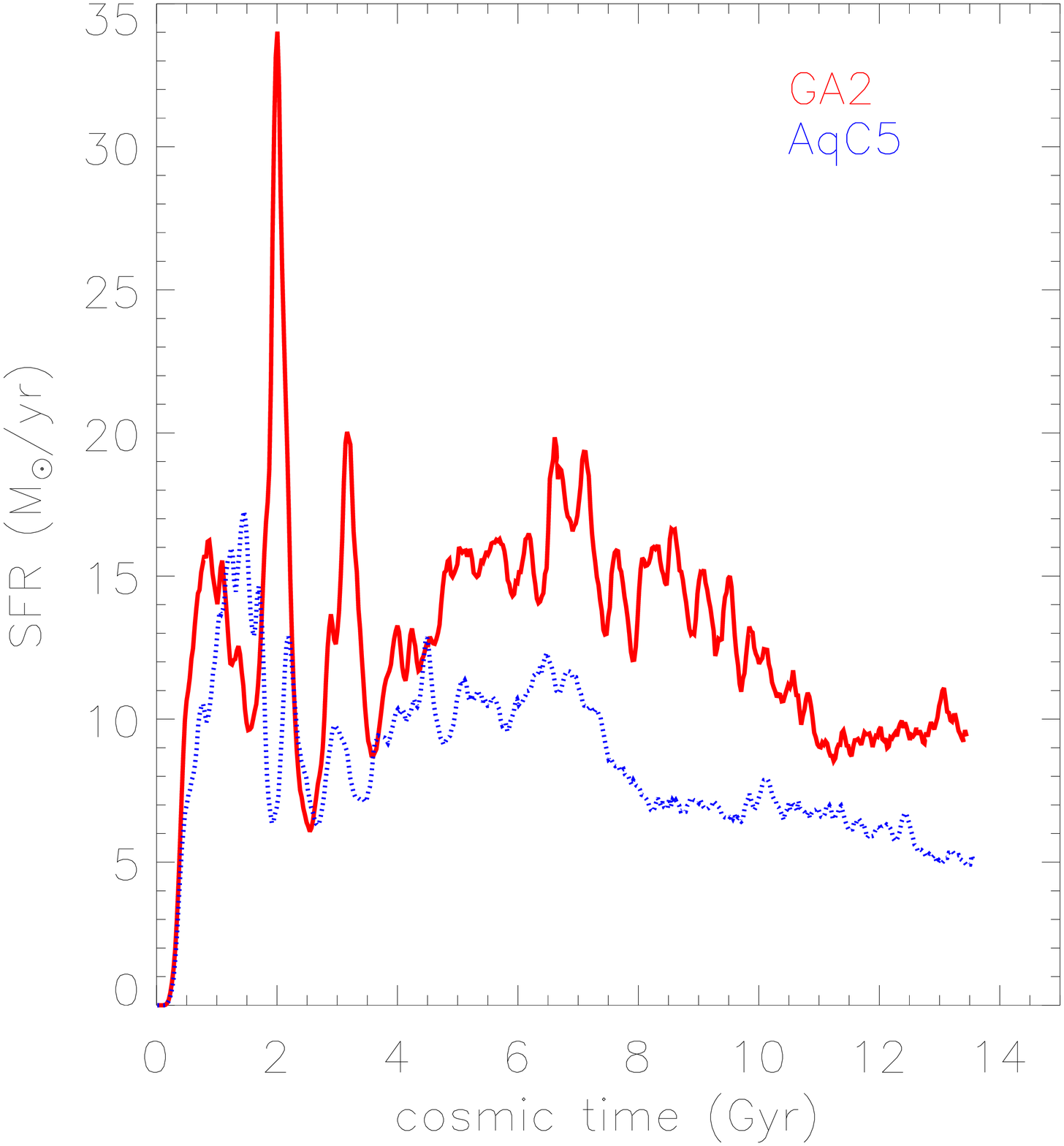}
}
\caption{
Star formation rates as a function of cosmic time, for simulations
AqC5 (blue dotted line) and GA2 (red continuous line). We plot the SFR relative to star particles which lie
inside the galaxy radius $R_{\rm gal}$ at redshift $z=0$.
}
\label{fig:sfr}
\end{figure}

\begin{figure*}
\centerline{
\includegraphics[scale=0.36]{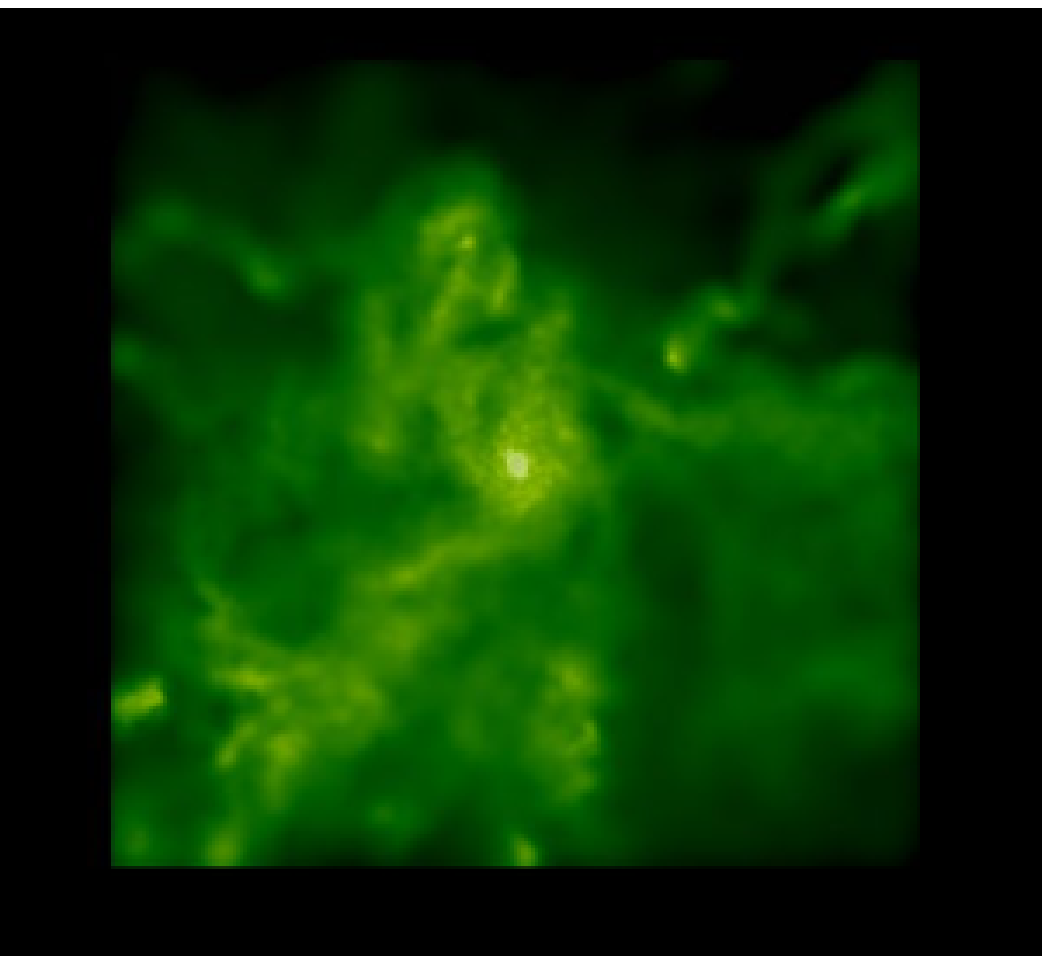}
\includegraphics[scale=0.36]{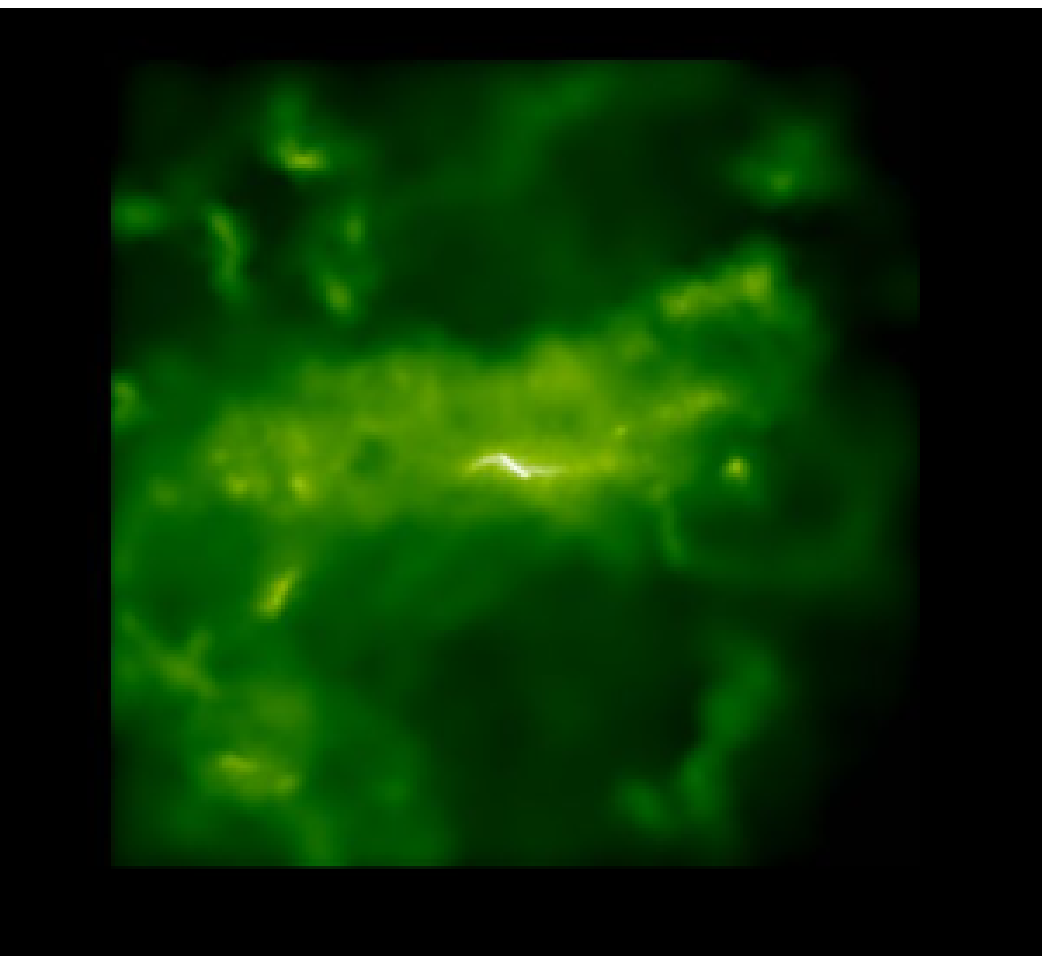}
\includegraphics[scale=0.36]{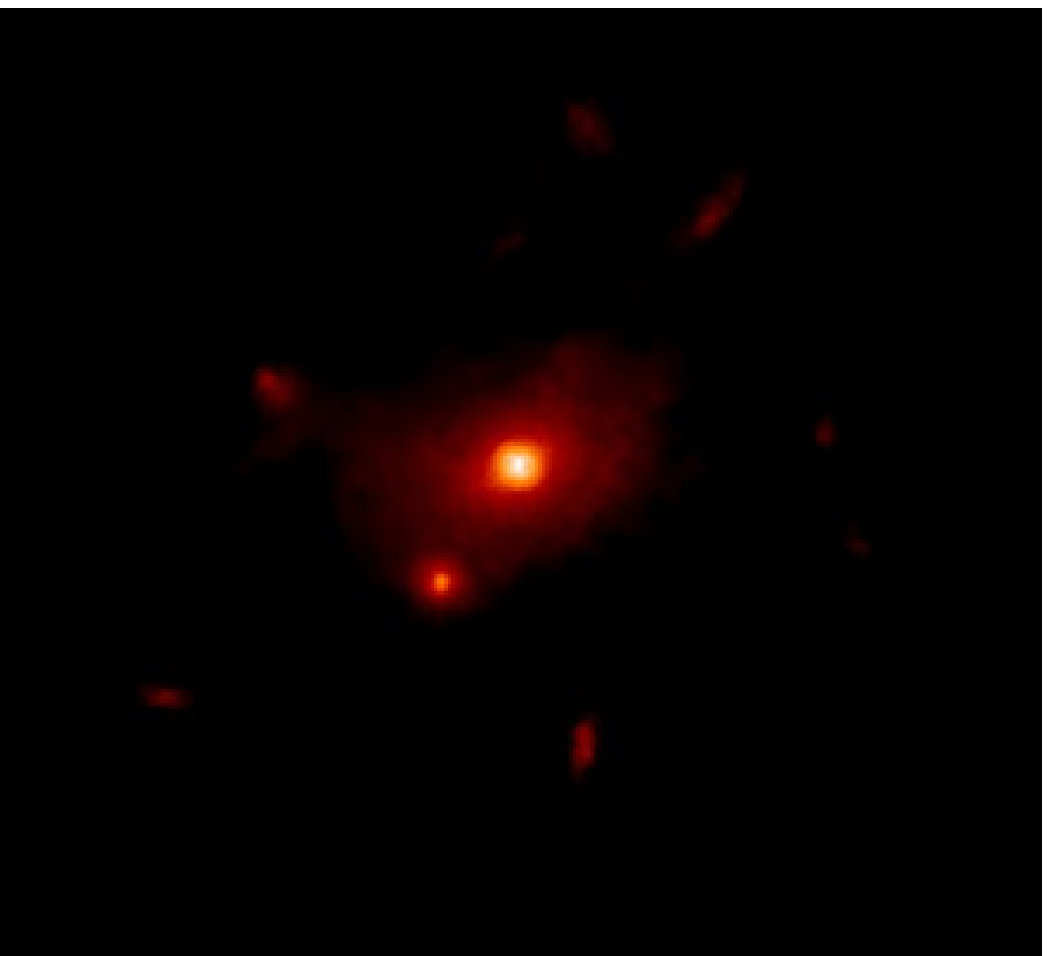}
\includegraphics[scale=0.36]{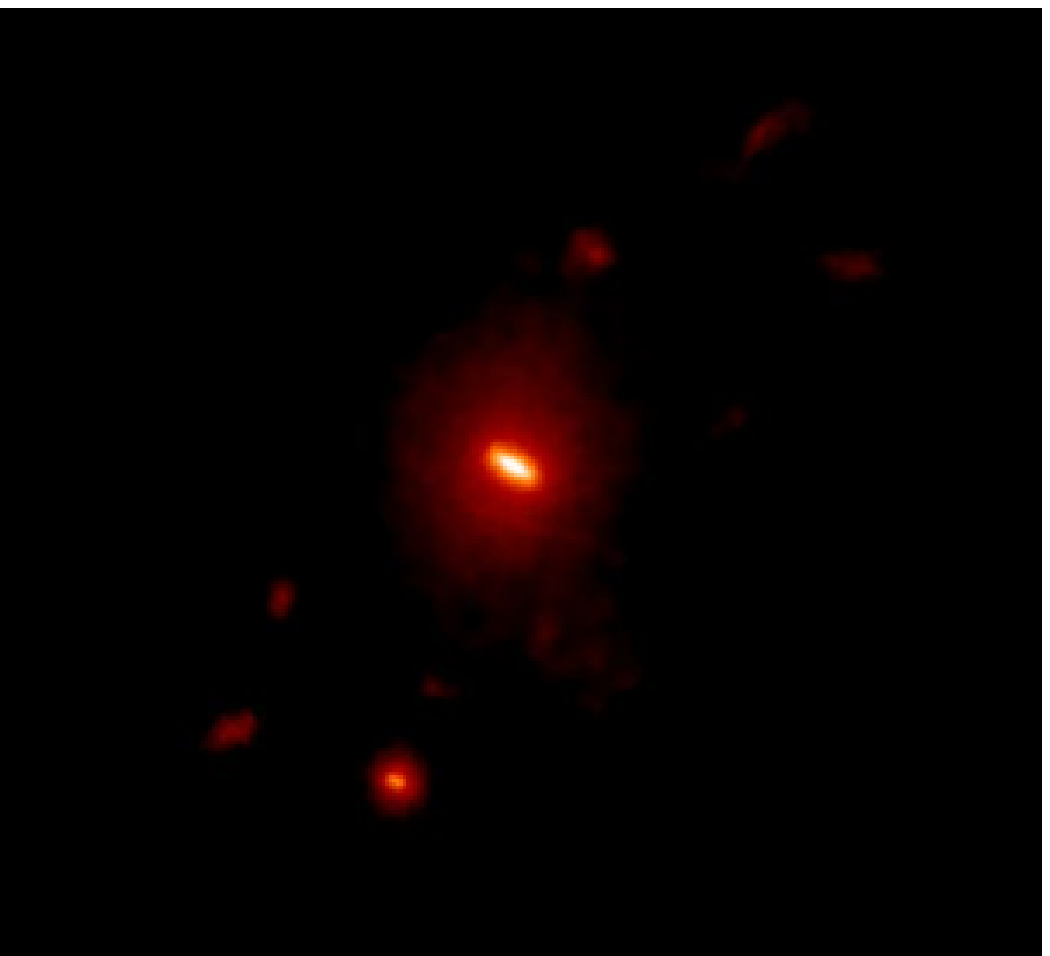}
}
\centerline{
\includegraphics[scale=0.36]{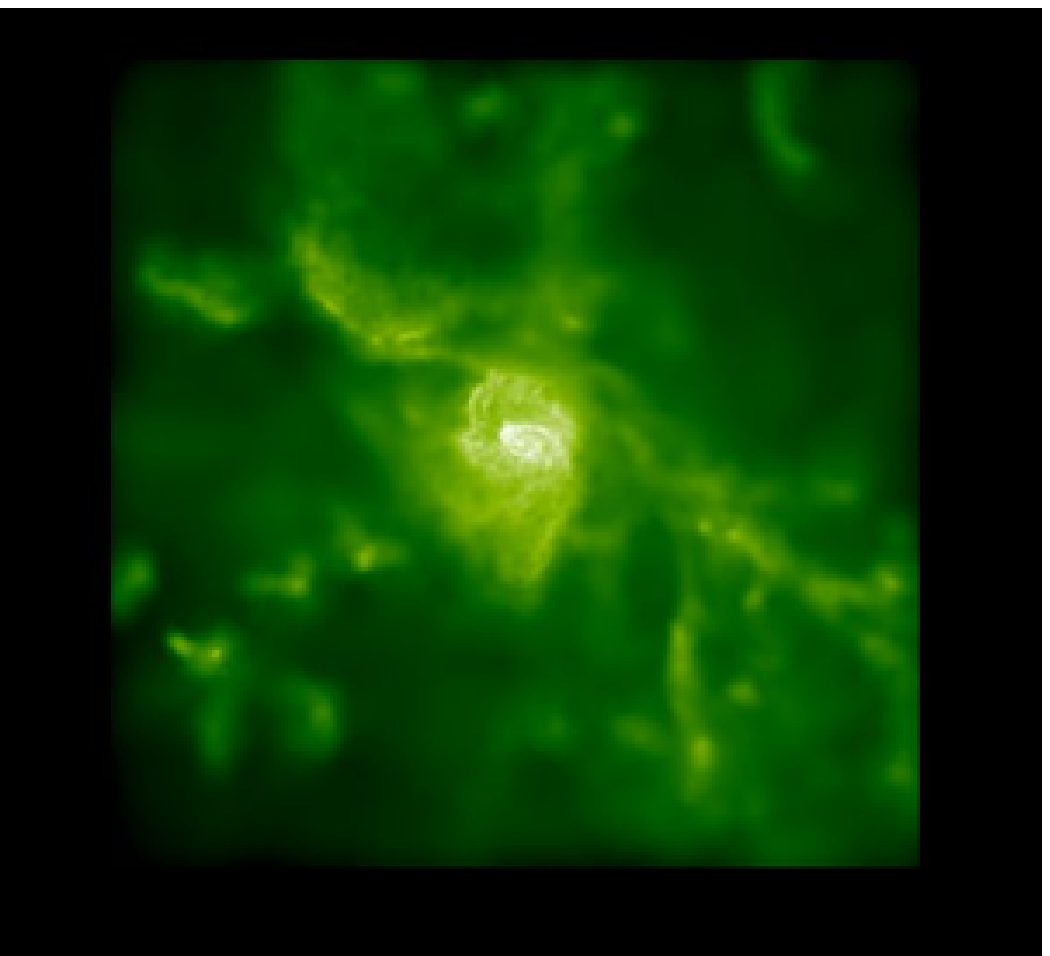}
\includegraphics[scale=0.36]{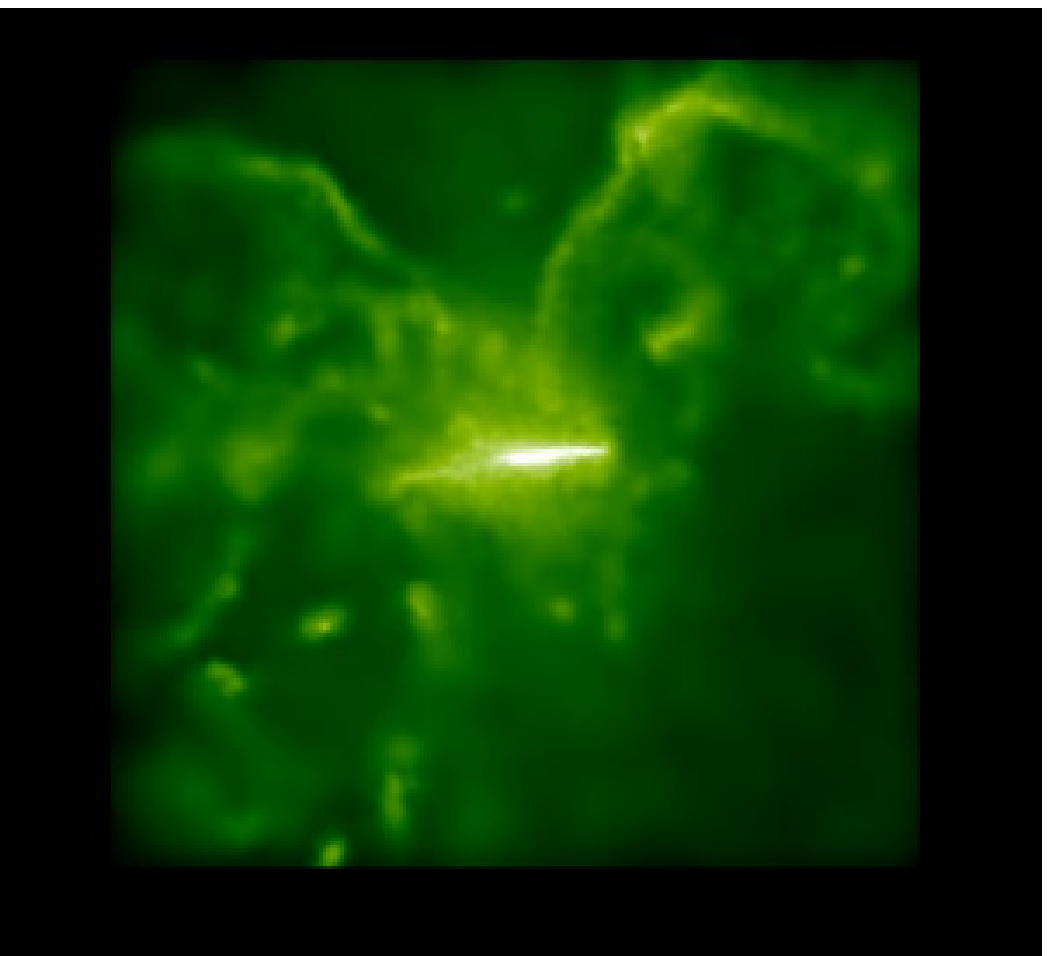}
\includegraphics[scale=0.36]{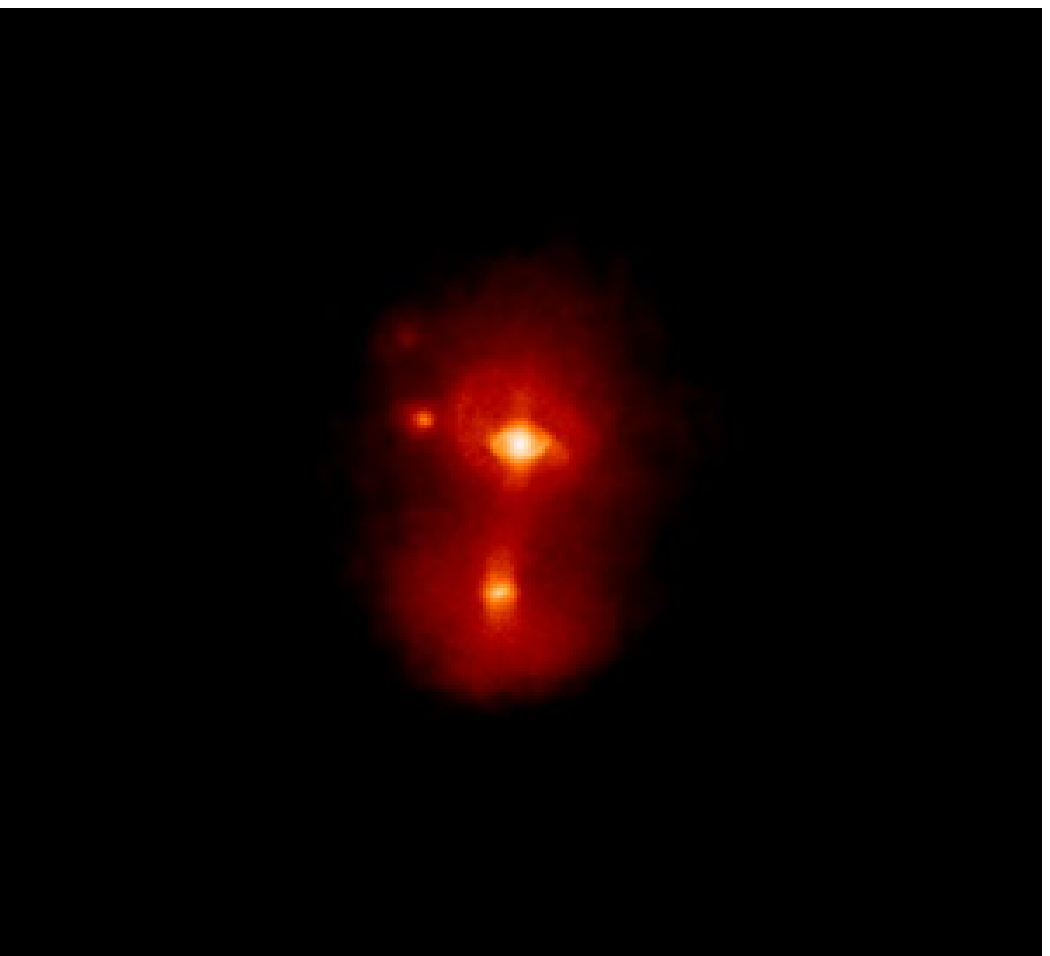}
\includegraphics[scale=0.36]{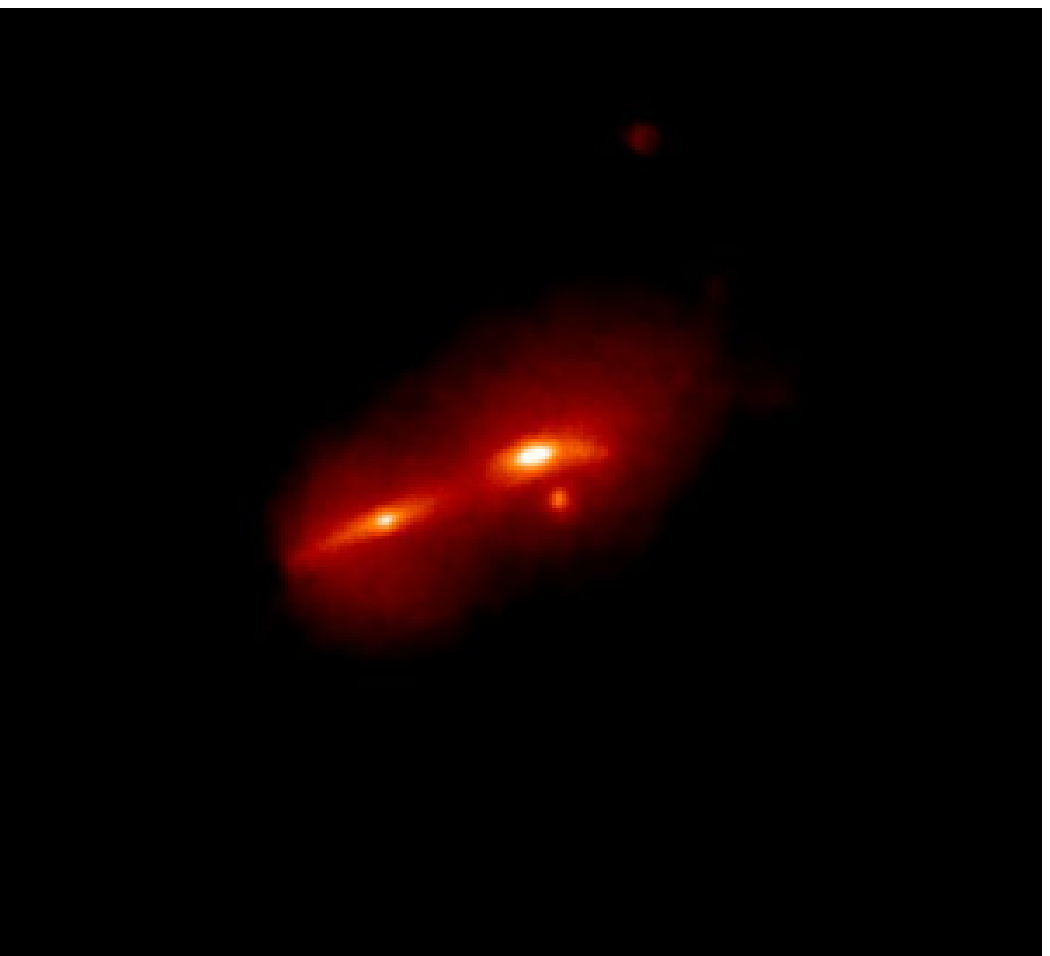}
}
\centerline{
\includegraphics[scale=0.36]{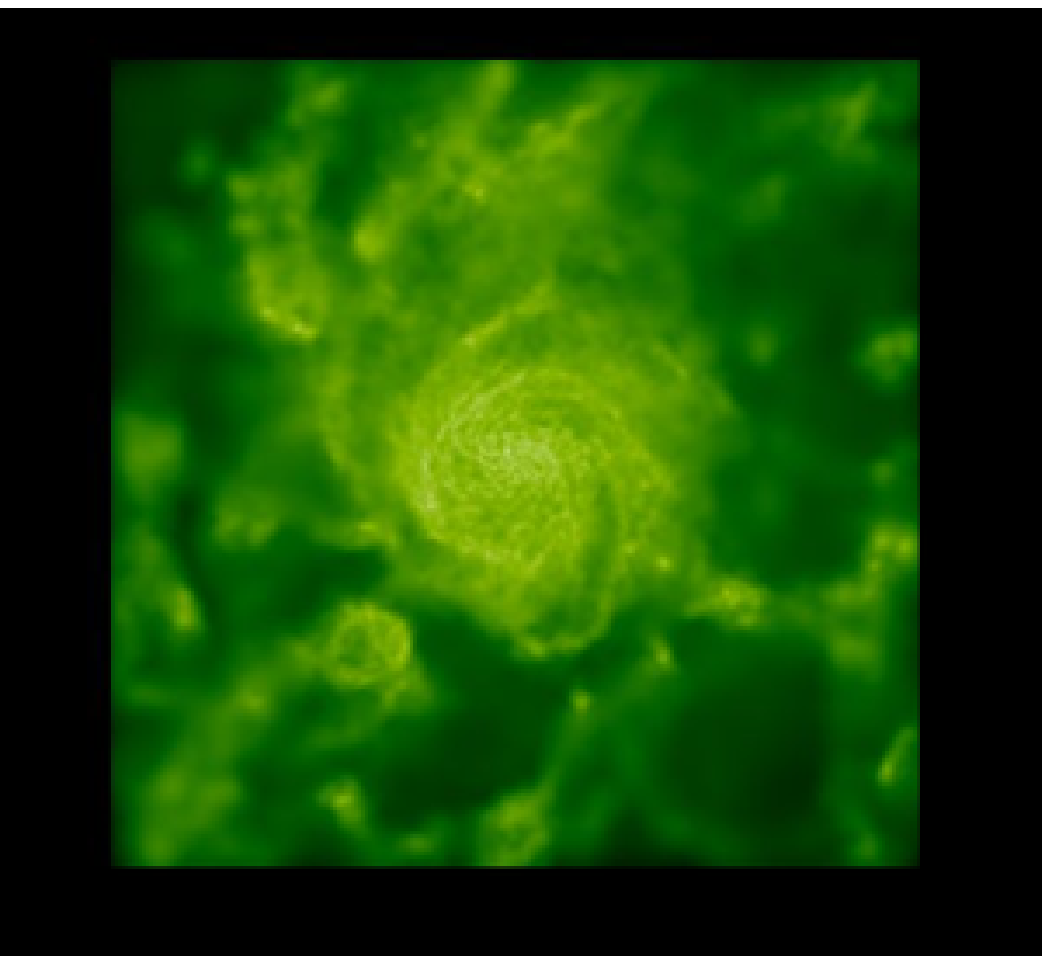}
\includegraphics[scale=0.36]{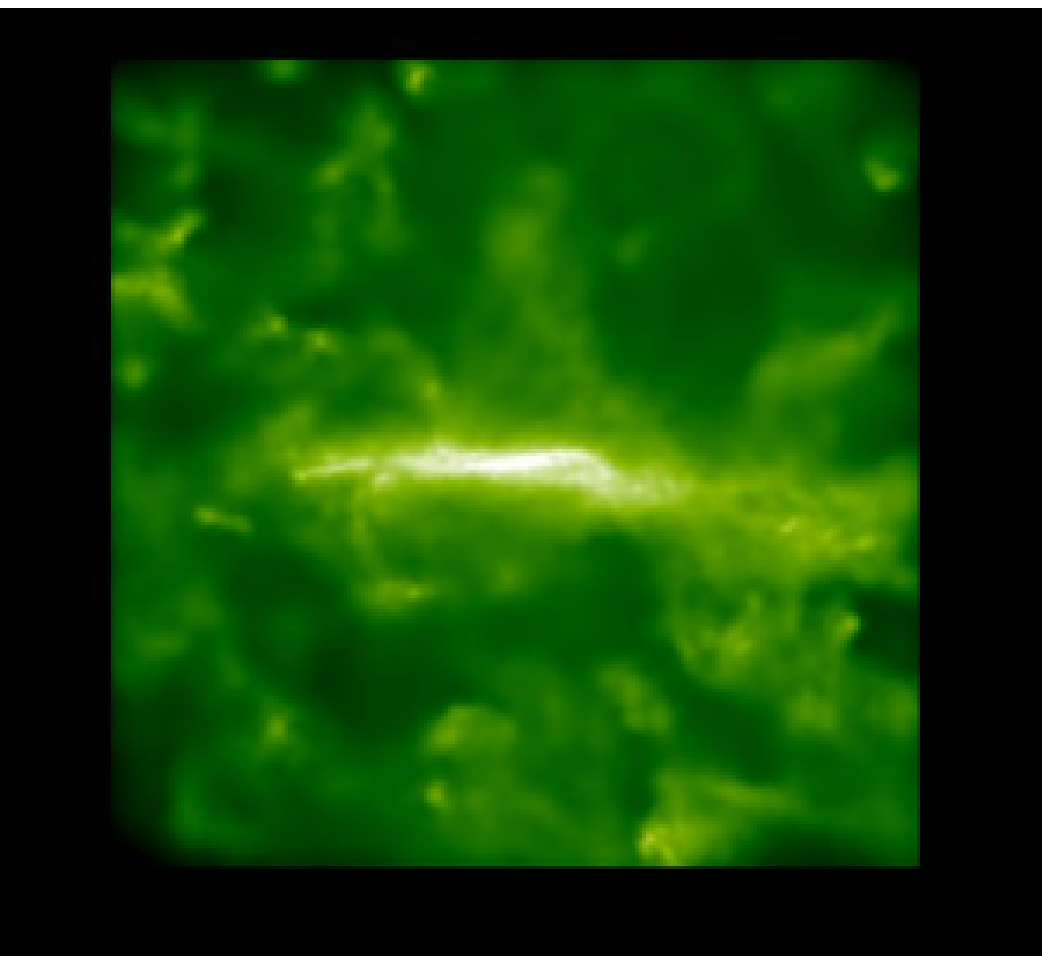}
\includegraphics[scale=0.36]{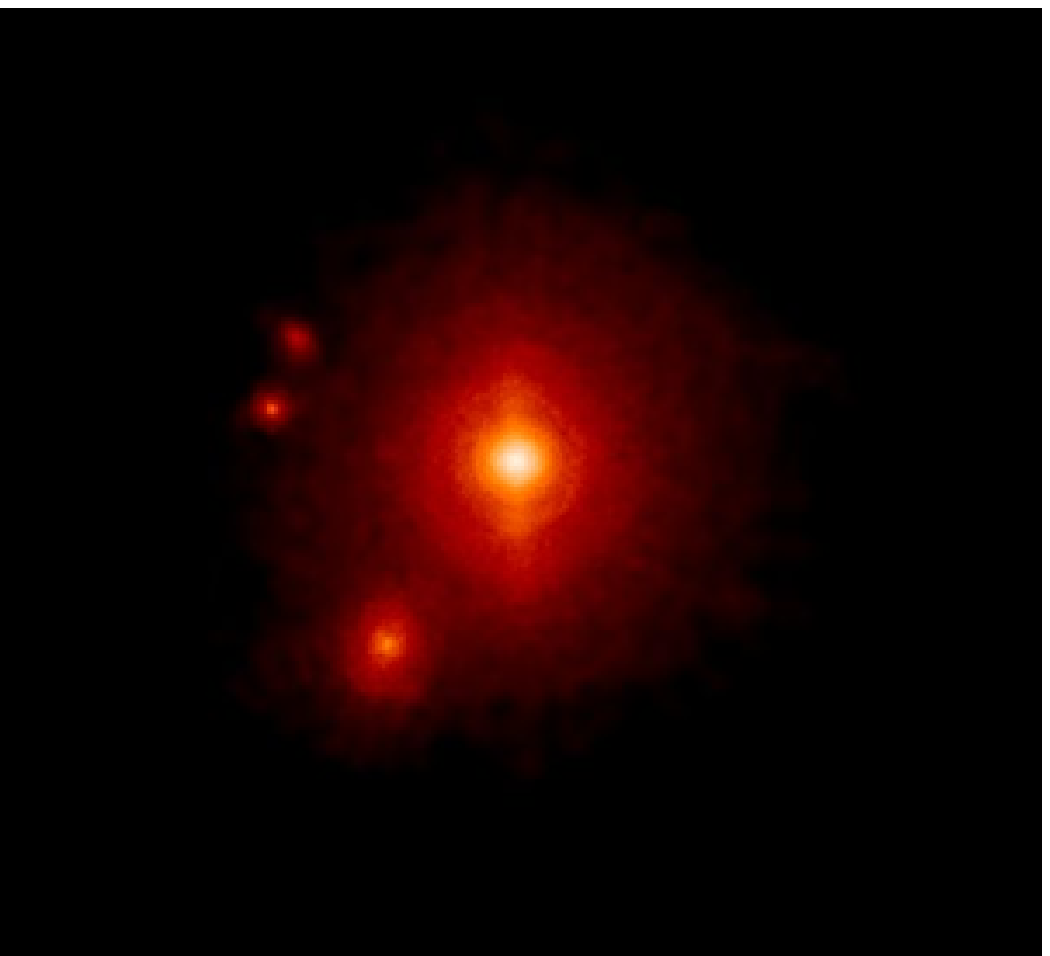}
\includegraphics[scale=0.36]{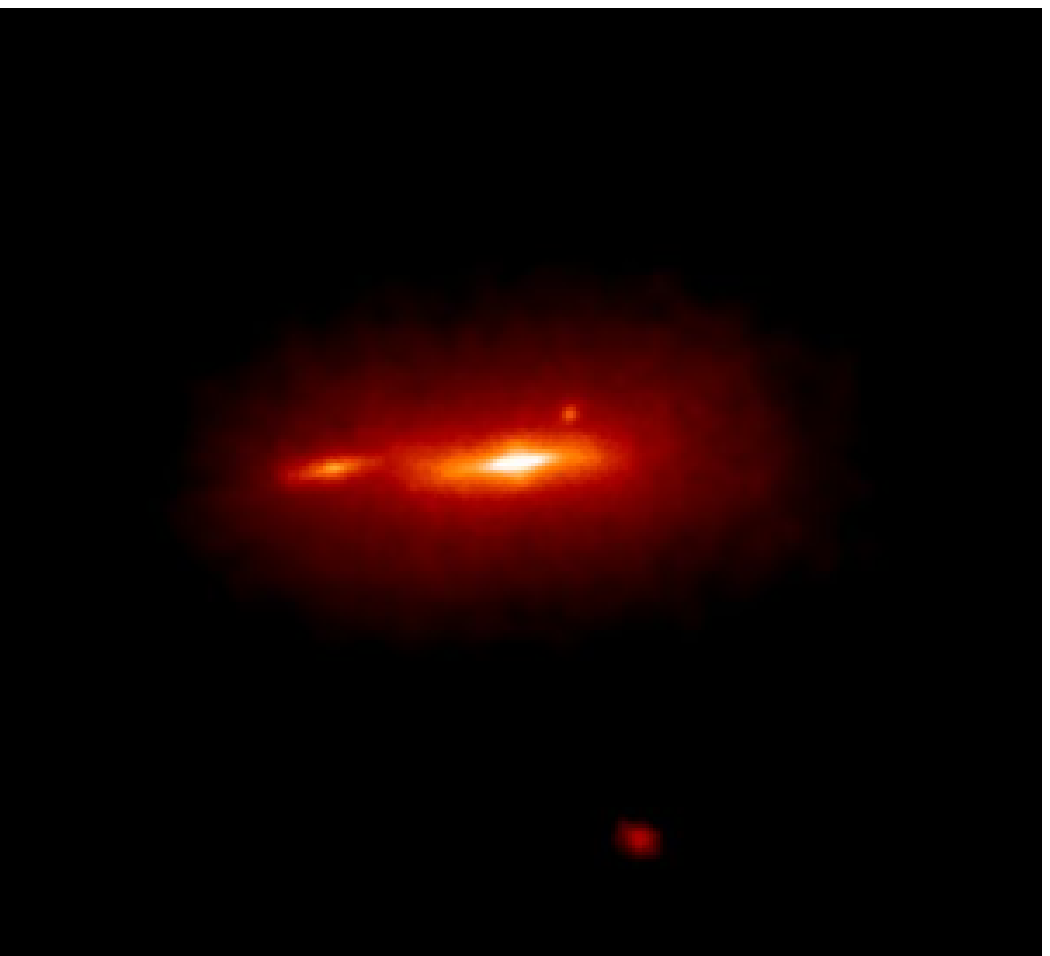}
}
\centerline{
\includegraphics[scale=0.36]{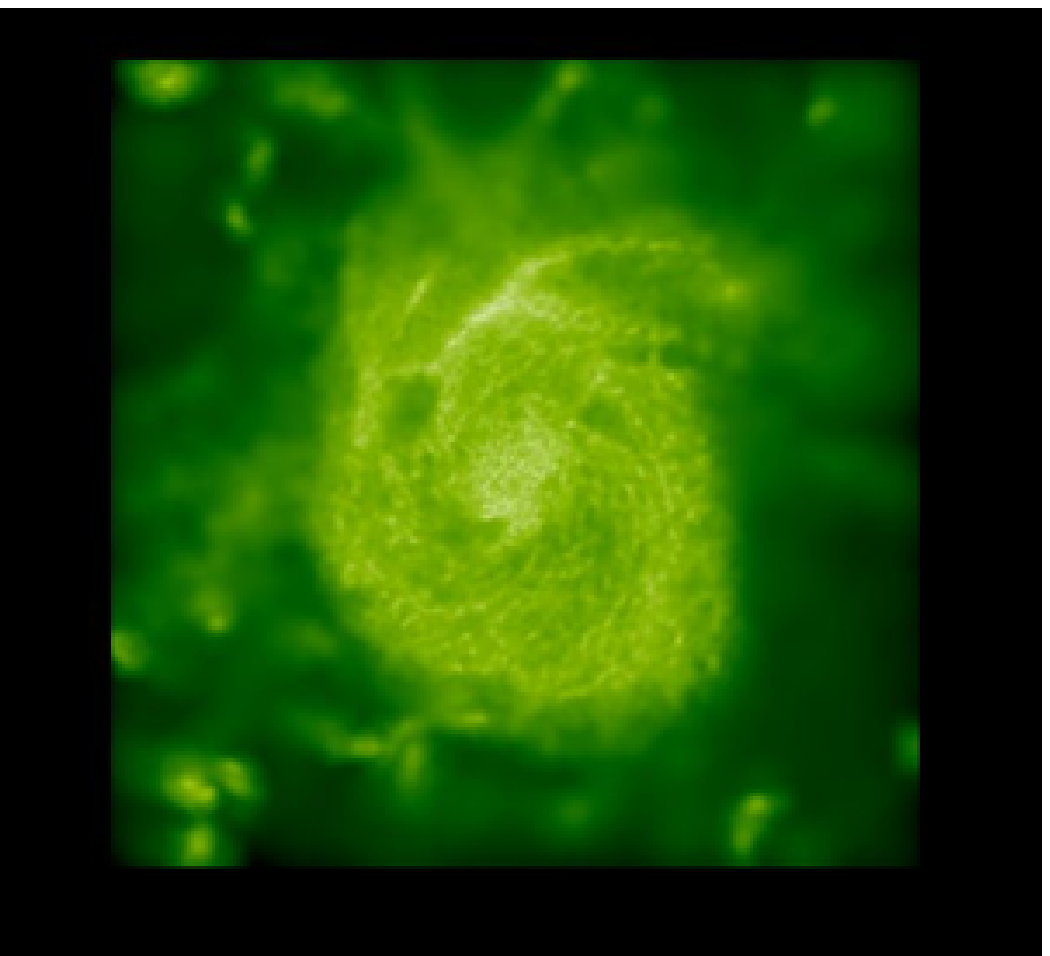}
\includegraphics[scale=0.36]{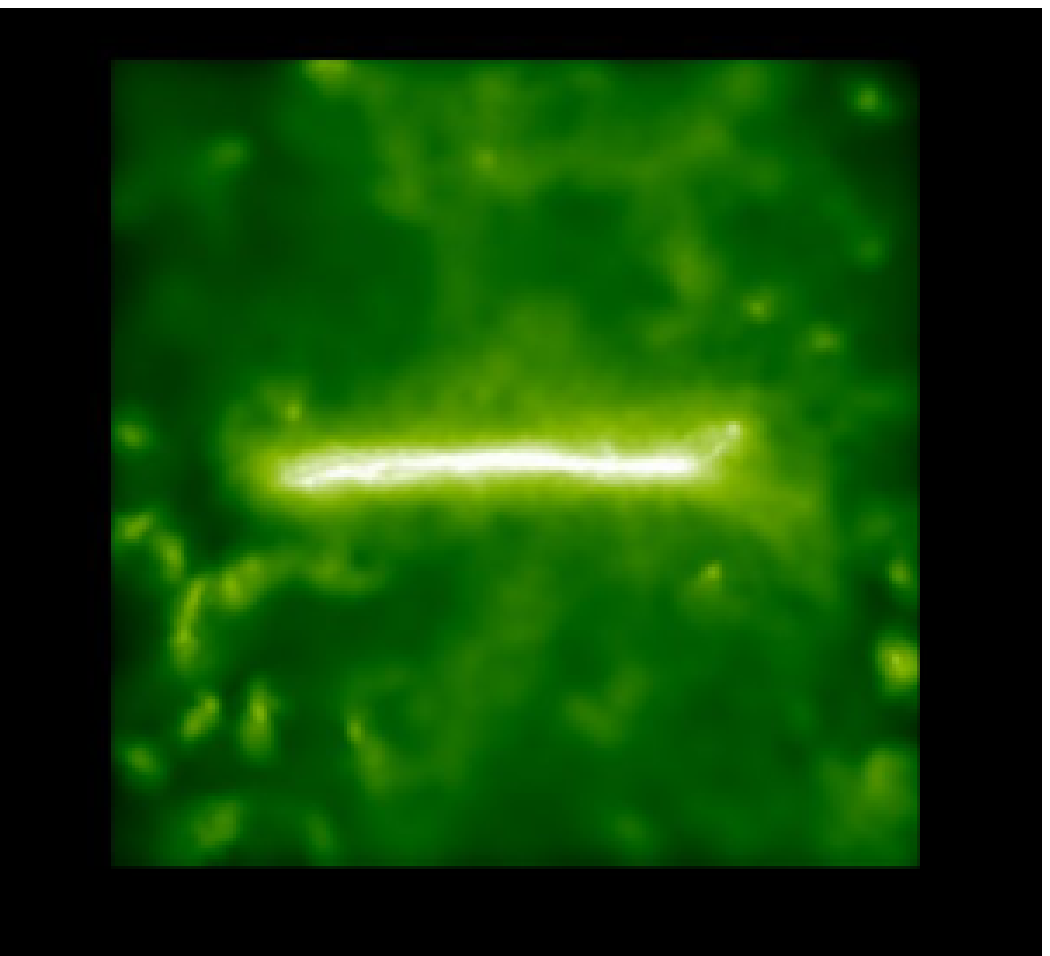}
\includegraphics[scale=0.36]{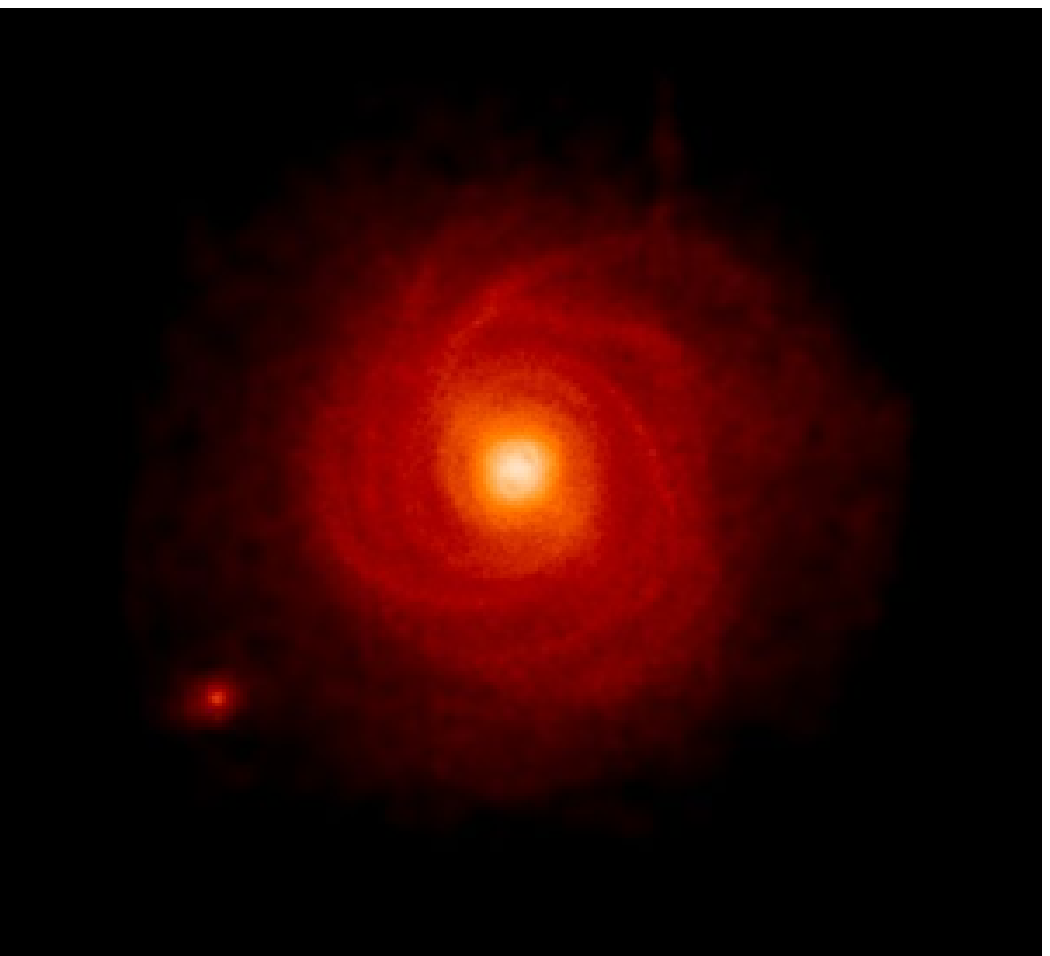}
\includegraphics[scale=0.36]{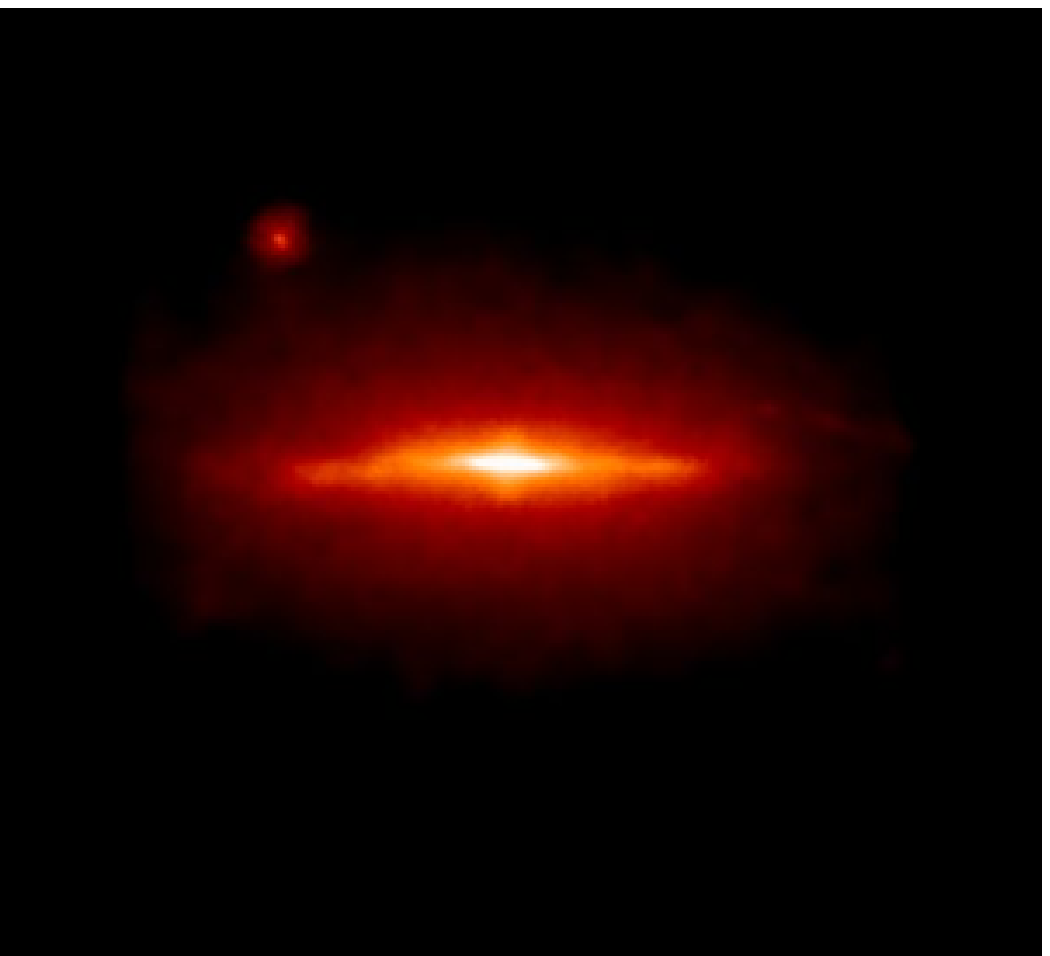}
}
\centerline{
\includegraphics[scale=0.36]{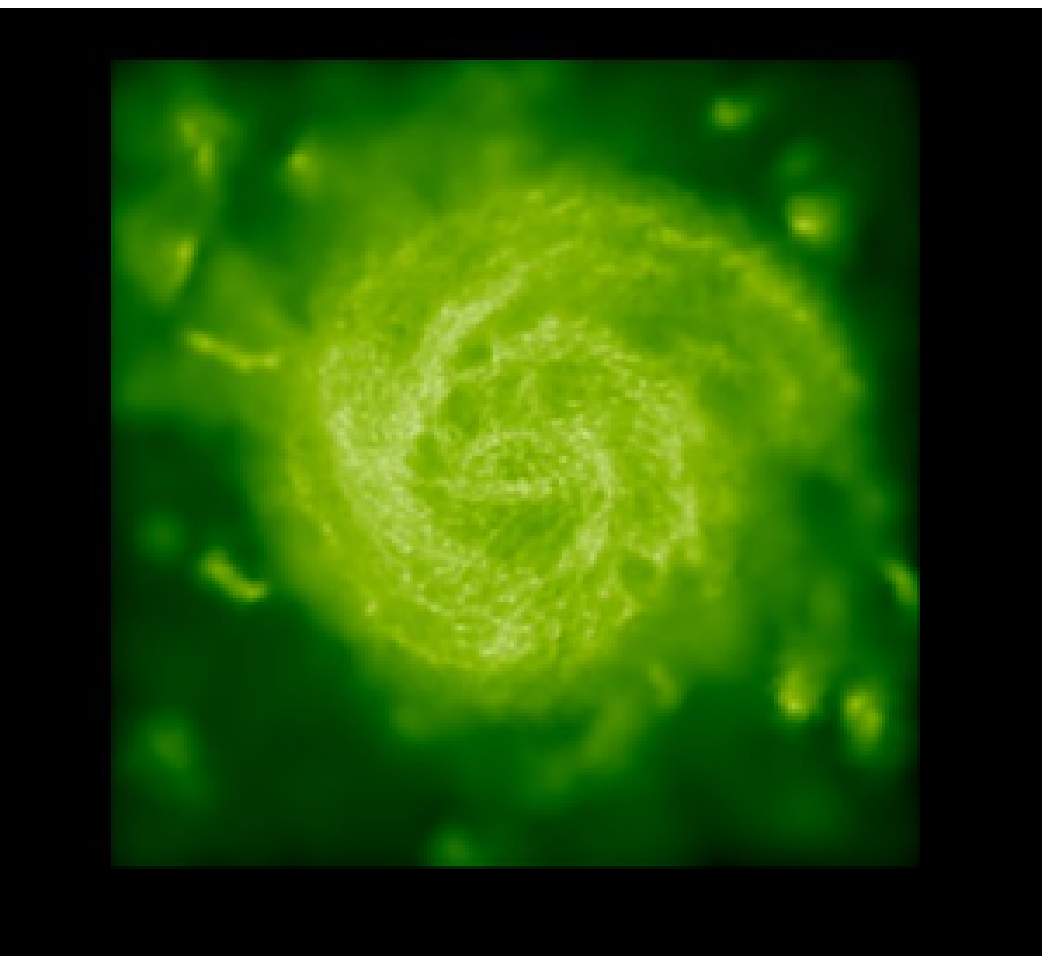}
\includegraphics[scale=0.36]{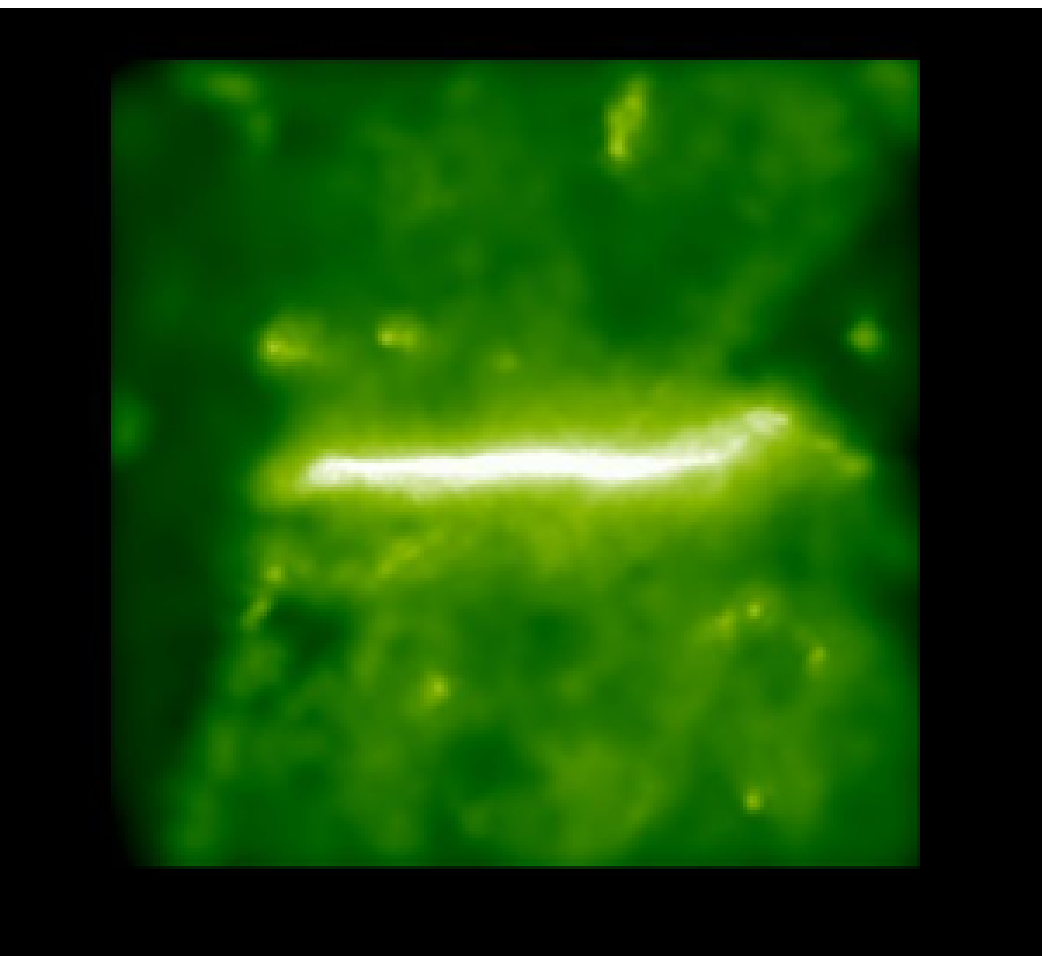}
\includegraphics[scale=0.36]{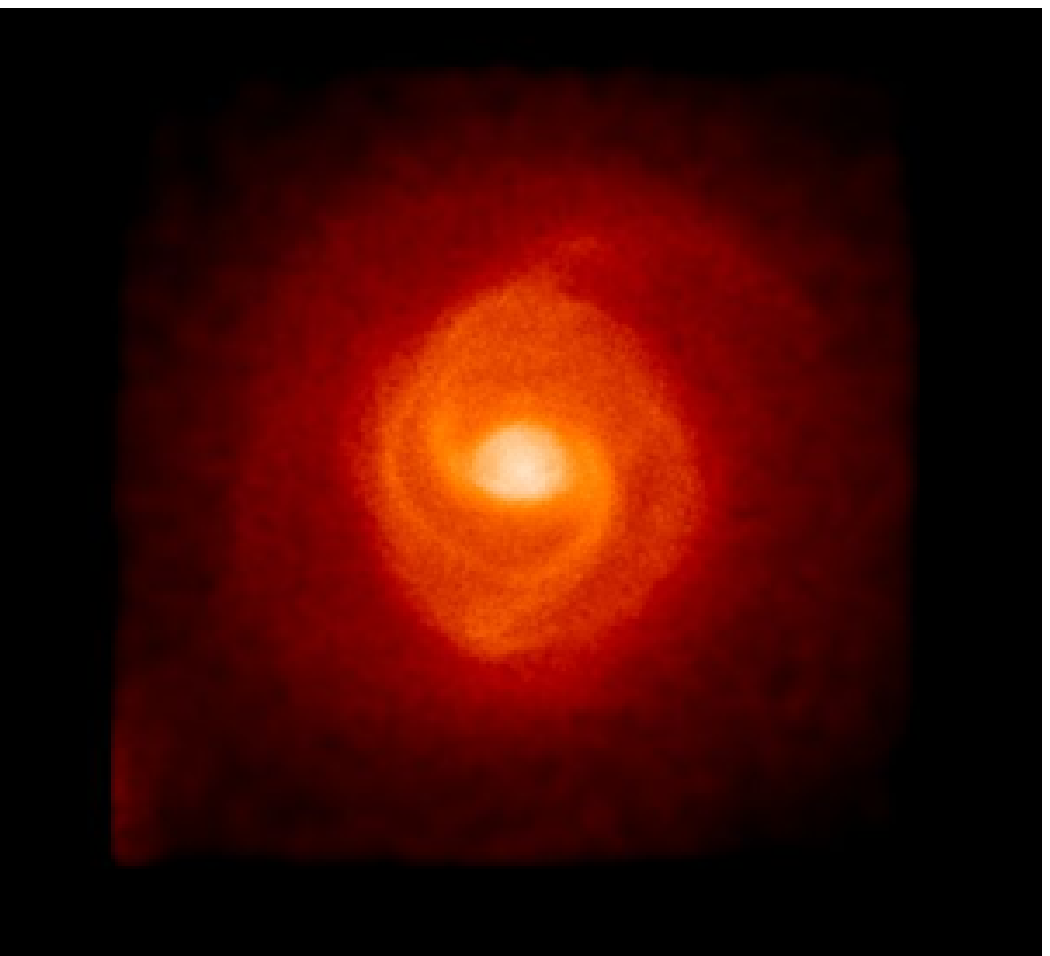}
\includegraphics[scale=0.36]{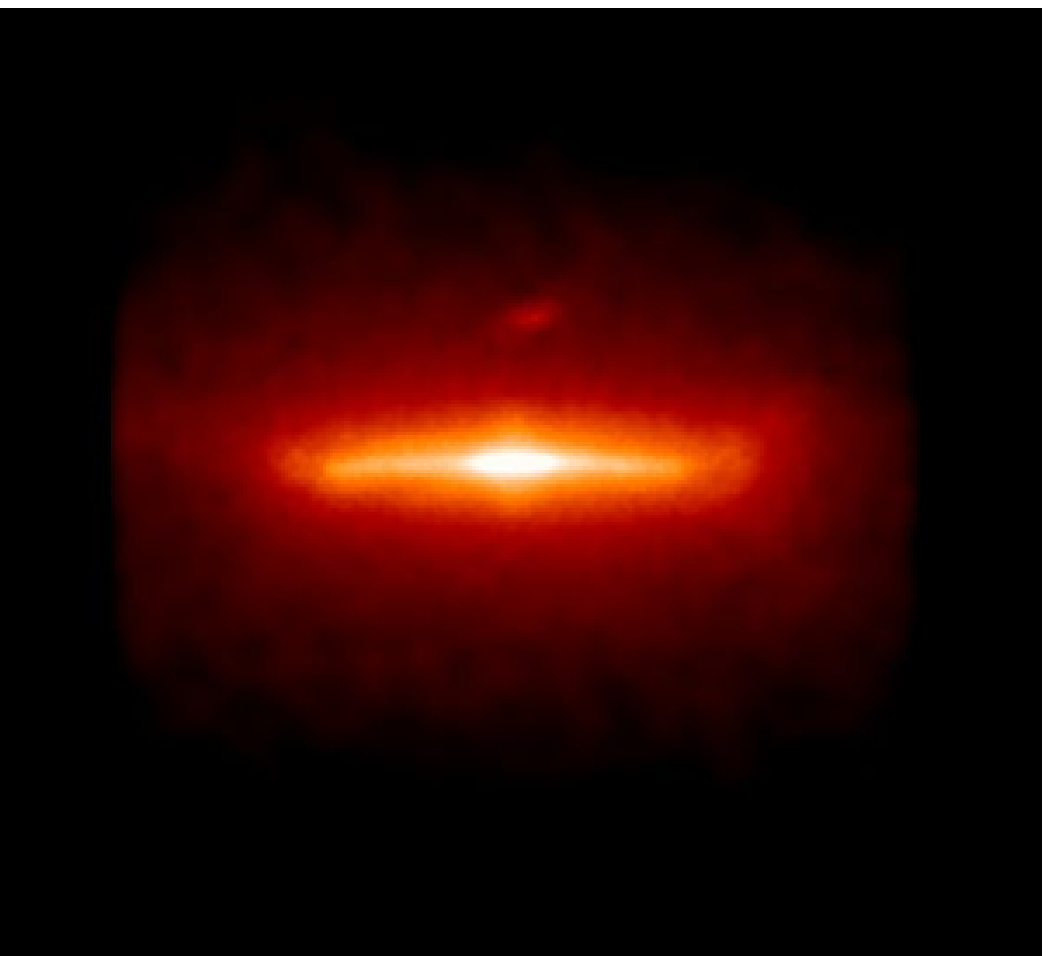}
}
\caption{From left to right columns: face-on and edge-on projected gas
  densities, face-on and edge-on stellar densities for the GA2
  simulation, at redshifts $z=2.48$, $2.02$, $1.50$, $1.01$ and $0.49$
  (from top to bottom panels). The z-axis of the coordinate system is
  aligned with the angular momentum vector of the gas enclosed in the
  inner 8 physical kpc (at all redshifts).  Box size is $57$ physical kpc}
\label{fig:MapGA2evol}
\end{figure*}

\begin{figure*}
\centerline{
\includegraphics[scale=0.36]{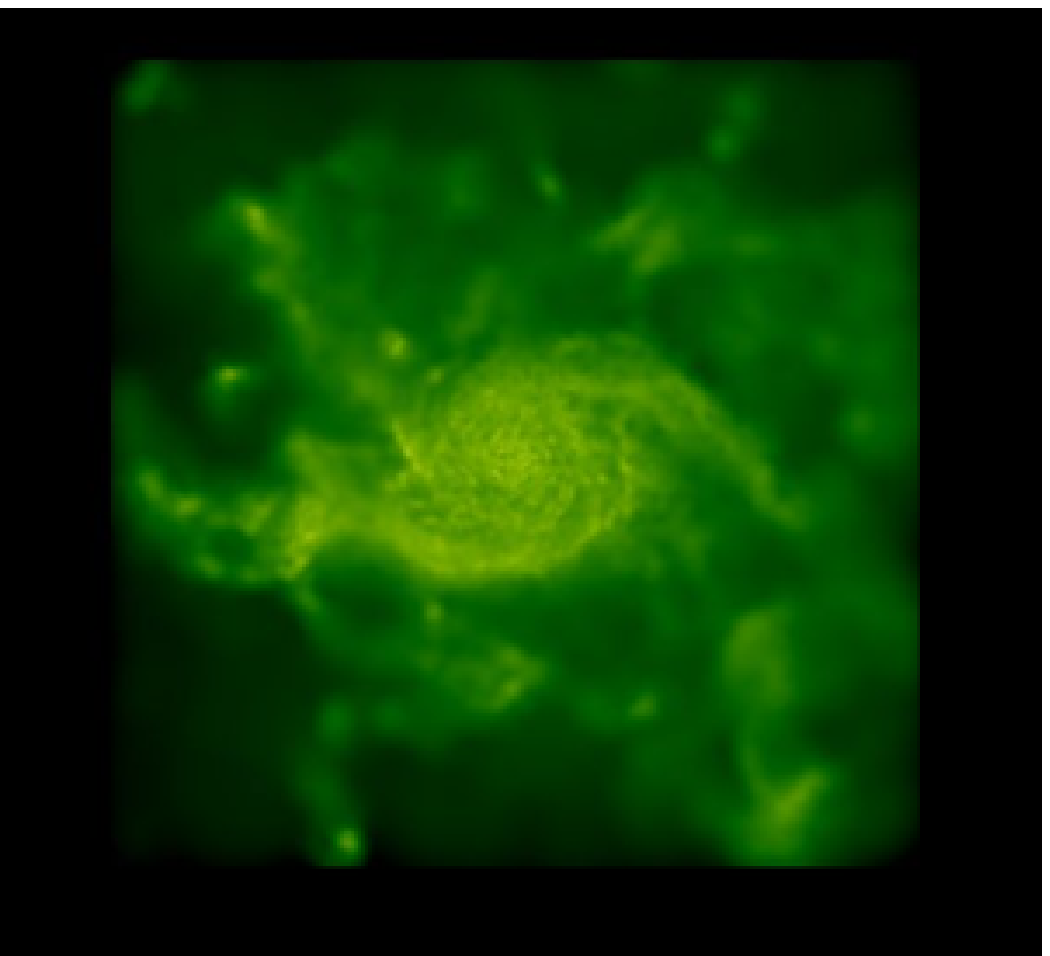}
\includegraphics[scale=0.36]{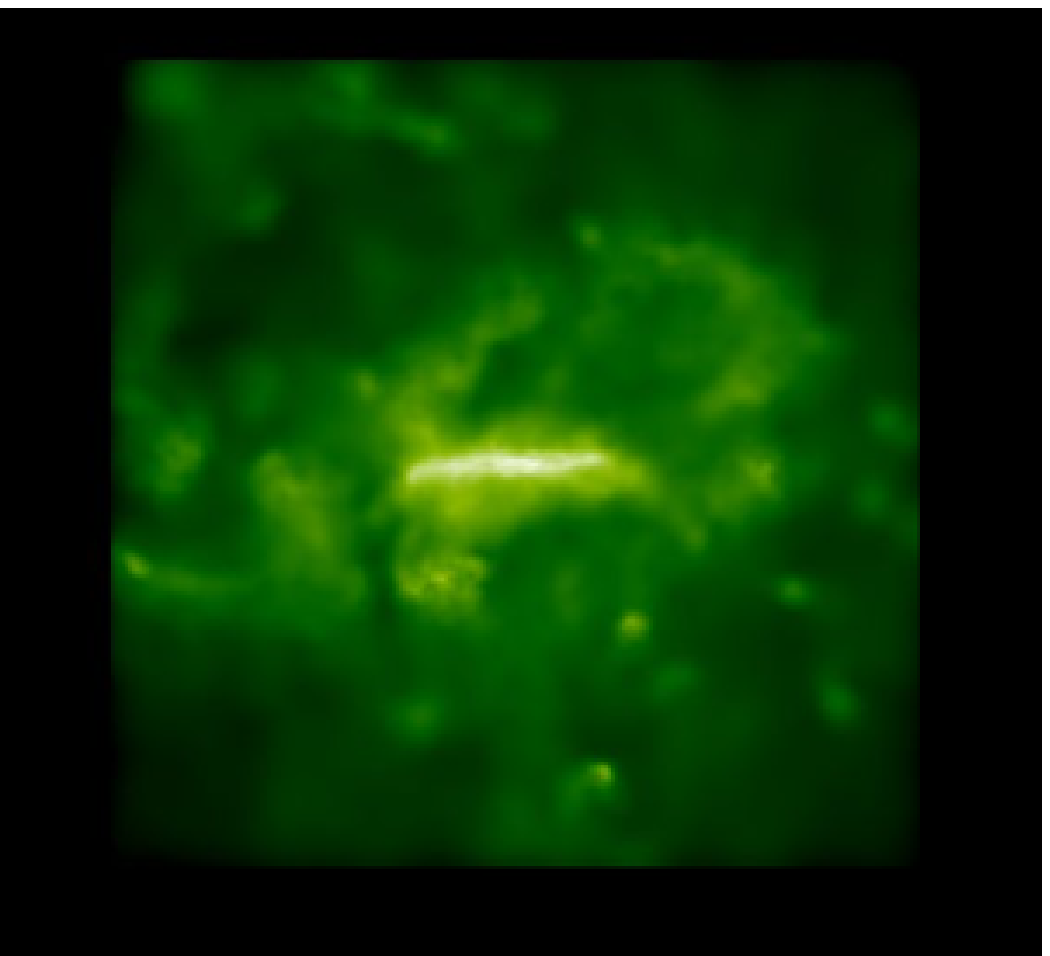}
\includegraphics[scale=0.36]{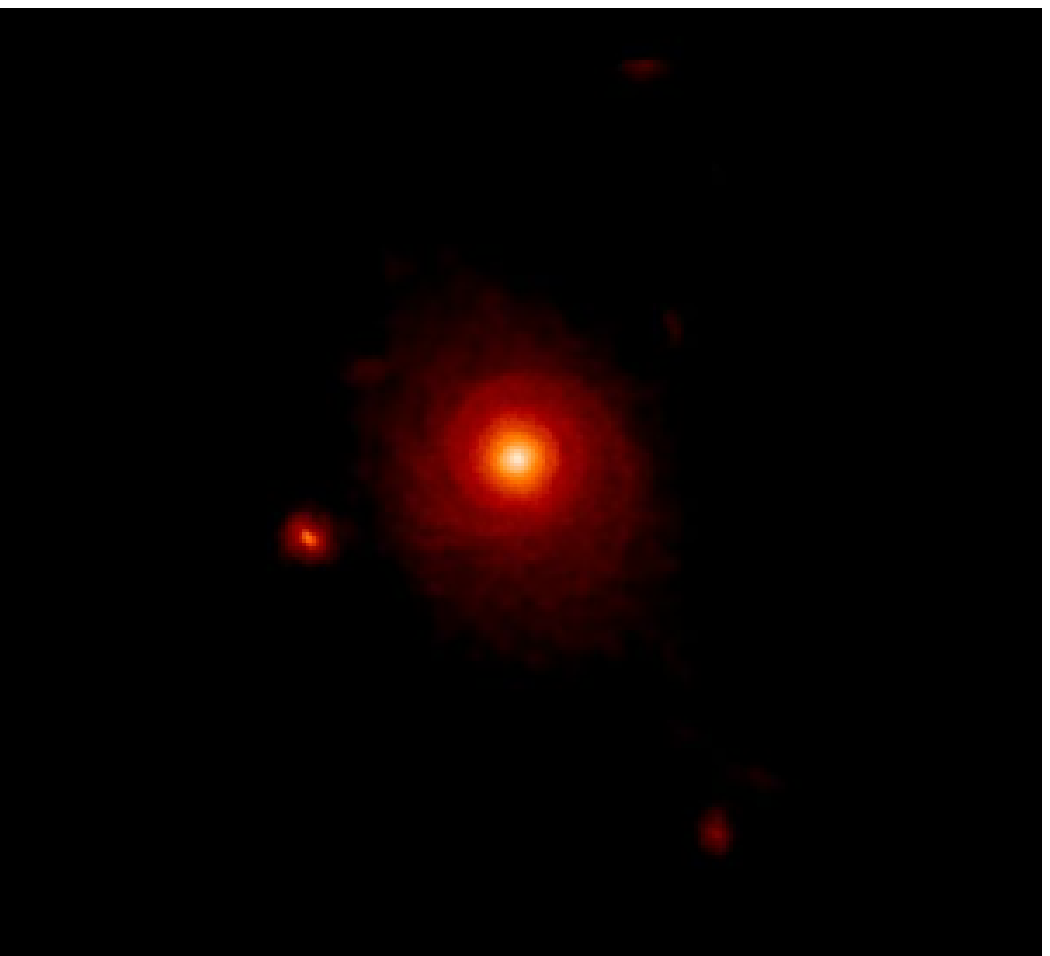}
\includegraphics[scale=0.36]{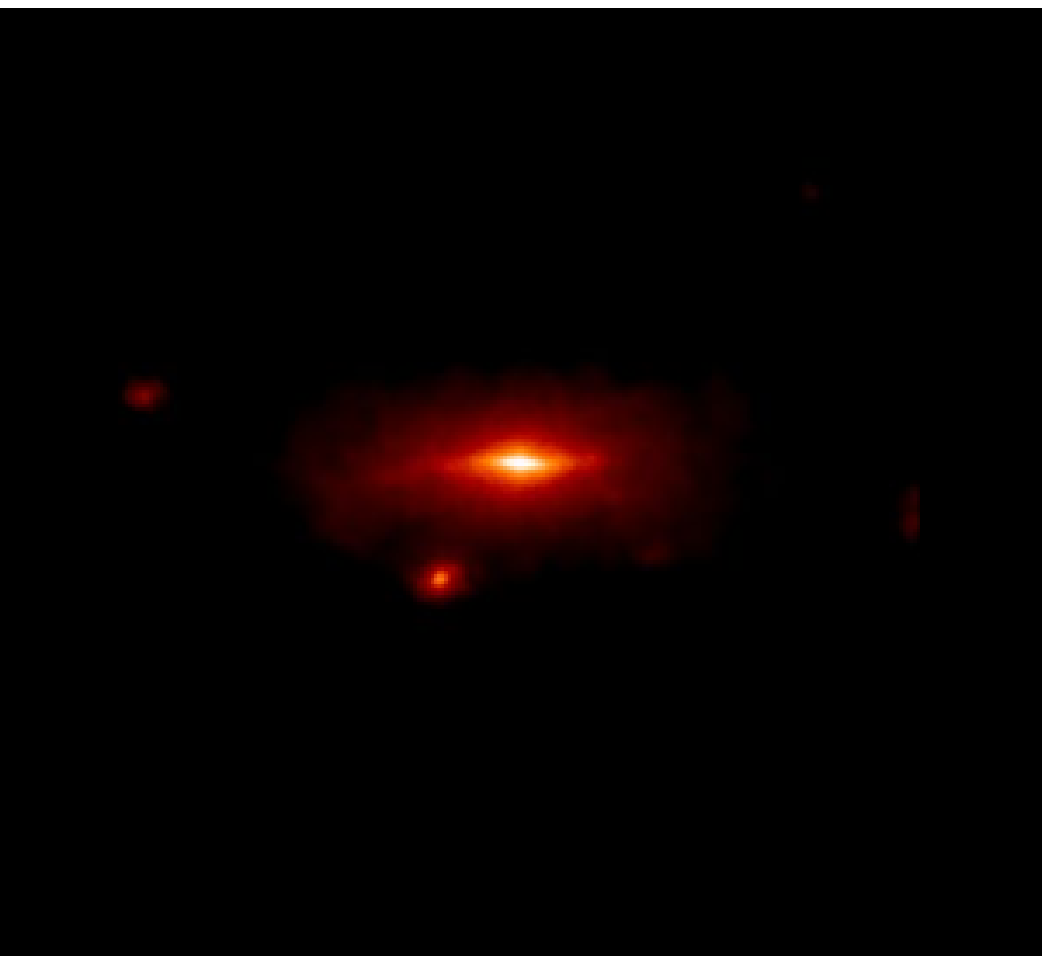}
}
\centerline{
\includegraphics[scale=0.36]{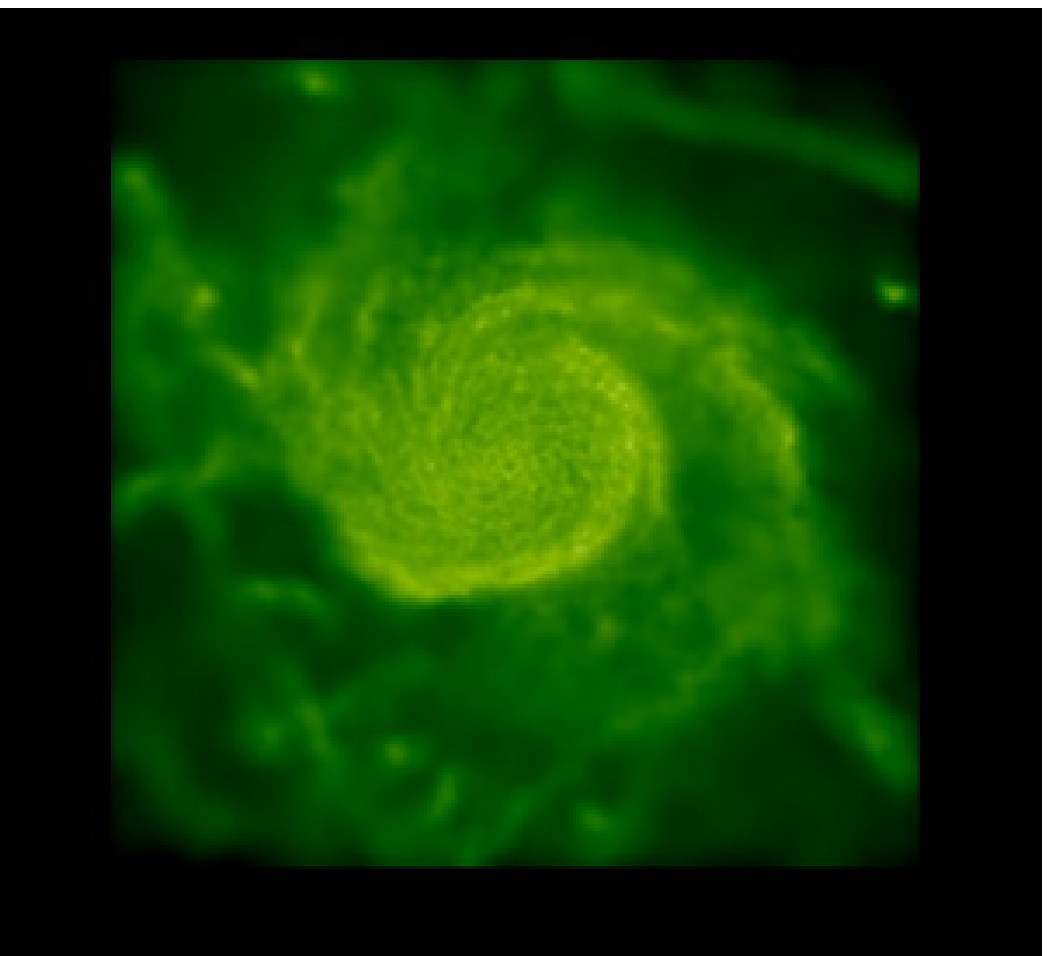}
\includegraphics[scale=0.36]{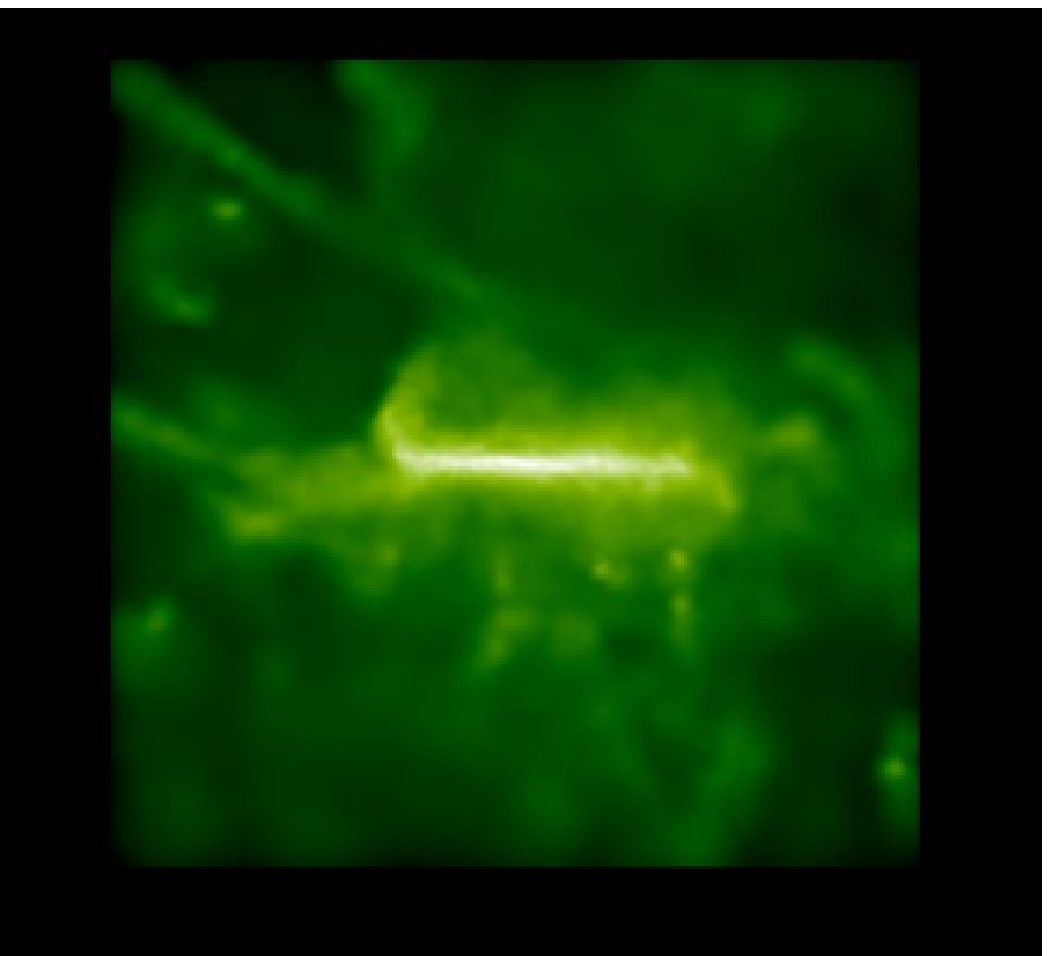}
\includegraphics[scale=0.36]{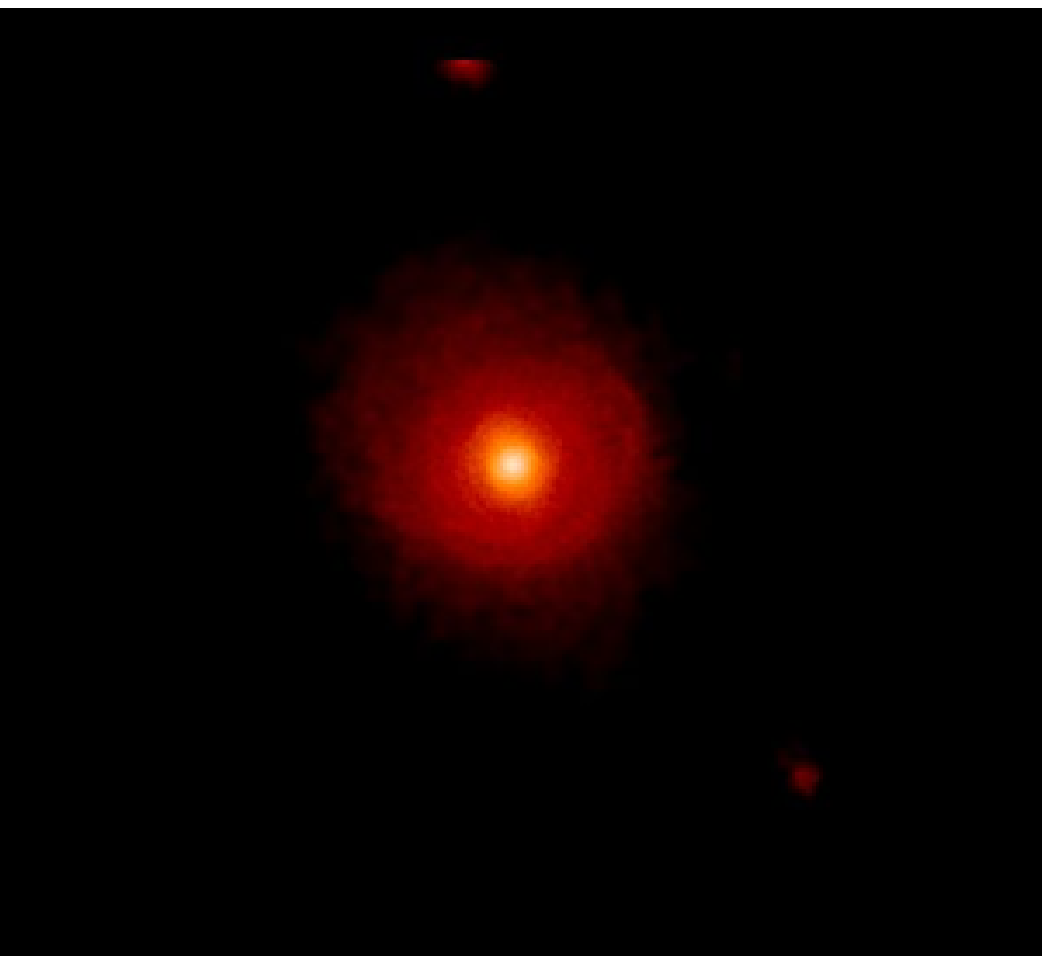}
\includegraphics[scale=0.36]{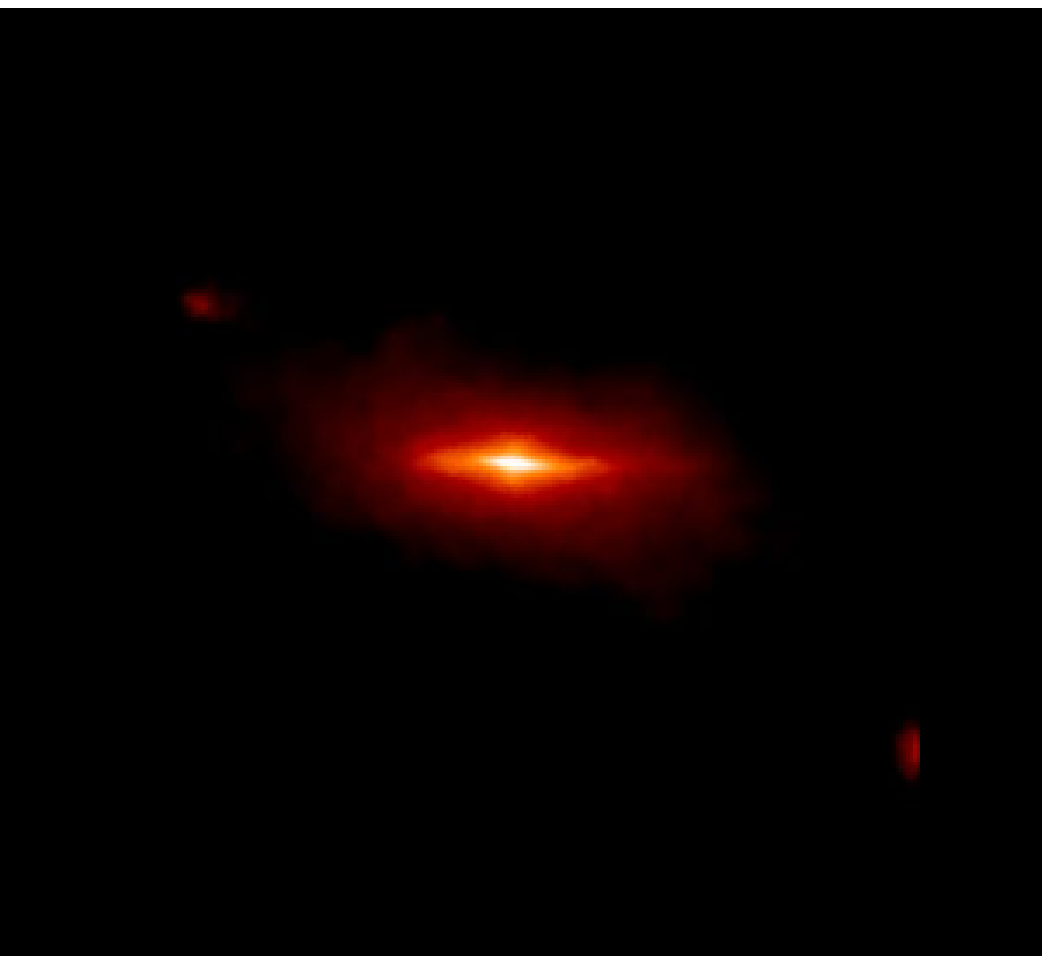}
}
\centerline{
\includegraphics[scale=0.36]{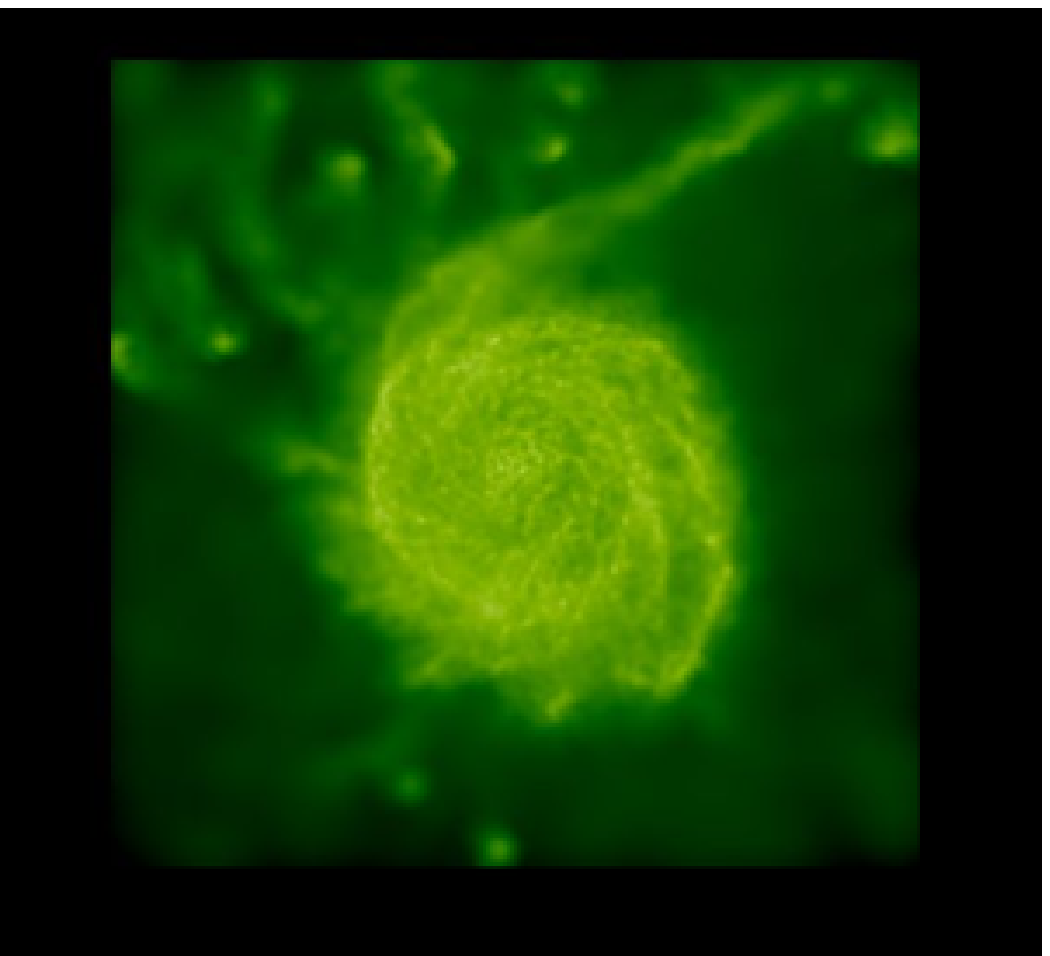}
\includegraphics[scale=0.36]{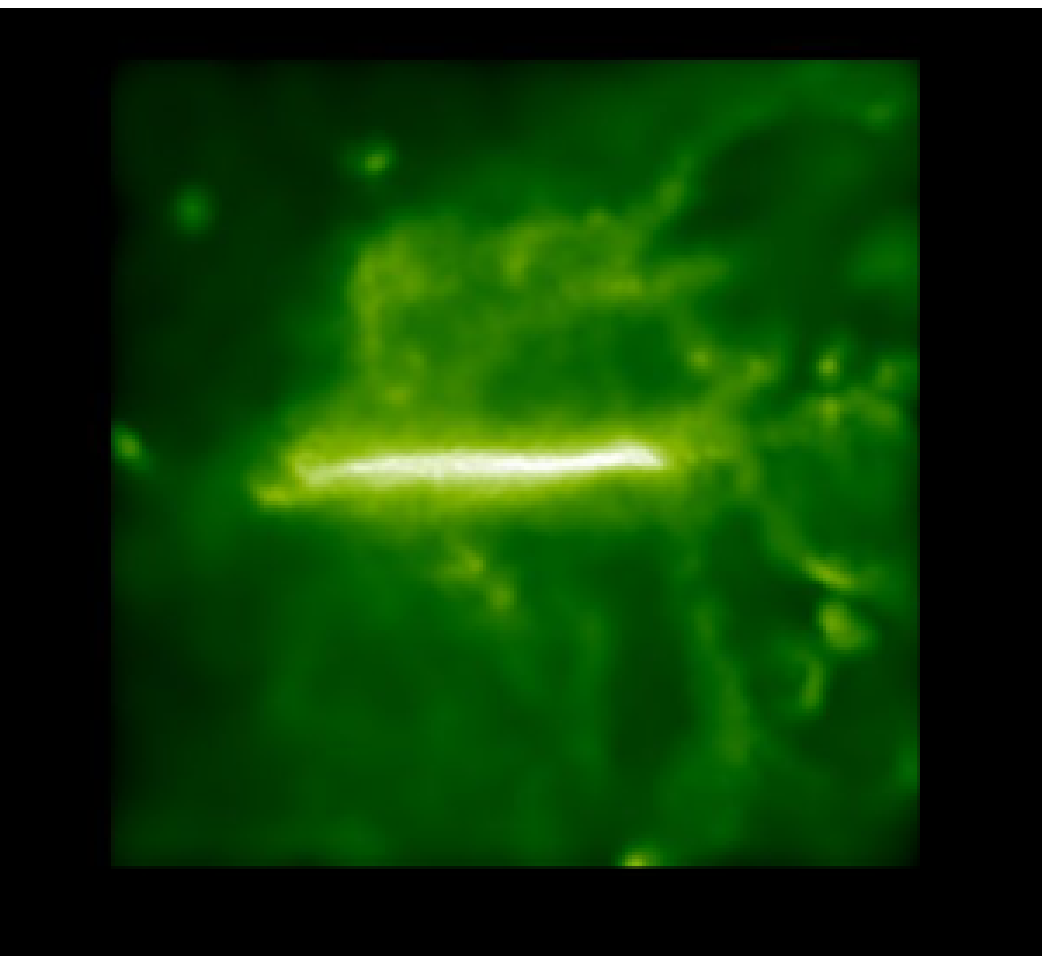}
\includegraphics[scale=0.36]{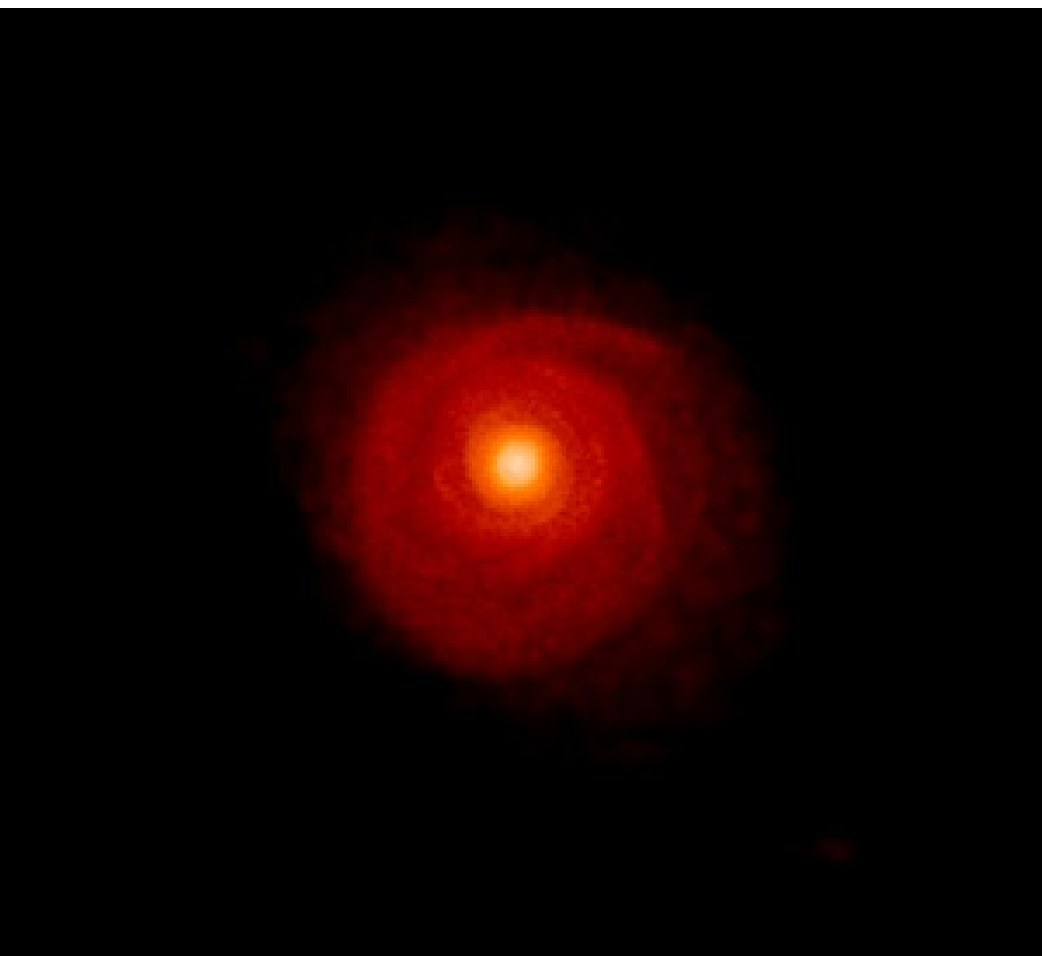}
\includegraphics[scale=0.36]{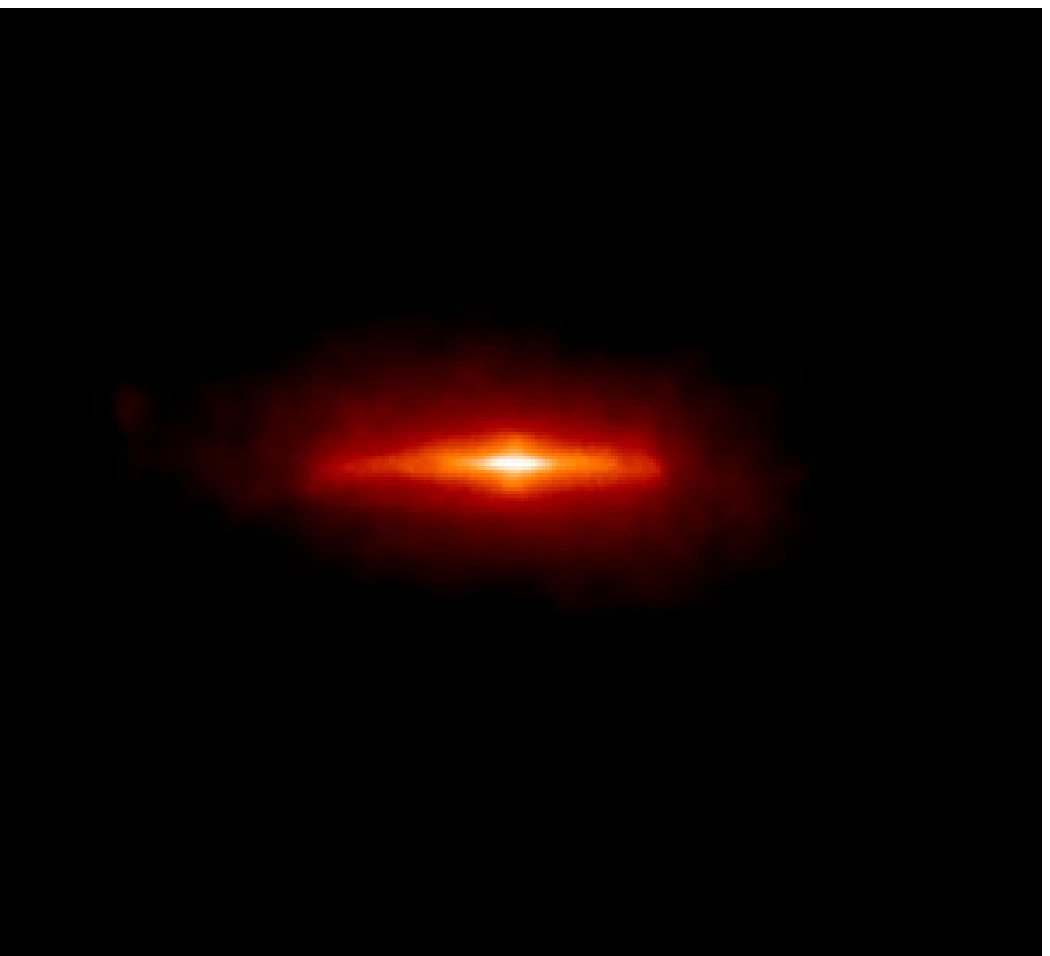}
}
\centerline{
\includegraphics[scale=0.36]{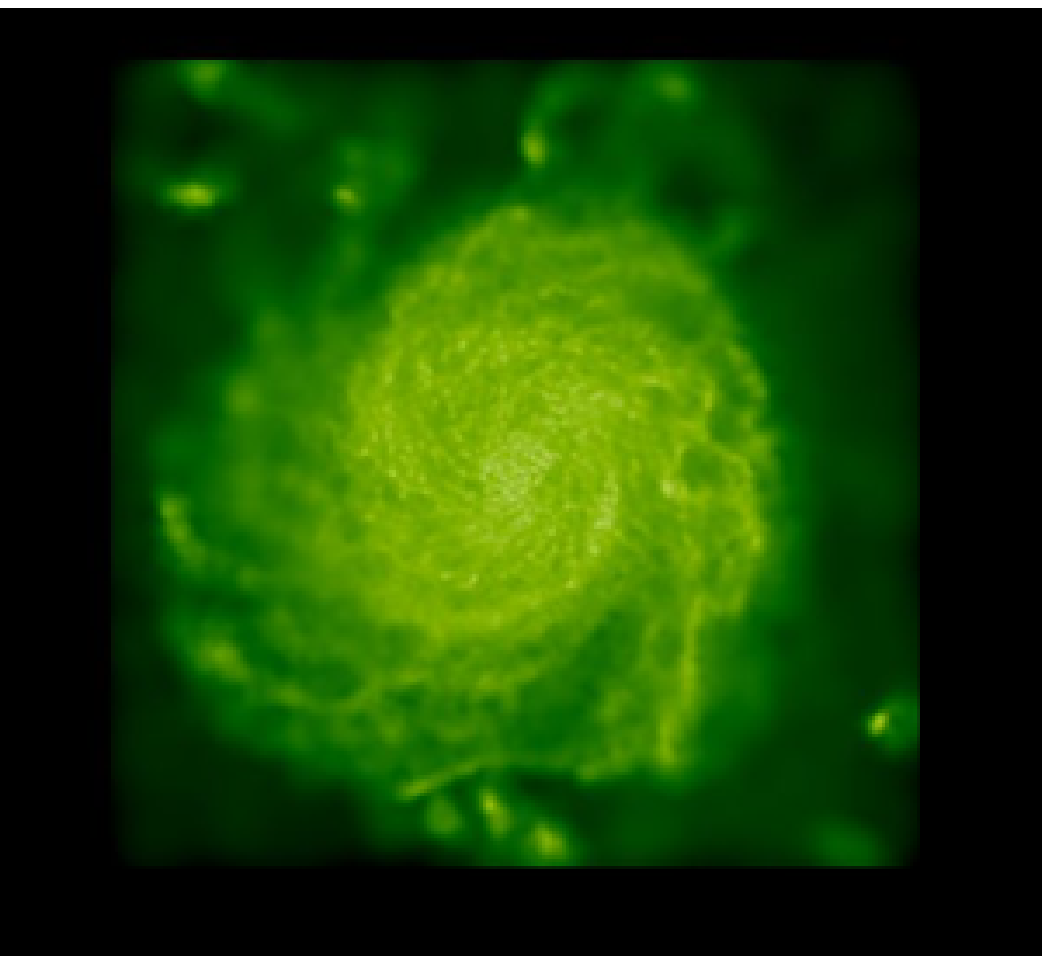}
\includegraphics[scale=0.36]{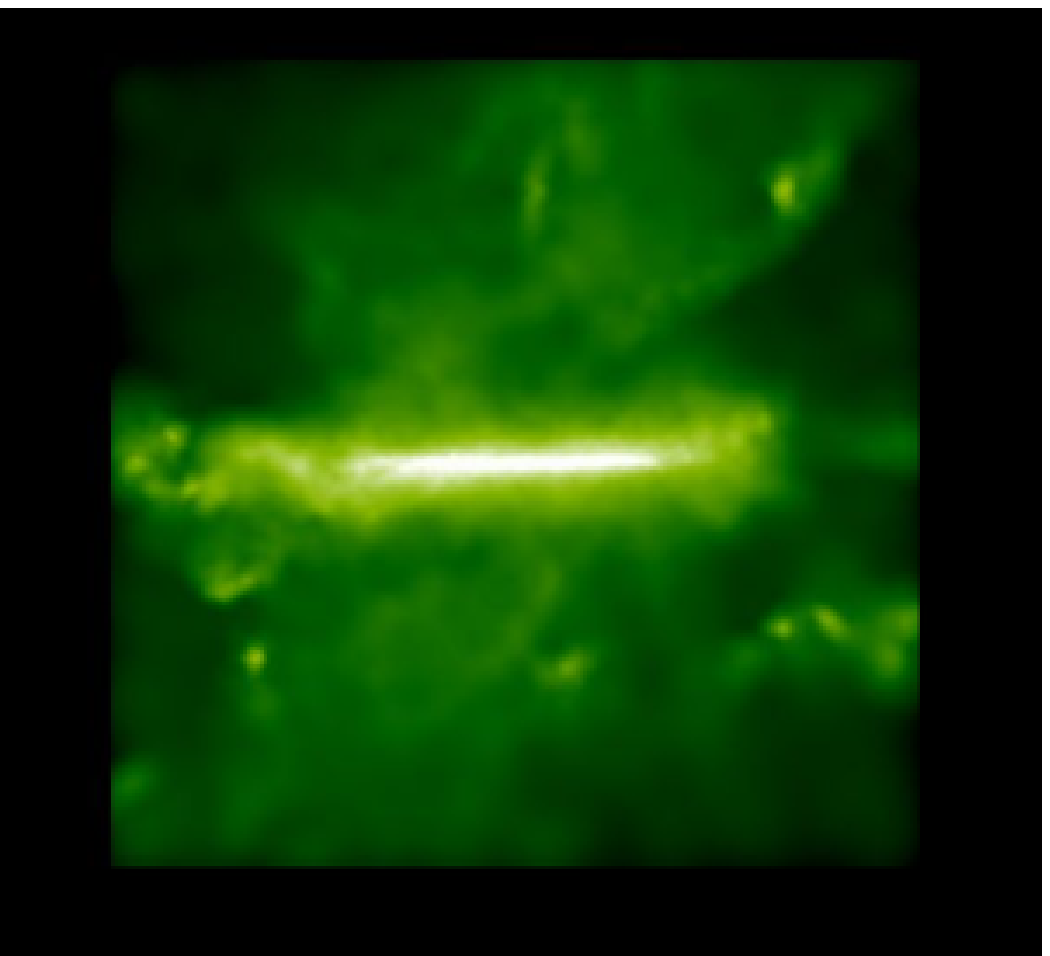}
\includegraphics[scale=0.36]{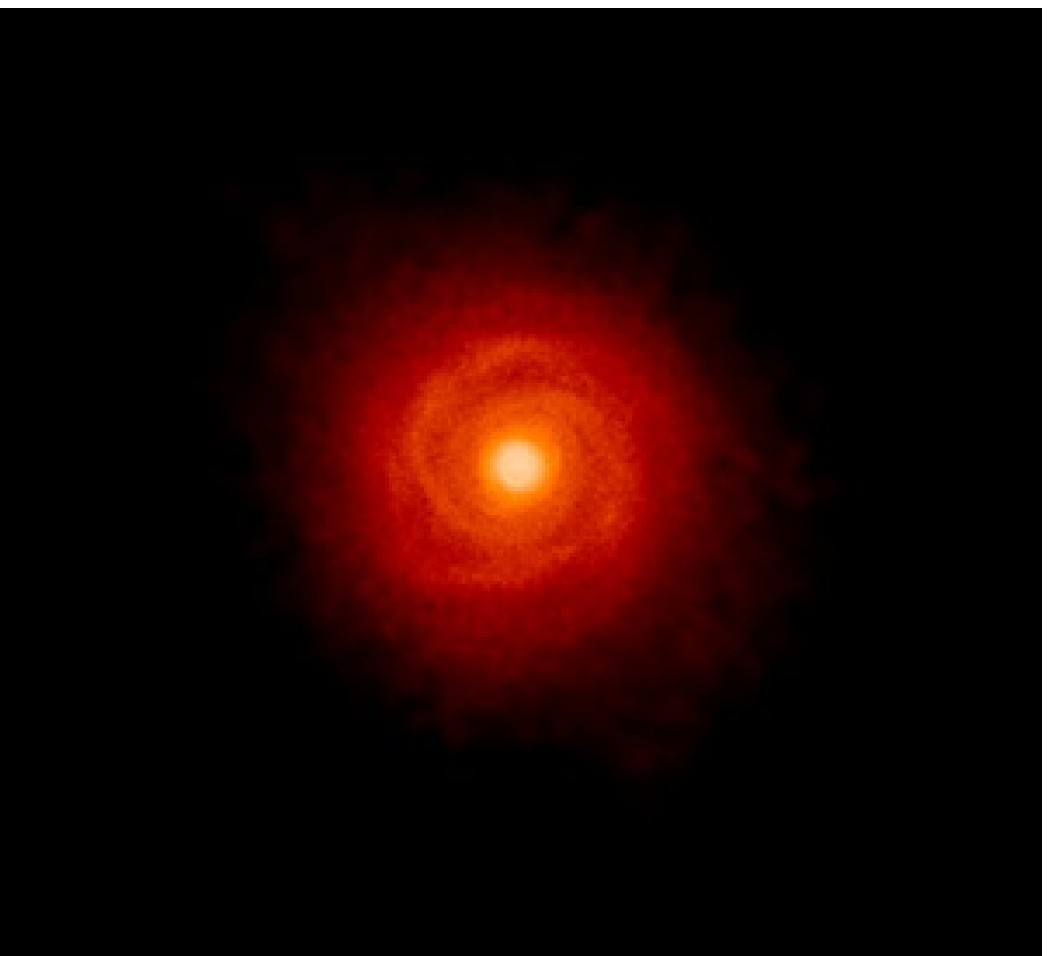}
\includegraphics[scale=0.36]{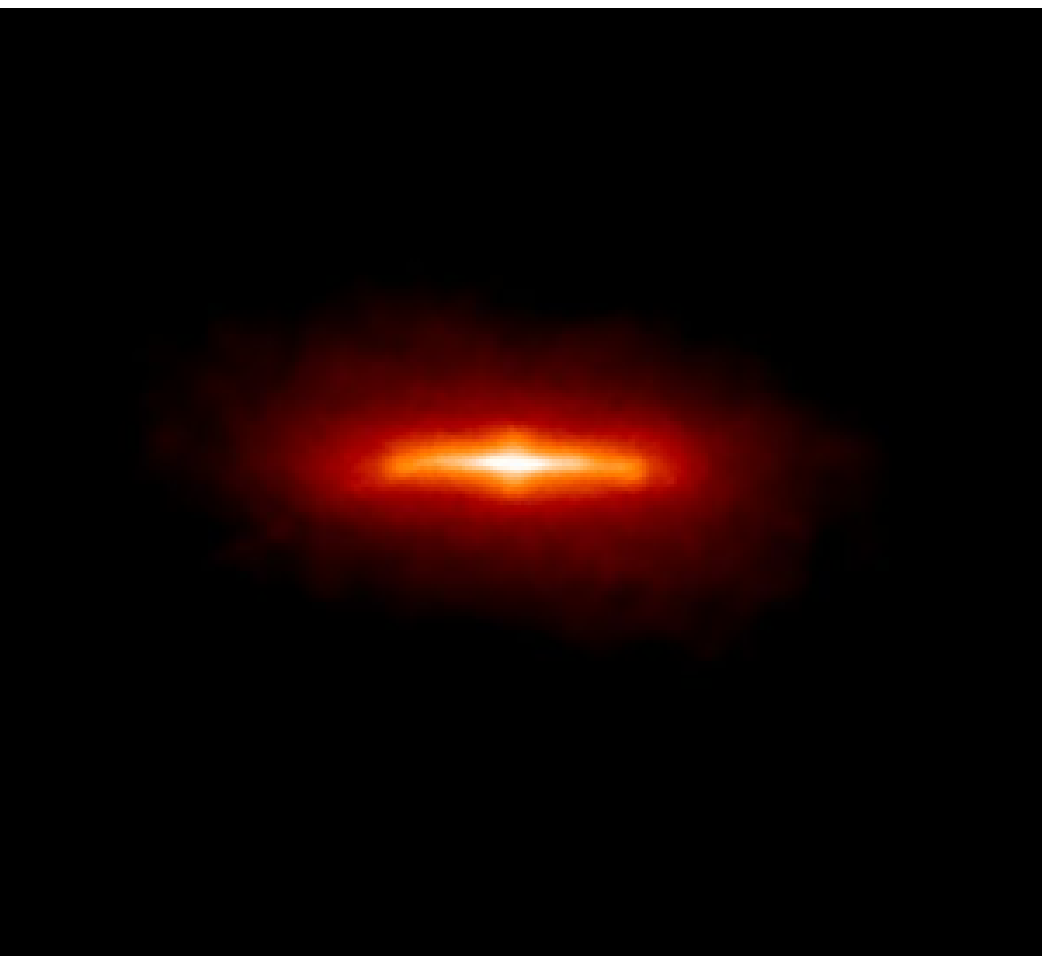}
}
\centerline{
\includegraphics[scale=0.36]{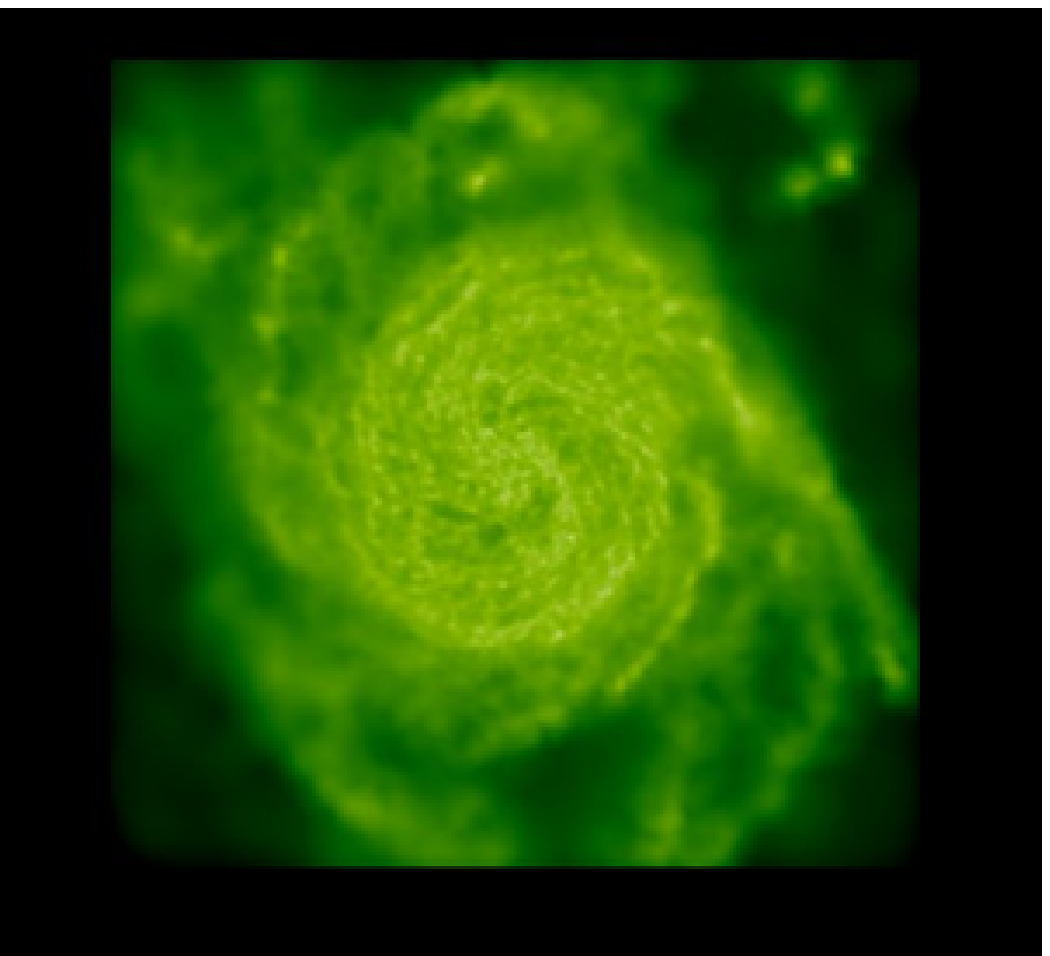}
\includegraphics[scale=0.36]{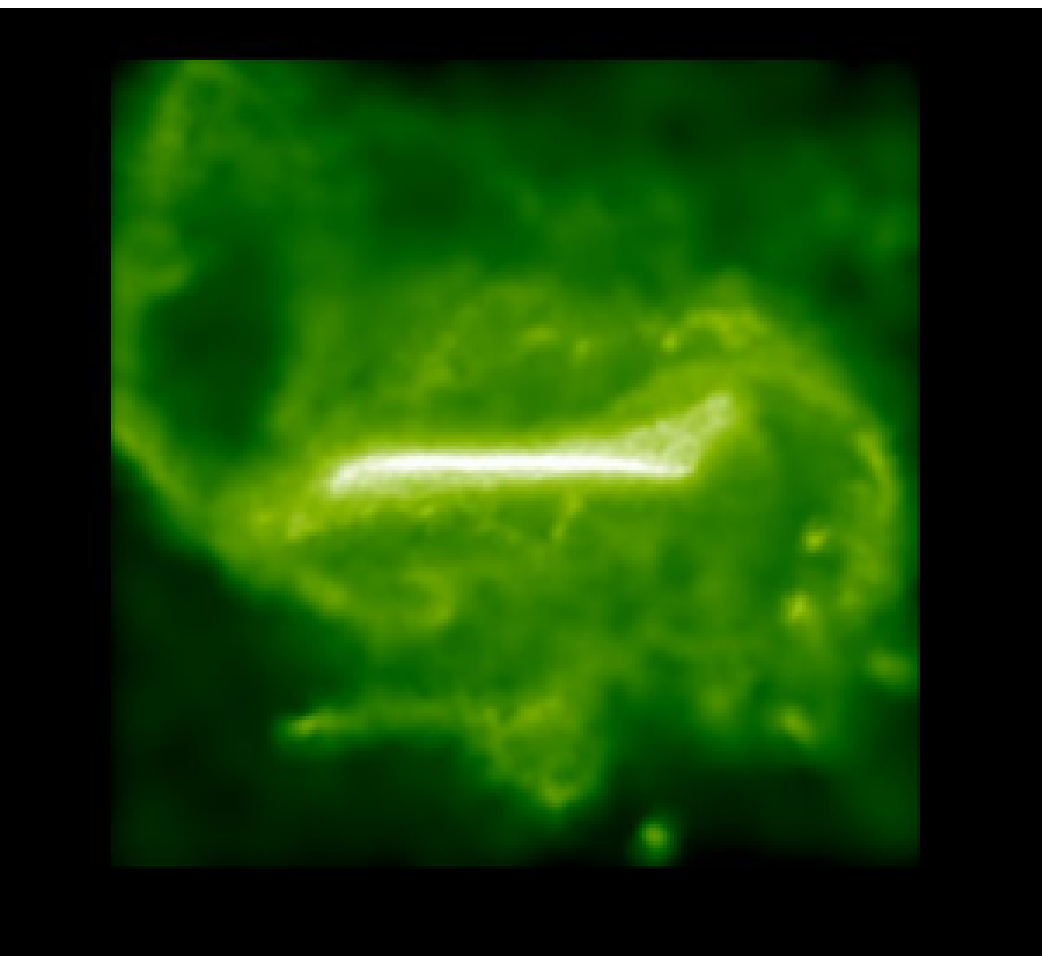}
\includegraphics[scale=0.36]{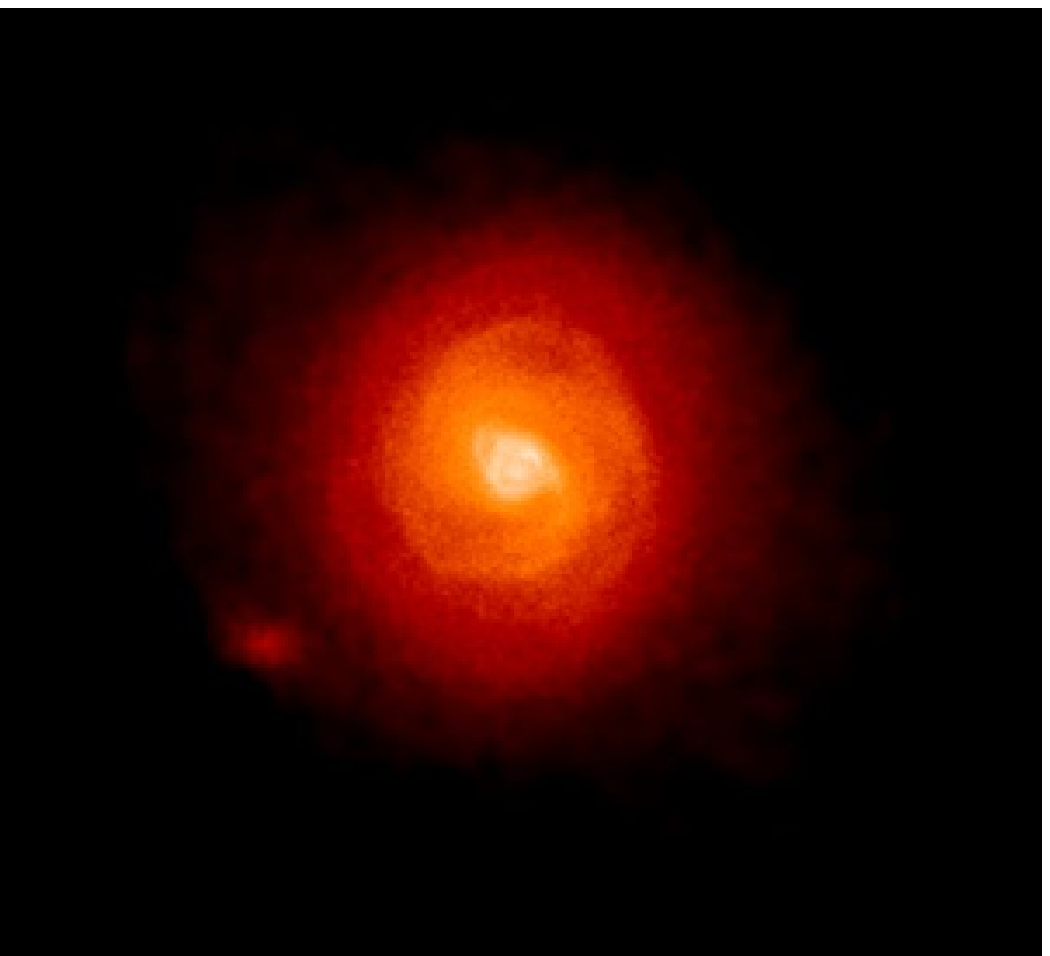}
\includegraphics[scale=0.36]{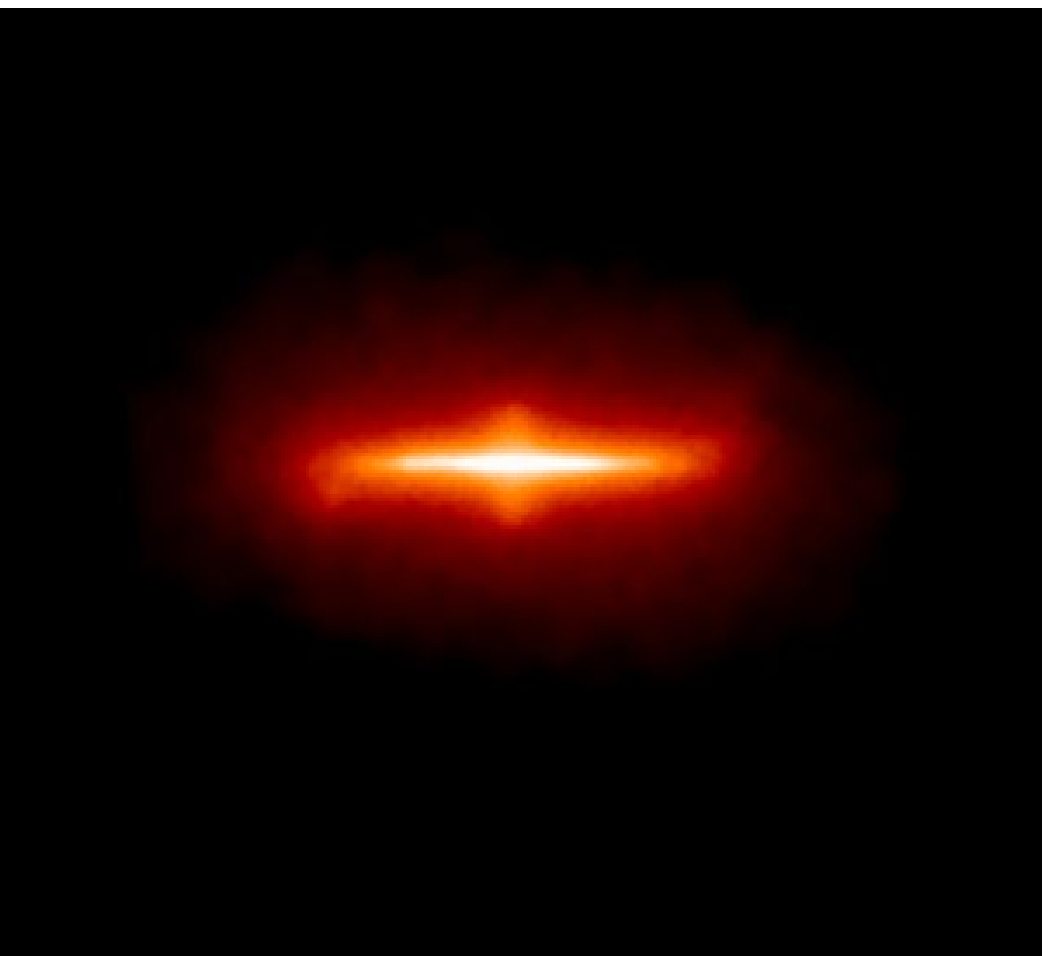}
}
\caption{The same as in Figure \protect\ref{fig:MapGA2evol} but for
  the AqC5 galaxy simulation.}
\label{fig:MapAqC5evol}
\end{figure*}

\begin{figure*}
\centerline{
\includegraphics[scale=0.5,angle=90]{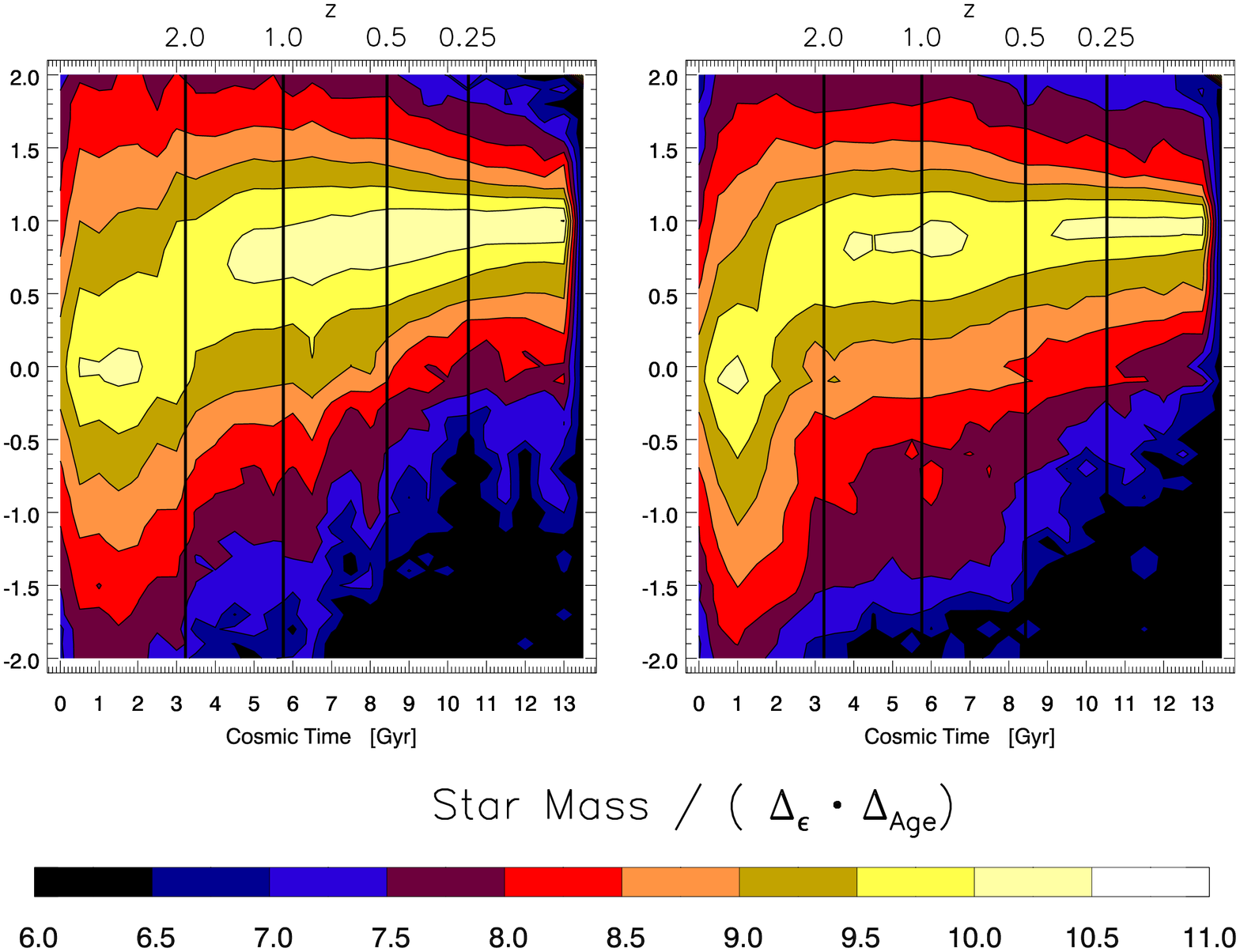}
}
\caption{
  Maps of circularity {\it vs} formation time of star particles that
  at $z=0$ belong to AqC5 (left panel)
  and GA2 (right panel). Each pixel of the map represents
  the average star formation in the corresponding interval of
  circularity. Formation time is binned in intervals of $\Delta t=0.5$
  Gyr and circularity in intervals of $0.1$.
  }
\label{fig:circ-age}
\end{figure*}

Figures \ref{fig:MapGA2evol} and \ref{fig:MapAqC5evol} show the
density of gas and stars of our simulated galaxies, in face-on and
edge-on projections, at redshifts $z=2.48$, $2.02$, $1.50$, $1.01$ and
$0.49$.  We always align the z-axis of our coordinate systems to the
angular momentum vector, evaluated in the inner 8 kpc.  In both cases,
the inside-out formation of the disk is evident. At the highest
redshift, no disk is visible in the GA2 simulation, while an ongoing
major merger appears at $z=2.02$.  Another minor merger is
perturbing the disc at $z=1.50$.  Then the accretion history becomes
more quiet, with the disk growing in size until the present time.  The
evolution of AqC5 does not show major merger events, and is overall
more quiet.  The accretion pattern of the gas of both galaxies is
quite complex, with filamentary structures directly feeding the
disk. As shown by \cite{Murante12}, gas accreting along these
filaments undergoes a significant thermal processing by stellar
feedback before it can reach the disc, thereby determining the
relative amounts of accreted gas in cold and in warm flows.

Figure \ref{fig:circ-age} shows 2D histograms of circularities and
cosmic time of formation of the star particles belonging to the two
galaxies at $z=0$.  
Colours correspond to stellar mass densities in bins of given
circularity and age.  The corresponding histograms of circularity and
radius are shown and commented in Goz et al. (2014).  For both
simulations, most stars belonging to the bulge and halo components
(circularity values around zero) form at high redshift, before $z
\approx 2$, corresponding to the first phase of star formation in
Figure \ref{fig:sfr}.  Stars that belong to the disc component 
(circularity values around unity) form at lower redshift.  These
maps illustrate that the formation history of these spiral galaxies
can be broadly  divided into two events: the formation of bulge and halo
components and the formation of the disc, separated by a relatively
quiet period.  This scenario is in line with the double infall
scenario of \cite{Chiappini97} for the formation of the Milky Way. A
similar result was obtained by \cite{Cook09}, who used a cosmological
Semi-Analitical Model of galaxy formation and modelled a two-phases
formation of disk galaxies.
Note that the force resolution of our simulations does not allow
  us to safely investigate the fine structure of the disk, like
  separating disk stars into a thin and a thick component. Therefore, a more
  detailed comparison with the result of galactic archeology models is
  not feasible at this stage.

\begin{figure*}
\centerline{
\includegraphics[scale=0.25]{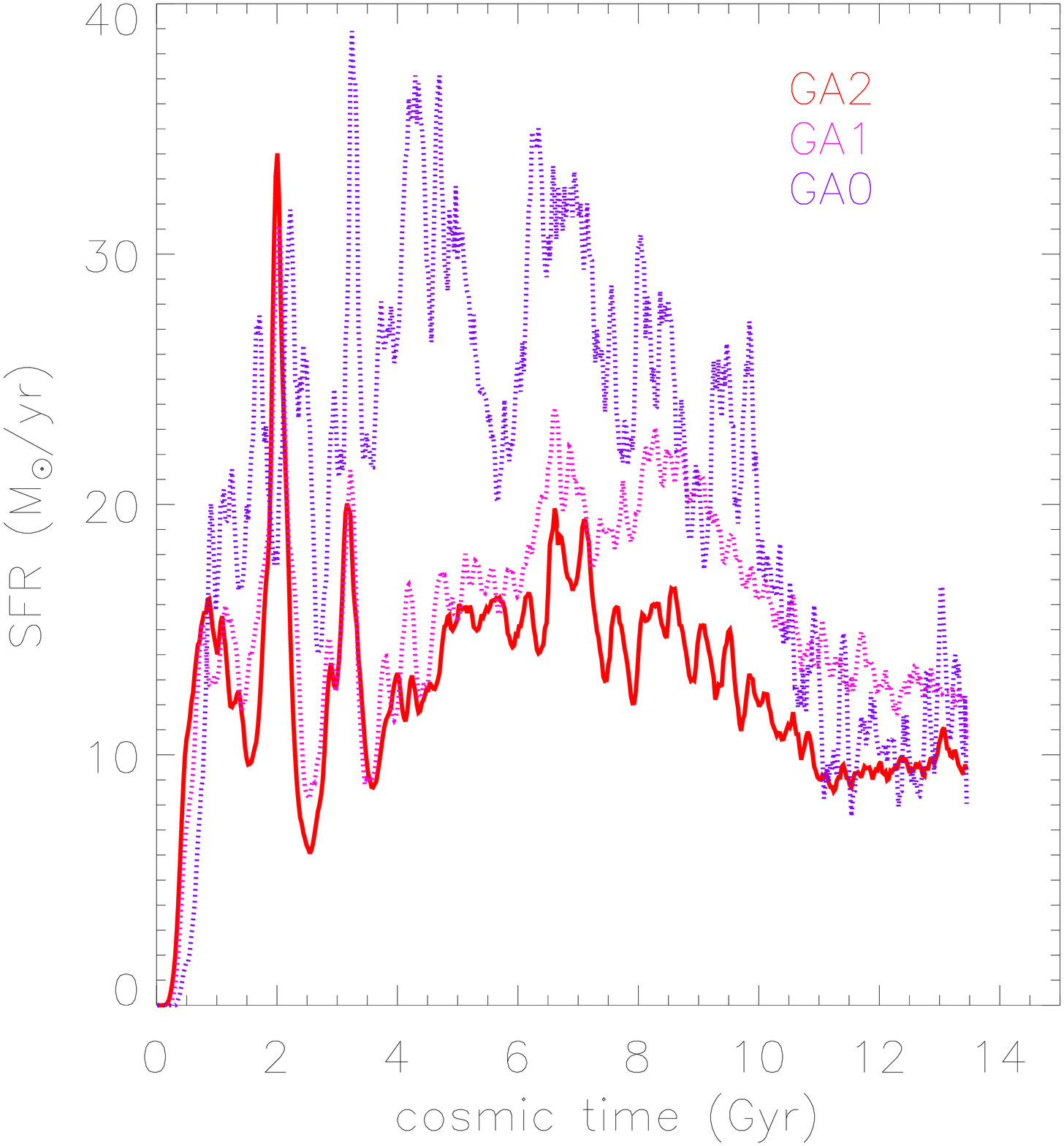}
\includegraphics[scale=0.25]{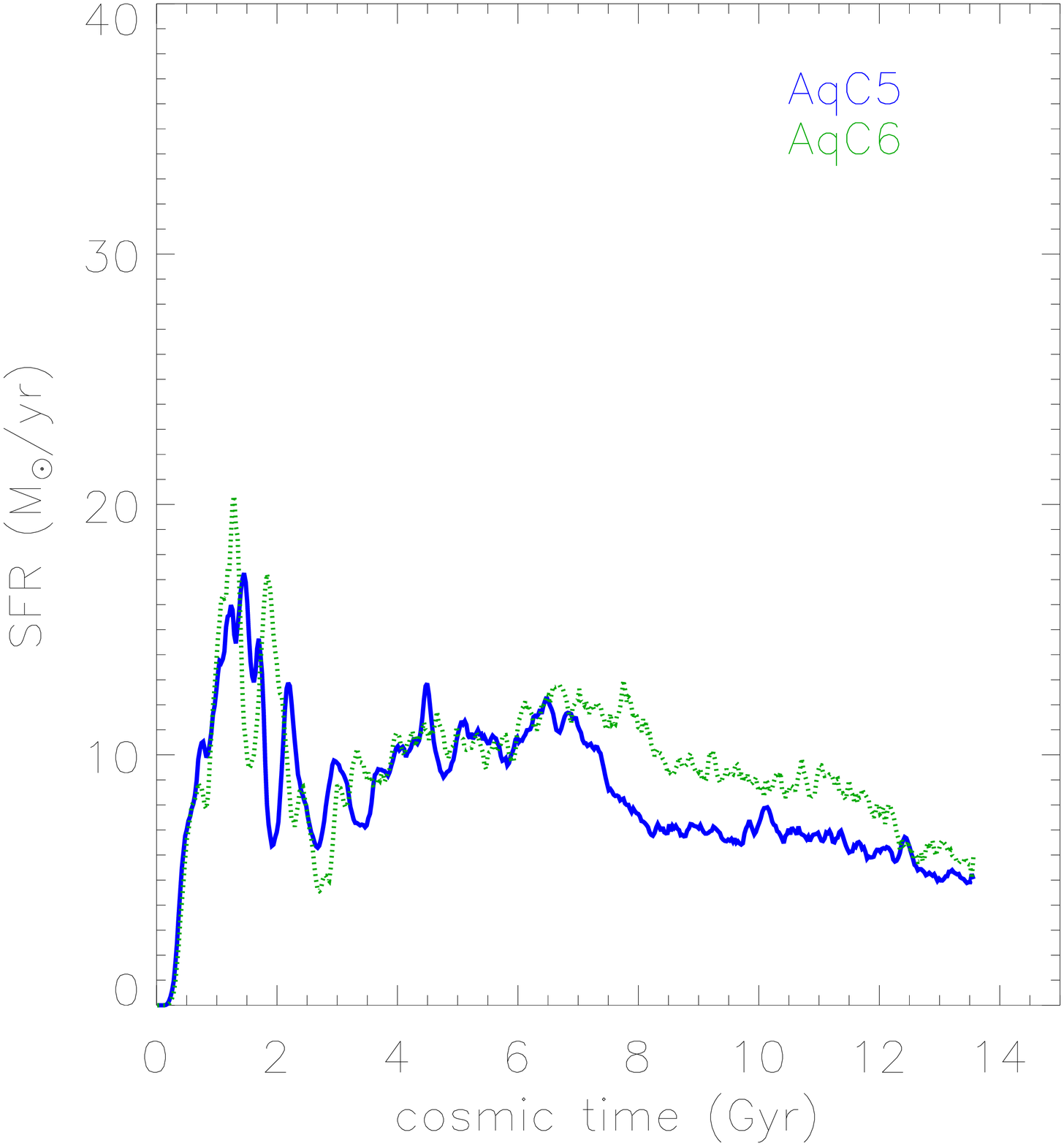}
}
\caption{
Star formation rate as a function of the cosmic time for GA (left
panel) and Aq (right panel) simulations at various resolutions. As for the GA series, purple
dashed line refers to GA0, pink dotted line to GA1 and red solid line
to GA2. In the right panel we show AqC6 (green dotted line) and AqC5
(blue solid line).
}
\label{fig:SFRresol}
\end{figure*}

\begin{figure*}
\centerline{
\includegraphics[scale=0.25]{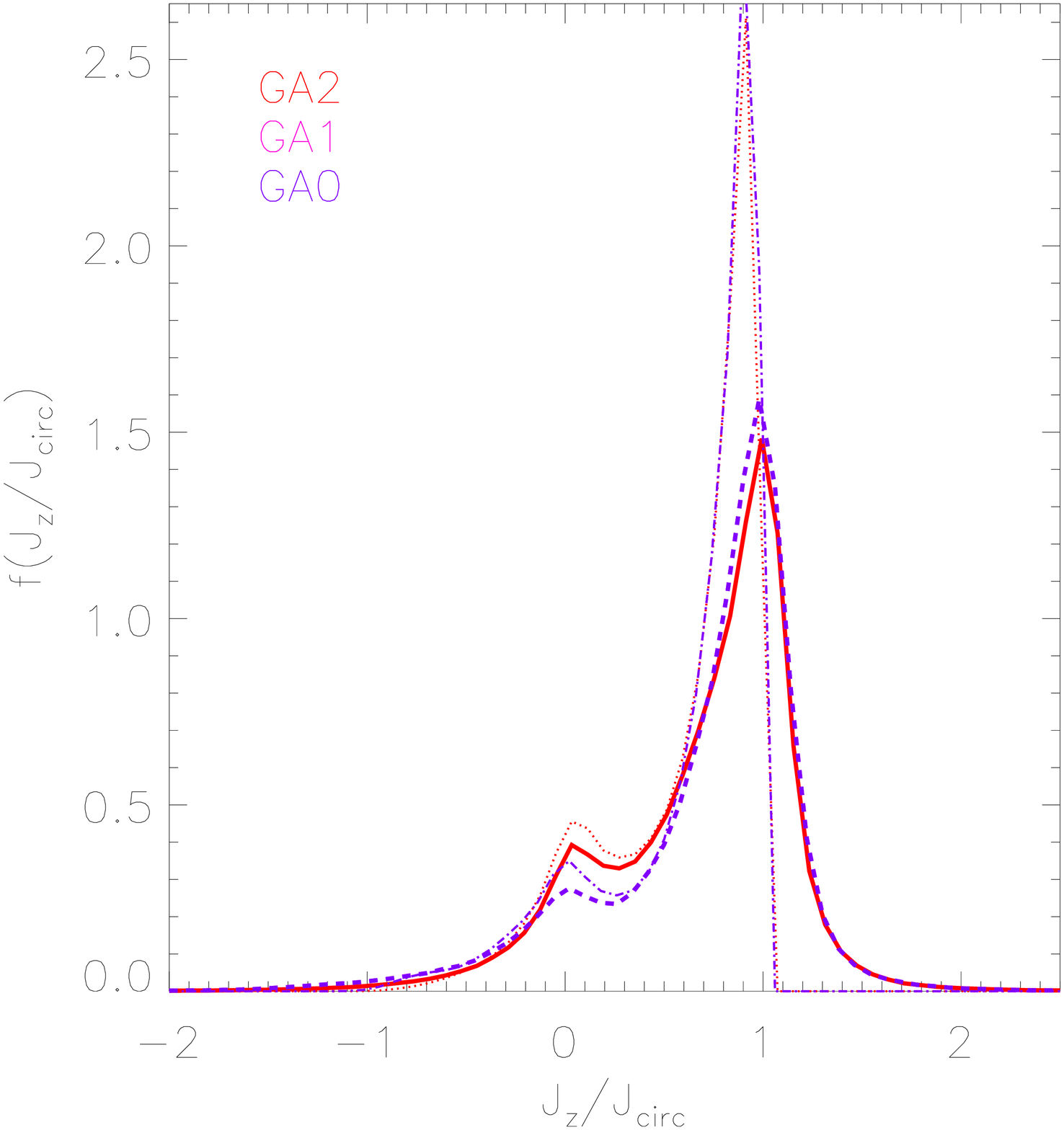}
\includegraphics[scale=0.25]{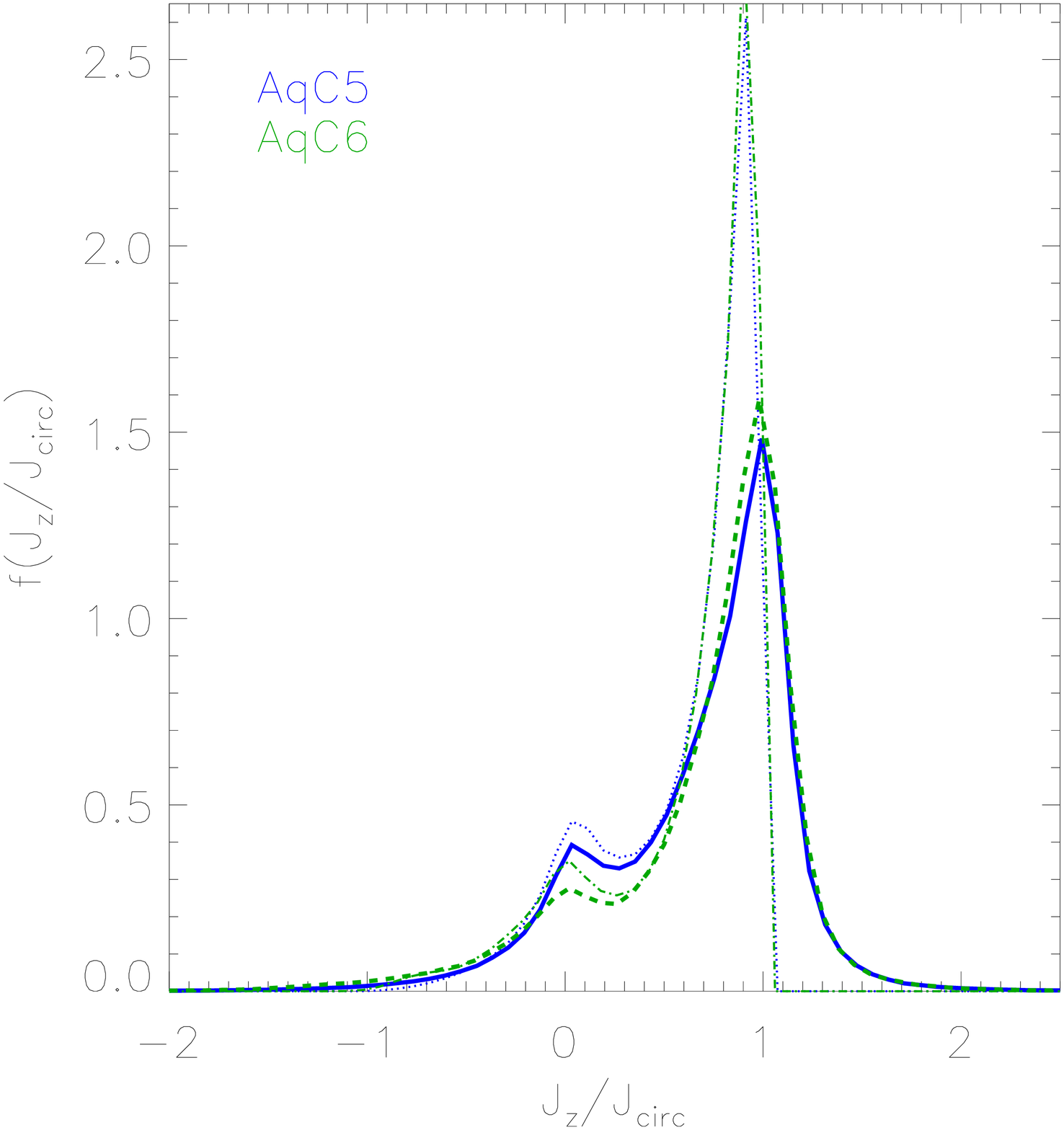}
}
\caption{
Effect of resolution on the circularity histograms. In the left panel
we show the fraction of stellar mass as a function of the circularity
for GA0 (purple dotted line), GA1 (pink dashed line)  and
GA2 (red solid line). In the right panel we show the same for AqC6
(green dotted line) and AqC5 (blue solid line).
}
\label{fig:jhistresol}
\end{figure*}

\begin{figure*}
\centerline{
\includegraphics[scale=0.25]{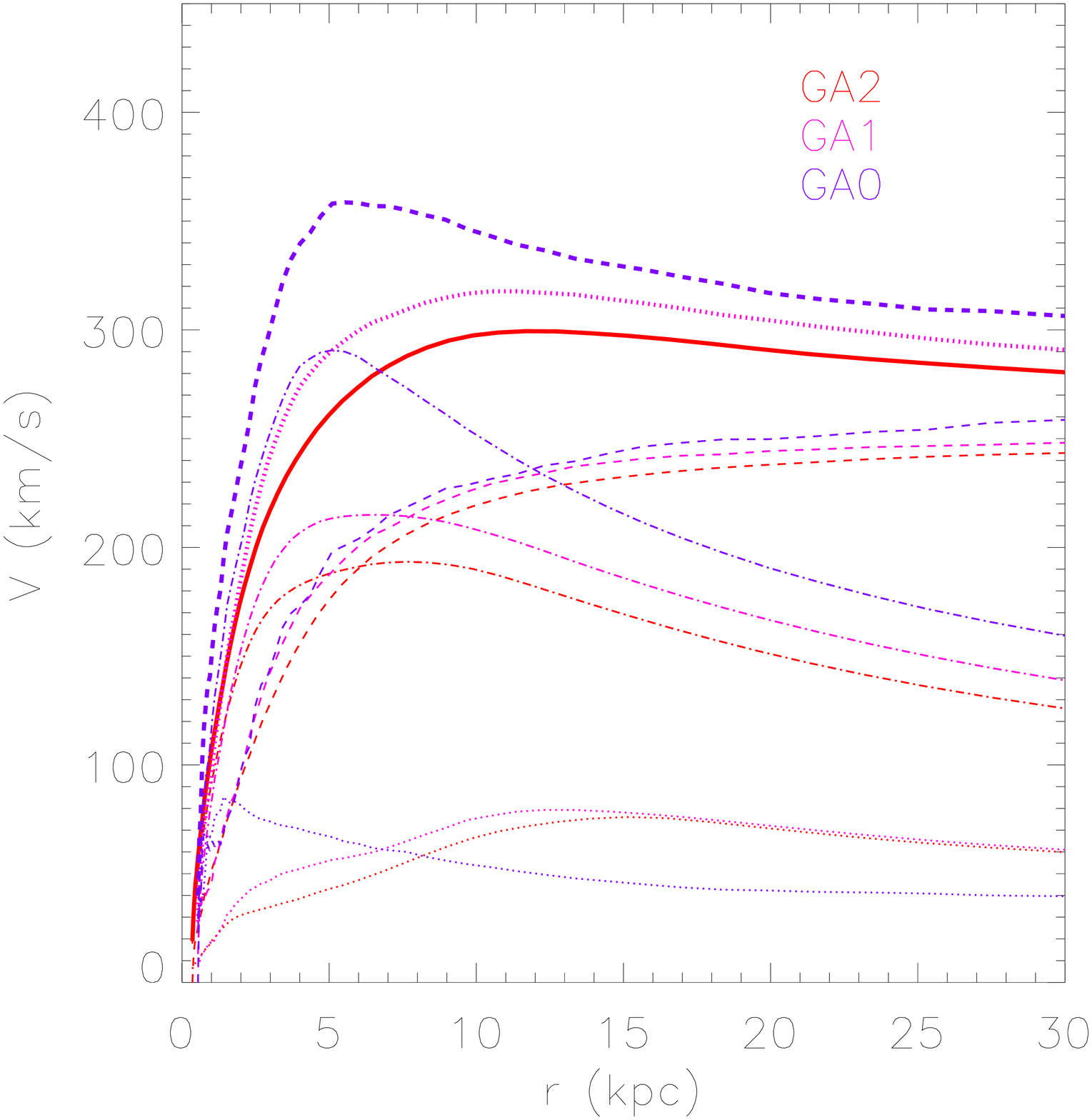}
\includegraphics[scale=0.25]{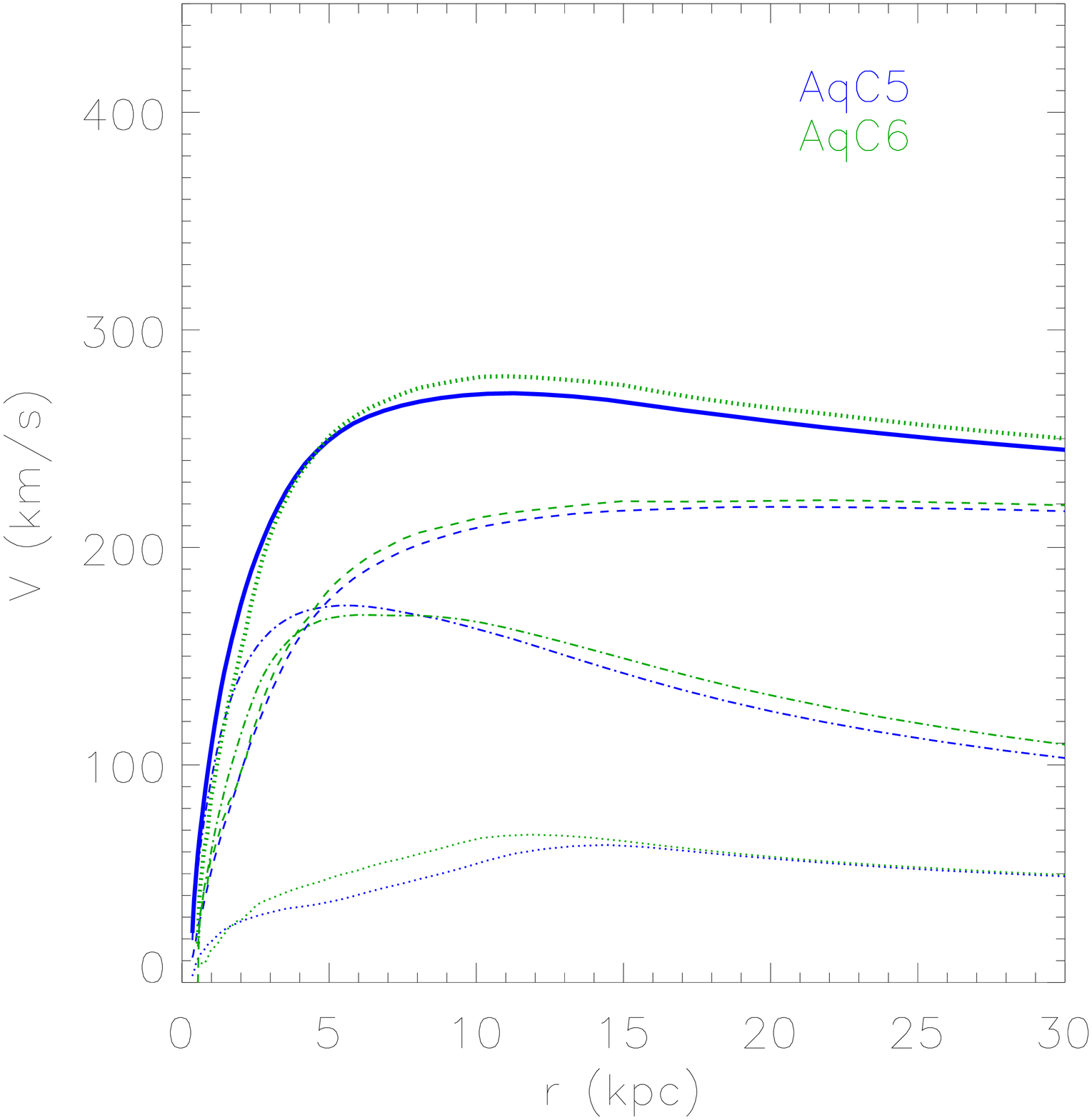}
}
\caption{
  Effect of resolution on the velocity rotation curve. In the left panel
we show the velocity rotation curve
for simulations GA0 (violet dotted line), GA1 (pink dashed line)  and
GA2 (red solid line). In the right panel we show the same for AqC6
(green dotted line) and AqC5 (blue solid line).
}
\label{fig:Vrotresol}
\end{figure*}

\subsection{Effect of resolution}
\label{section:resolution}
In this section, we show how our main results change by varying the
mass resolution by a factor $\approx 10$ in the Aq series, and by a
factor up to $\approx 100$ in the GA series.

The multi-phase model for star formation and feedback that we
implemented depends the physical properties of the gas in the
simulation.  These properties will in general depend on resolution. As
a consequence, the resulting feedback efficiency will in turn show a
degree of resolution dependence.  This was already discussed in M10,
based on simulations of idealised spiral galaxies.  In cosmological
simulations, increasing the resolution does not only imply resolving
smaller masses and scales. Initial conditions themselves change as the
resolution varies, since more small-scale power is added. Therefore,
an exact convergence when resolution is changed is not expected.  We
also point out that our sub-resolution model has a range of validity,
outside which its results are not necessarily reliable.  In fact, to
properly represent the average density and temperature of gas, we
cannot use force resolution that is much larger than the disk height.
Because the temperature of the disc is relatively high, a force
resolution of several 100 pc is sufficient to properly represent
density and pressure. However, when softening increases to $>1$ kpc,
density and pressure in a gaseous disk are no longer resolved.  As a
consequence, we find that the feedback is less efficient. On the other
hand, the physical motivation of the model breaks down when the
resolution is so accurate to properly resolve single molecular clouds.
In this case, a different physical approach is required and additional
physical processes must be modelled in order to appropriately
describe, at these scales, star formation and SNe feedback.

This said, it is quite remarkable that the results of our model are
relatively insensitive to force and mass resolution.  In
Figure~\ref{fig:SFRresol} we show the star formation rate as a
function of the cosmic time for the GA and Aq simulation series. As in
Figure~\ref{fig:sfr}, we use only stars within the galactic radius to
evaluate the SFR. The GA0 simulation, our lower resolution case, has a
remarkably different SFR from GA1 and GA2. On the other hand, GA1 and
GA2 produce very similar SFRs for the first seven Gyrs; later, GA1
shows a SFR that is higher by $\approx 40$ per cent at worst, at 9
Gyr, and by less than $30$ per cent at the final epoch.  AqC6 and AqC5
show a much better convergence in the SFR, that is almost equal at the
end, and differ by less than $\approx 30$ per cent between $t=7$ and
$t=12$ Gyr.  This is likely to be an effect of the quieter merger
history of this halo. We note in general that the SFR is higher when
resolution is decreased. In fact, when higher densities
and pressures are resolved, feedback is more effective. The effect
of enhanced feedback efficiency at higher resolution is to
over-compensate the increased amount of available gas, thus resulting
in a lower star formation.

In Figure \ref{fig:jhistresol} we compare the circularity histograms of
our simulated galaxies at $z=0$. The resolution convergence of the GA
and Aq galaxies is striking. Only the lowest resolution GA0 shows
a very different circularity distribution, with no disk component.
Based on this test, we can conclude that already at the resolution of
the GA1 simulation, our model is able to produce realistic disk
galaxies.
From Table \ref{table:baryons_z0}, it is clear that almost all galaxy
properties are very similar between the two resolutions (GA1/AqC6 and
GA2/AqC5), with the Aq serie showing better convergence. In
particular, the mass of the disk, bulge and gas in these two
simulations do not differ by more than $\sim 20$ per cent, while the
halo baryon fraction and the fraction of baryons in the galaxy stays
within $15$ per cent.  The bulge mass of the GA series changes more,
but this difference is due to the different action of the stellar bar,
which forms at different epochs. In fact, changing the resolution
also the timing of mergers and fly-bys may change, thereby changing the
epoch of formation of a tidally triggered bar.

In Figure \ref{fig:Vrotresol} we compare the rotation curves of the
two simulation sets, at different resolutions. Also in this case,
the Aq series show results that are almost independent of 
resolution. GA0 shows the largest difference with respect to the other
GA curves, with a too strong mass concentration at the centre.  GA1
and GA2 show rotation curves that never differ by more than $10$ per cent: at
the peak, the difference between the two curves is about $20$ km
s$^{-1}$, while it decreases to $10$ km s$^{-1}$ at 30 kpc.

Given the stability of stellar masses and rotation curves with
resolution (except for the case of GA0), also the positions of our
simulated galaxies in the Tully-Fisher plot and in the stellar {\it
  vs.} halo mass results to be quite insensitive to numerical
resolution.

\begin{figure*}
\centerline{
\includegraphics[scale=0.2]{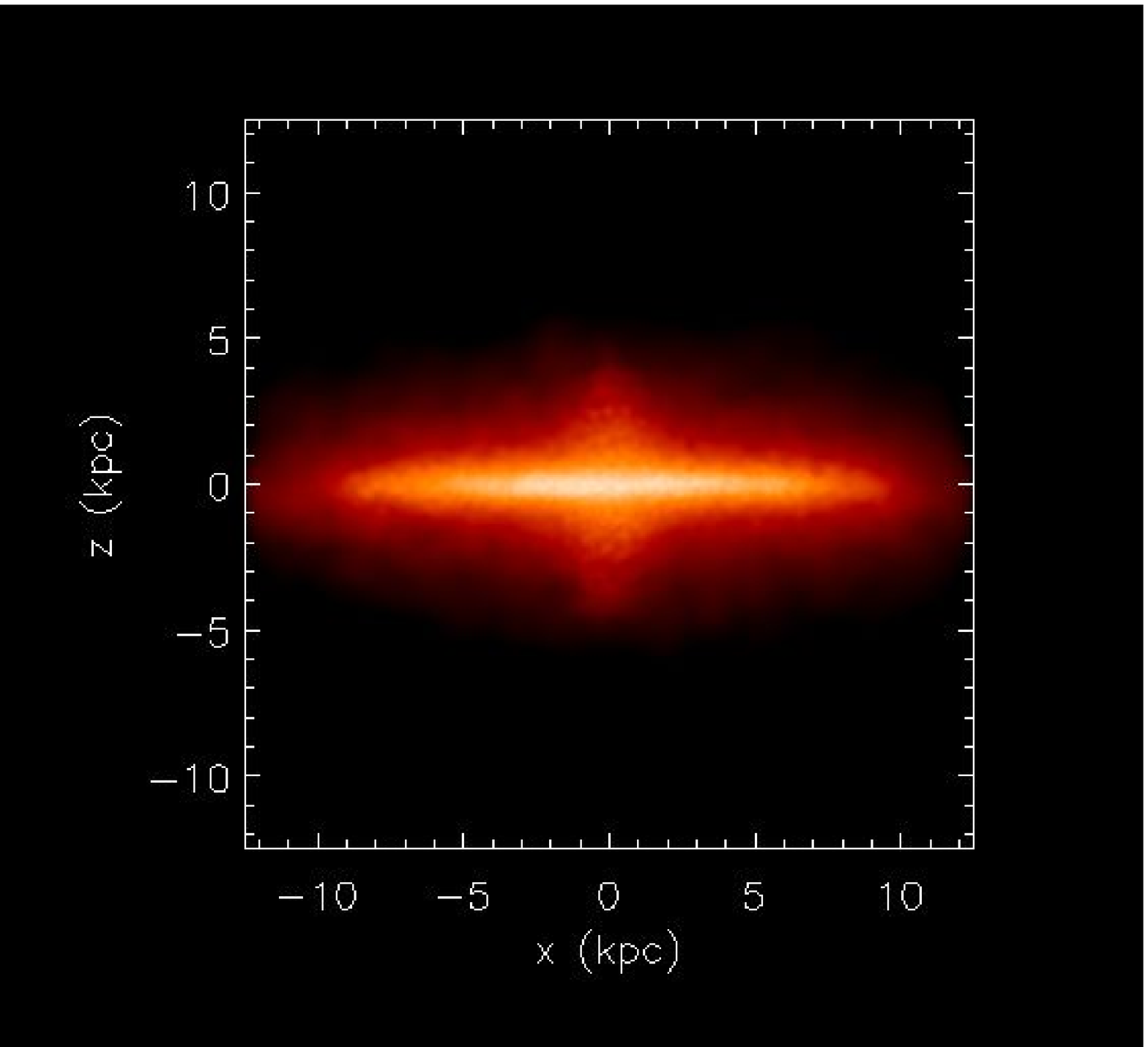}
\includegraphics[scale=0.2]{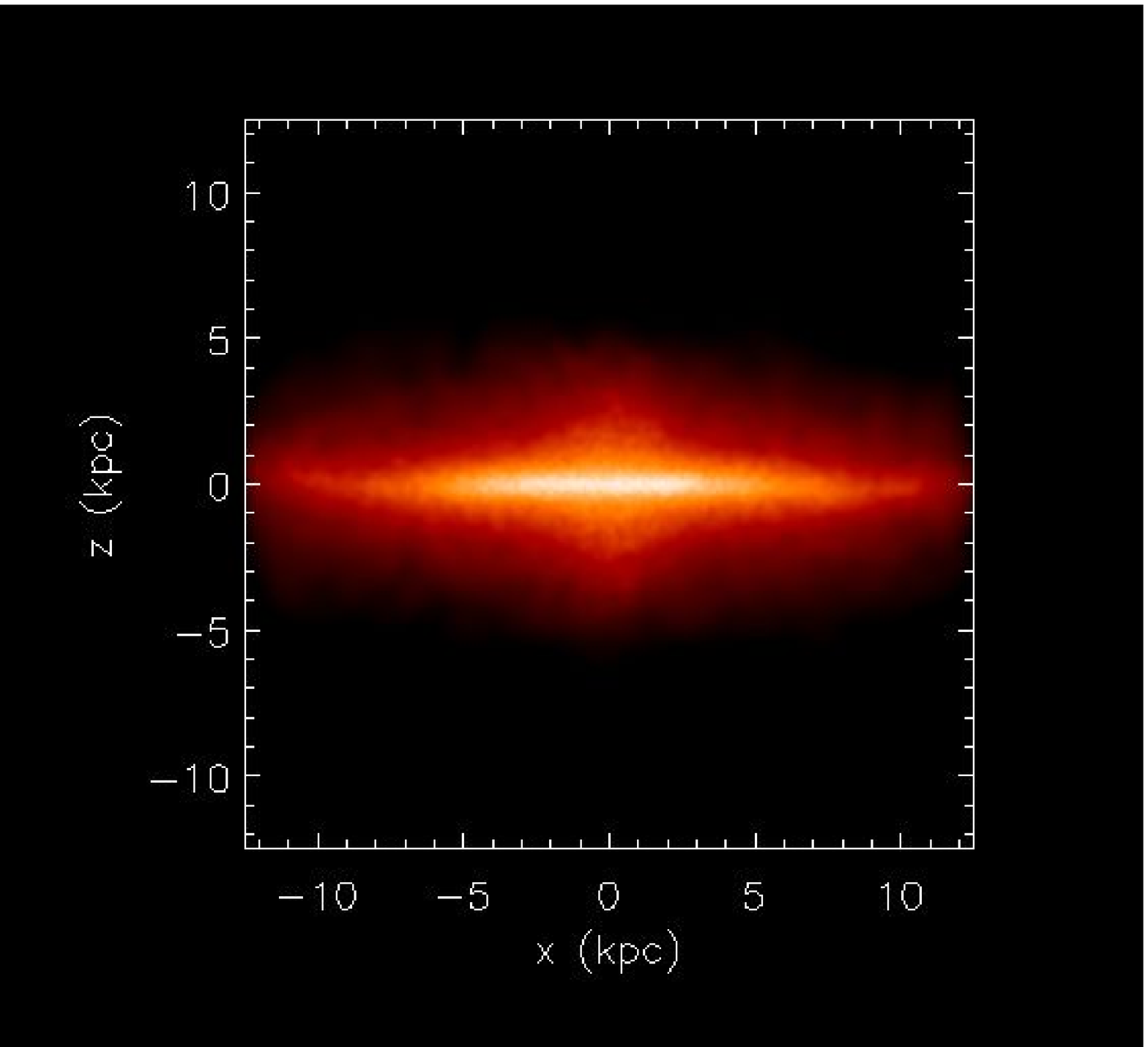}
\includegraphics[scale=0.2]{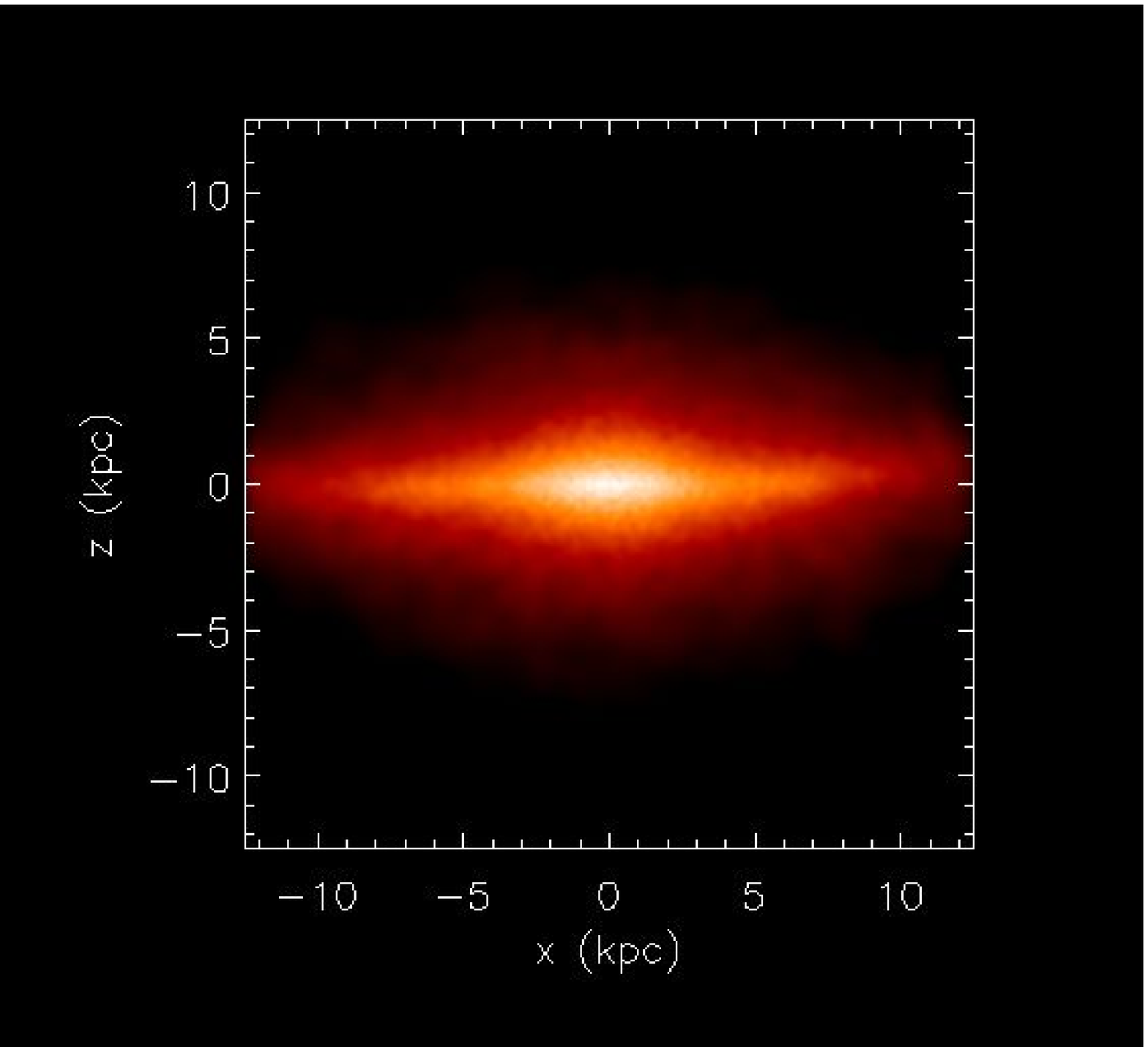}
\includegraphics[scale=0.2]{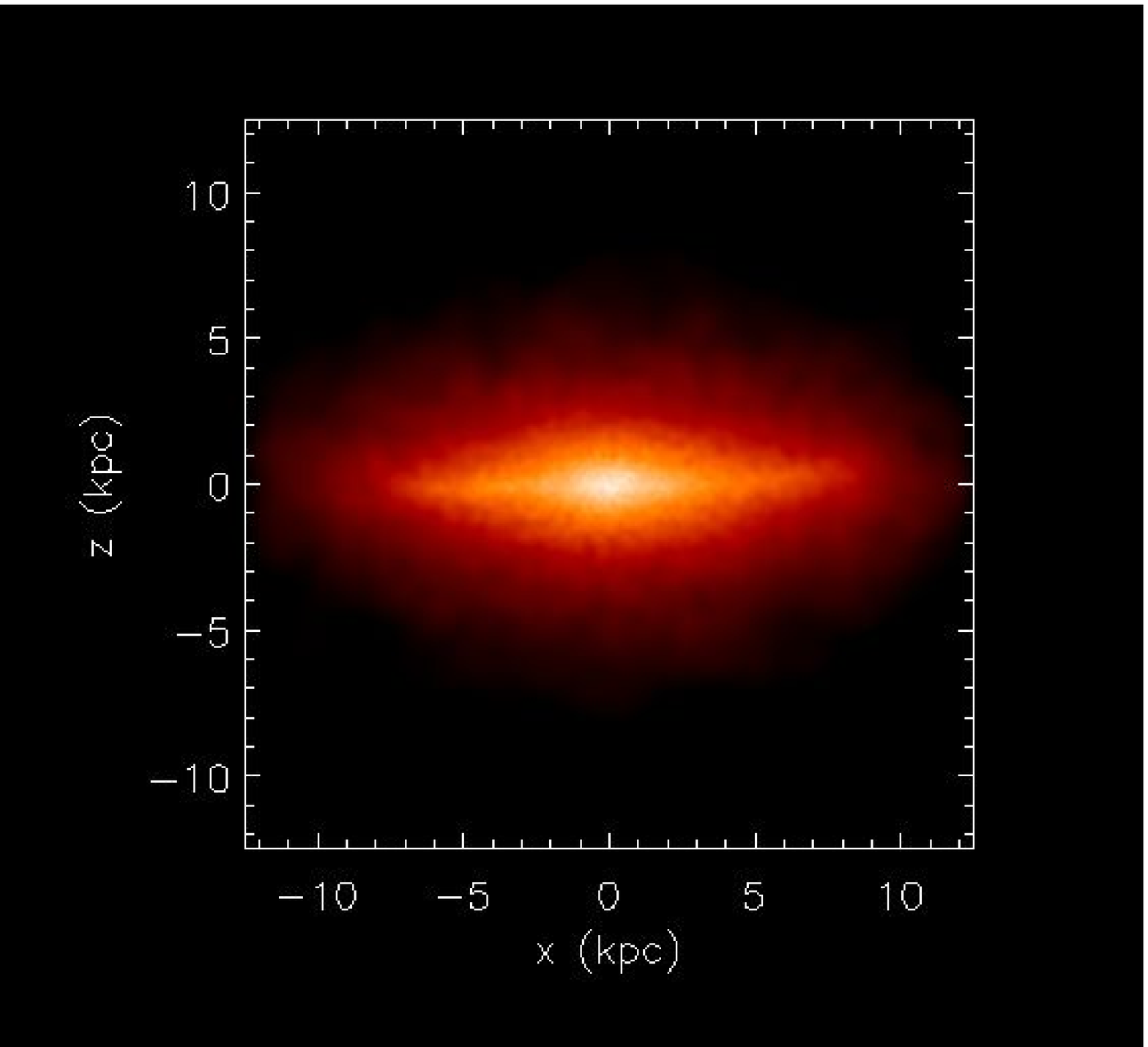}
}
\caption{Edge-on view of gas density maps for different values of the
  Plummer-equivalent gravitational softening: $\epsilon_{Pl}=0.1$,
  $0.3$, $0.65, 1.3$ kpc/h (from left to right panels).  }
\label{fig:softMaps}
\end{figure*}

\begin{figure}
\centerline{
\includegraphics[scale=0.2]{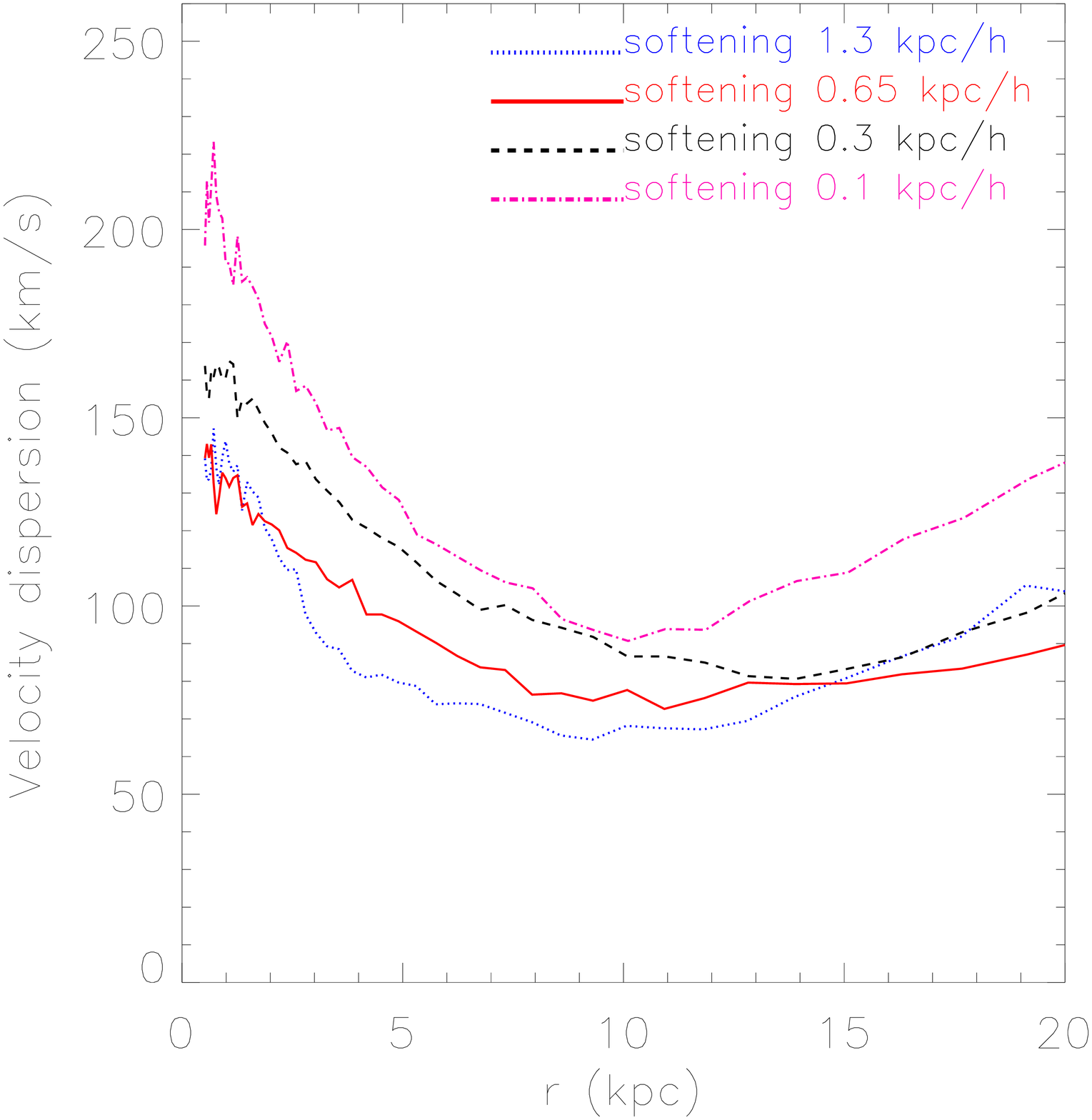}
}
\caption{ Stellar velocity dispersions along the vertical direction
  for simulated GA1 galaxies when the softening is changed from $0.1$
  to $1.3$ h$^{-1}$ kpc, all the other details of the simulation being
  kept fixed.}
\label{fig:softveldisp}
\end{figure}

Finally, we studied the dependence of our result on the force
resolution. To this aim, we repeated the GA1 simulation by changing
only the softening scale for the gravitational force. We used
$\epsilon_{Pl}=0.1$, $0.3$, $0.65$ (our standard value) and $1.3$
kpc/h (runs sA, sB, sC, SD), for high resolution DM, gas and star
particles.  In Figure~\ref{fig:softMaps} we show edge-on stellar
density maps of the corresponding four simulations. Quite remarkably,
our reference force resolution sC gives the best defined disk
structure. A larger softening produces a well defined disk, but its
scale length is significantly smaller. Decreasing the softening has
the effect of heating up the disk, due to increased numerical heating.
The disk structure is almost lost at our smaller softening value. 
A more quantitative description of this trend is given in
  Figure~\ref{fig:softveldisp}, where we show radial profiles of
  velocity dispersions of stars in the vertical direction.  These
  dispersions are computed as the {\it root means square} of the
  $z$-components of stellar velocities in the galaxy reference frame.
  In the companion paper by Goz {\it et al.} (2014) we show that, due
  to the existence of broad tails in the velocity distribution, this
  quantity overestimates the velocity dispersion of disk stars. It is
  presented here not to quantify the thickness of the disk, but to
  show how this quantity changes with the softening.  There is a clear
  trend of increasing velocity dispersion with decreasing softening,
  and this illustrates well how the thickness of the stellar disk is
  affected by numerical effects, most importantly by 2-body heating.
  The difference among our reference choice of 0.65 kpc/h and a
  softening twice as large is relatively small ($\sim$20 per cent),
  thus indicating that only some residual numerical heating is present
  at our standard force resolution.

\begin{figure}
\centerline{
\includegraphics[scale=0.25]{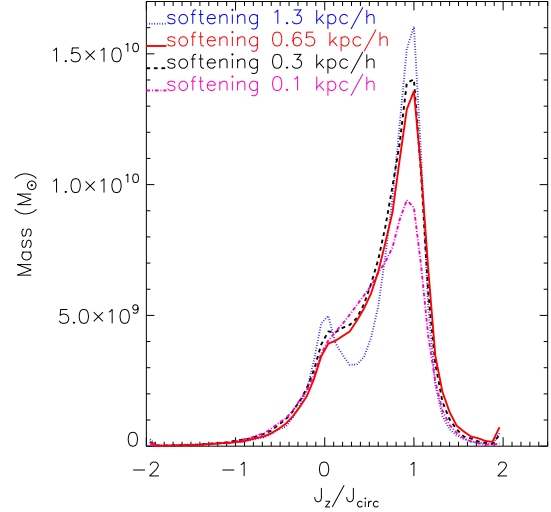}
}
\caption{ Effect of changing the softening, at fixed mass resolution,
  on the circularity histograms.  We show the stellar mass as a
  function of the circularity for GA1 simulations when we use a
  Plummer-equivalent softening length $\epsilon_{Pl}=0.1$
  (dotted-dashed magenta line), $0.3$ (dashed black line), $0.65$
  (continuous red line), $1.3$ (dotted blue line) kpc/h.}
\label{fig:softCirc}
\end{figure}

This visual impression is quantitatively confirmed by the analysis of
the stellar circularities, shown in Figure~\ref{fig:softCirc}.  The
simulation run with our reference value for the softening, and that
run with about one half of such a value, have similar circularity
histograms with a well defined disk component and a small bulge. 
The smallest softening produces a drastic change in  galaxy
properties, with a strong decrease in the disk component.  We
therefore caution about the use of an aggressive softening in
simulations of galaxy formation: numerical heating can damage the disk
structure to an amount that exceeds the advantage of a higher force
resolution.

\section{Conclusions}
\label{section:conclusions}

In this paper we presented results of simulations of disk galaxies,
carried out with the GADGET-3 code where we implemented our
sub-resolution model for star formation and feedback {\it MUPPI}
(MUlti Phase Particle Integrator). In our model, differential
equations describing the evolution a multi-phase system, that
describes the ISM at unresolved scales, are integrated for each
multi-phase gas particle.  Because cooling times are computed using
the density of the local hot phase, thermal energy from SNe is not
quickly radiated away. Moreover, no assumption of self-regulation of
the multi-phase system is made. As a result, with {\muppi} gas
particles can hydrodynamically respond to the injection of thermal
energy from SNe.  With respect to the first version of the code
described in \cite{Murante12}, we included the chemical evolution
model by \cite{Tornatore07}, metal cooling as in \cite{Wiersma09a},
and a prescription for kinetic feedback.

Our simulations started from zoomed-in cosmological initial conditions
of two DM haloes, GA and AqC, at different resolutions. The AqC halo
was also simulated by several other groups
\citep[e.g.][]{Scannapieco09,Scannapieco12,Aumer13,Marinacci13}.  In
our highest resolution run, we use a Plummer-equivalent softening of
$\epsilon_{\rm Pl}=0.325$ h$^{-1}$ kpc (fixed in physical coordinates
since $z=6$), of the same order of magnitude of a typical galaxy disk
scale height; our simulation can therefore be described as ``moderate
resolution'' ones.

Our main results can be summarised as follows.

\begin{itemize} 

\item We obtain disk galaxies with a bulge-over-total stellar mass
  kinematic ratio $B/T \approx 0.2$, typical of late-type
  galaxies. The rotation curves are gently rising for 1-2 disk scale
  radii, and then remain flat. Our simulated galaxies lie within the
  scatter of the observed Tully-Fisher relation, although
    on the high side of the allowed range of stellar masses. Our baryon conversion
    efficiencies are consistently slightly too large, being however
    within $3 \sigma$ from the \cite{Moster10} relation, and in better
    agreement with that of \cite{Guo10} (based on the same cosmology
    used for AqC5).  The fraction of stellar mass in the central
  galaxy over the total halo mass is of the order of 5 per cent, with
  a baryon fraction within the virial radius of $0.1$. Stellar density
  profiles are exponential, with a scale radius of $4.45$ kpc (GA2)
  and $3.45$ kpc (AqC5). Therefore we consider our simulated galaxies
  to be ``realistic''.

\item Our model also provides predictions for several sub-resolution
  ISM properties: e.g., the amount of molecular gas, and of hot and
  cold gas.  The profiles of such quantities, as well as the total
  surface density profile of the gas, are rather flat within the inner
  10 kpc. Also metallicity profiles are relatively flat. Our disks are
  gas-rich: the mass of gas is about 30 per cent of the total baryonic
  disk mass. We obtain a good fit to the observed Schmidt-Kennicutt
  relation, that is not imposed in the star formation model.

\item The evolution of our simulated galaxies is similar to that
  reported by other authors \citep[e.g.][]{Marinacci13,Aumer13}. The
  bulge component forms first, while the disk starts to form after
  $z=2$ and begins to dominate at $z\sim1$, showing a clear inside-out
  formation.  The star formation rate peaks at a cosmic time of about
  2 Gyr, then has a minimum and reaches a new maximum and a mildly
  decreasing rate. At $z=0$ it reaches values of of 5-10 M$_\odot$
  yr$^{-1}$, well within the main sequence of star-forming galaxies.
  The overall behaviour of bulge formation followed by a minimum of
  the SFR history and then by the gradual formation of the disc is in
  line with the two-infall model for the formation of the Milky Way
  \citep{Chiappini97,Cook09}.

\item Our results are rather stable as resolution is decreased by a
  factor of eight. In particular, morphology-related quantities, such
  as rotation curves and circularity histograms, vary by less than 10
  per cent. Also the SFR varies approximately by the same amount.  At
  lower resolutions, convergence is lost.We note that reducing the
  softening by a factor six (at fixed mass resolution) induces effects
  related to numerical heating, thereby severely changing the disk
  properties.  Also, doubling the softening parameter results in a
  thicker and less extended disk and increases the bulge mass.

\end{itemize}

It is interesting to compare our results for the AqC5 case with the
works by \cite{Scannapieco09}, \cite{Scannapieco12}, \cite{Aumer13}
and \cite{Marinacci13}, who simulated the same halo starting from the
same initial conditions.  \cite{Scannapieco09} found that their
simulated AqC5 was bulge-dominated: they reported a disk-over-total
stellar fraction of 0.21. In the Aquila comparison project
\citep{Scannapieco12}, using the same IC, several group obtained a
wide range of galactic properties.  An earlier version of MUPPI was
included in the comparison and also produced a bulge-dominated galaxy
at redshift $z=0$.  \cite{Aumer13} further developed the {\sffb} model
of \cite{Scannapieco09} by including kinetic feedback and the effect
of radiation pressure by young massive stars.  With these
implementations, they obtained a disk-dominated galaxy with a flat
rotation curve. The circularity diagram that they reported is very
similar to that shown here in Fig. \ref{fig:jhist} However, the baryon
conversion efficiency of our simulation is higher than that obtained
by \cite{Aumer13}.  With the present choice of feedback parameters,
the agreement between our results and theirs does not extend to the
stellar mass, and, as a consequence, to the position of the AqC5
simulations on the Tully-Fisher relation.

Using AREPO \citep{AREPO}, \cite{Marinacci13} showed a similar result
on their simulated AqC5.  In their case, the bulge component is
slightly more prominent than in our simulations, but also their
simulation produced a galaxy which is clearly disk-like and has a
dimension similar to what we obtain: their disk scale-length is
$R_d=3.11$ kpc and the disk mass is $6.07 \cdot 10^{10}$ M$_\odot$,
while our AqC5 has $R_d=3.42$ kpc and a disk mass of $5.66 \cdot
10^{10}$ M$_\odot$. The agreement between results obtained in this
work and those presented in \cite{Marinacci13} extends to the position
on the simulated galaxy in the TF relation and in the stellar {\it vs}
halo mass one, the shape and normalisation of the rotation curve, and
the stability with the resolution. It is worth noticing that the
{\sffb} models of the three works are quite different, and
\cite{Marinacci13} one also includes black hole feedback.

It is worth pointing out that our results have been obtained at
moderate resolution, without resorting to early stellar feedback, to a
high threshold for star formation or to a delayed cooling.  The ERIS
simulation \citep{Guedes11} also has realistic properties, in the
sense defined above. However, their mass resolution is more than 200
times better than ours for the AqC6 case, that is already
``realistic''.
Therefore, the possibility of obtaining realistic disk galaxies at
moderate resolution and the remarkable numerical convergence of our
results open the perspective of simulating large volumes of the
universe, containing a representative population of galaxies, with a
high, but affordable, computational cost. This is the natural
  prosecution of the current work and will be the subject of
  forthcoming papers. At the same time, we must be
aware that our simulations still tend to produce stellar disks that
are still somewhat too massive. This suggests that we may need to
introduce in our simulations other feedback sources which are able to
further regulate star formation. In this respect, AGN feedback
resulting from gas accretion onto super-massive BHs appears as an
obvious missing ingredient.

The agreement of results presented here with those obtained, using the
same IC, by other groups, is promising.  It is interesting that now,
several different models are able to produce realistic disk galaxies,
and to obtain, using the same initial condition, very similar results.
\sffb prescriptions share the ability to suppress star formation at
high redshift, and favour late-time, high angular momentum gas
accretion (see e.g. \cite{Ubler14}). On the other hand, this also
  means that we are still unable to precisely state which 
  {\it ``micro-physical''} mechanisms are relevant in galaxy
  formation and responsible for producing simulated properties in
  agreement with observations. In fact, the various {\sffb}
  prescriptions used in literarature are {\it not} simply different
  ways to implement the {\it same} unresolved processes, but
    include significantly different combinations and implementations
    of sub-resolution physics.

However, modern simulations of galaxy formation start being predictive enough that future
observations, for instance on the circumgalactic medium, will possibly
be able to distinguish among them and, indirectly, tell us what
physical processes are needed to form galaxy disks and what other
processes are not necessarily involved.

As a final remark, we want to stress again that it is currently not
possible to simulate observational properties of galaxies, embedded in
a cosmological context, from first physical principles. Therefore,
there is no a-priori reason to prefer any sub-resolution prescription
upon any other, or a given value for a model parameter, e.g. the
density threshold for the onset of star formation. Sub-resolution
models can only be judged on the basis of being physically
motivated\footnote{This is not even a stringent requirement. A
  hypothetical cellular automata, physically unmotivated but able to
  recover the effects of unresolved physics at the resolved scales,
  would perfectly fit our current needs.}  and falsifiable through a
detailed comparison with observational data of ever-increasing
quality.
{\it Codesto solo oggi possiamo dirti, cio'
  che non siamo, cio' che non vogliamo.}\footnote{We can only tell
  this, what we aren't, what we don't want.} \citep{Montale}.

\section*{Acknowledgements}
We would like to thank the anonymous referee for a careful reading of
the manuscript and for constructive comments that helped improving the
presentation of the results.  We are highly indebted with Volker
Springel who provided us with the non-public version of the GADGET-3
code, and with Felix Stoehr who provided the initial condition for the
GA series.  We acknowledge useful discussions with Federico Marinacci,
Gabriella De Lucia and Debora Sijacki.  The simulations were carried
out at the ``Centro Interuniversitario del Nord-Est per il Calcolo
Elettronico'' (CINECA, Bologna), with CPU time assigned under
University-of-Trieste/CINECA and ISCRA grants, and at the CASPUR
computing center with CPU time assigned under two standard grants.
This work is supported by the PRIN MIUR 2010-2011 grant ``The dark
Universe and the cosmic evolution of baryons: from current surveys to
Euclid', by the PRIN-MIUR 2012 grant ``Evolution of Cosmic Baryons'',
by the PRIN-INAF 2012 grant ``The Universe in a Box: Multi-scale
Simulations of Cosmic Structures'', by the INFN ``INDARK'' grant, by
the European Commissions FP7 Marie Curie Initial Training Network
CosmoComp (PITN-GA-2009-238356)', and by a FRA2012 grant of the
University of Trieste. We acknowledge financial support from
``Consorzio per la Fisica'' of Trieste.  KD acknowledge the support by
the DFG Cluster of Excellence 'Origin and structure of the universe'.

\bibliographystyle{hapj}
\bibliography{master}

\label{lastpage}

\end{document}